\documentclass[11pt]{article}
\usepackage{braket}
\usepackage{tikz}		

\usepackage{amsmath,amsfonts,amssymb,graphicx,epsfig,pdflscape,multirow}
\usepackage{arydshln,cite,enumitem}

\usepackage{amsthm}

\usepackage{stackengine}
\newcommand\xrowht[2][0]{\addstackgap[.5\dimexpr#2\relax]{\vphantom{#1}}}

\numberwithin{equation}{section}
\graphicspath{{./eps/}}
\usepackage{cite}
\usepackage{pstricks}
\usepackage{caption}
\usepackage[backref=false]{hyperref}
\usepackage[capitalise,noabbrev]{cleveref} 
\hypersetup{
colorlinks=true,
citecolor=red,
linkcolor=darkblue,
urlcolor=darkblue
}

\long\def\ignore#1{}
\definecolor{darkblue}{rgb}{0,0,.8}
\definecolor{red}{rgb}{1,0,0}
\definecolor{purple}{rgb}{1,0.4,1}
\definecolor{coloroflink}{rgb}{0.7,0,1}
\definecolor{darkpink}{rgb}{.7,0,.7}
\definecolor{pink}{rgb}{1,.7,.7}
\definecolor{lightblue}{rgb}{.61,.61,1}
\definecolor{midblue}{rgb}{.7,.7,1}
\definecolor{lightlightblue}{rgb}{.9,.9,1}
\definecolor{lightestblue}{rgb}{.96,.96,1}
\definecolor{pinot}{rgb}{0.784 0.082 0.031}
\definecolor{lightpurple}{rgb}{1,.65,1}
\definecolor{darkgreen}{rgb}{0.180392, 0.545098, 0.341176}

\newtheoremstyle{smallcaps}{5pt}{5pt}{\itshape}{}{}{}{.5em}
{\scshape\thmname{#1}~\thmnumber{#2}.\thmnote{~\textnormal{(#3)}}}
\theoremstyle{smallcaps}

\numberwithin{equation}{section}

\newcommand{\nc}{\newcommand}
\nc{\bib}{\bibitem}
\nc{\be}{\begin{equation}}
\nc{\ee}{\end{equation}}
\nc{\Mod}{\textrm{\,mod\,}}
\nc{\Tt}{\textrm{t}}	
\nc{\Db}{\mbox{\boldmath $D$}}
\nc{\Ib}{\mbox{\boldmath $I$}}
\nc{\eps}{\epsilon}
\nc{\llangle}{\langle\!\langle}
\nc{\tXX}{\textrm{\tiny$X\!X$}}
\nc{\tCDP}{\textrm{\tiny \it CDP}}
\nc{\ir}{\mathrm{i}}
\nc{\eE}{\mathsf{e}} 
\nc{\dd}{\mathrm{d}}   
\newcommand{\alphaC}{q}

\nc{\notelegant}{1.5pt}
\nc{\elegant}{1.5pt}
\nc{\moyen}{1.0pt}
\nc{\mince}{0.5pt}

\nc{\Tb}{\mbox{\boldmath $T$}}
\nc{\tl}{\mbox{$\mathsf {TL}$}}
\nc{\eptl}{\mathsf{\mathcal EPTL}}
\nc{\stanV}{\mathsf{V}}
\nc{\stanW}{\mathsf{W}}
\nc{\chit}{\protect\raisebox{0.25ex}{$\chi$}}

\def\facegrid#1#2{
\psframe[fillstyle=solid,fillcolor=lightlightblue,linewidth=0pt]#1#2
\psgrid[gridlabels=0pt,subgriddiv=1]#1#2}

\def\loopa{
\psframe[linewidth=.25pt](0,0)(1,1)
\psarc[linewidth=1.25pt,linecolor=blue](1,0){.5}{90}{180}
\psarc[linewidth=1.25pt,linecolor=blue](0,1){.5}{-90}{0}
}
\def\loopb{
\psframe[linewidth=.25pt](0,0)(1,1)
\psarc[linewidth=1.25pt,linecolor=blue](0,0){.5}{0}{90}
\psarc[linewidth=1.25pt,linecolor=blue](1,1){.5}{180}{270}
}

\def\bloopa{
\psarc[linewidth=1.25pt,linecolor=blue](0,0){.5}{-100}{0}
\psarc[linewidth=1.25pt,linecolor=blue](0.833,-1){.5}{80}{180}
}

\def\bloopb{
\psarc[linewidth=1.25pt,linecolor=blue](1,0){.5}{180}{260}
\psarc[linewidth=1.25pt,linecolor=blue](-0.166,-1){.5}{0}{80}
}

\def\cloopa{
\psarc[linewidth=1.25pt,linecolor=blue](0,0){.5}{-80}{0}
\psarc[linewidth=1.25pt,linecolor=blue](1.166,-1){.5}{100}{180}
}

\def\cloopb{
\psarc[linewidth=1.25pt,linecolor=blue](1,0){.5}{180}{280}
\psarc[linewidth=1.25pt,linecolor=blue](0.166,-1){.5}{0}{100}
}

\def \||{|}

\renewcommand{\le}{\leqslant}
\renewcommand{\geq}{\geqslant}
\renewcommand{\leq}{\leqslant}

\usepackage{authblk}

\begin{document}

\topmargin -15mm
\oddsidemargin 05mm

%
%

\title{\mbox{}\vspace{-.2in}
\bf 
\huge Bipartite fidelity for models \\[0.5cm] with periodic boundary conditions
}


\author[1]{\textsc{Alexi Morin-Duchesne}}

\author[2]{\textsc{Gilles Parez}}
\author[2]{\textsc{Jean Li\'enardy}}

\affil[1]{\em Max Planck Institut f\"ur Mathematik, 53111 Bonn, Germany}
\affil[2]{\em Universit\'e catholique de Louvain \\ Institut de Recherche en Math\'ematique et Physique\\ Chemin du Cyclotron 2, 1348 Louvain-la-Neuve, Belgium}

\date{}
\maketitle

\vspace{-1.3cm}


\begin{center}
{\tt alexi.morin.duchesne\,@\,gmail.com}
\qquad
{\tt gilles.parez\,@\,uclouvain.be}
\qquad
{\tt jean.lienardy\,@\,uclouvain.be}
\end{center}
\medskip

%
%
 
\begin{abstract}
For a given statistical model, the bipartite fidelity $\mathcal F$ is computed from
the overlap between the groundstate of a system of size $N$ and the tensor product of the groundstates of the same model defined on two subsystems $A$ and $B$, of respective sizes $N_A$ and $N_B$ with $N = N_A + N_B$. In this paper, we study $\mathcal F$ for critical lattice models in the case where the full system has periodic boundary conditions. We consider two possible choices of boundary conditions for the subsystems $A$ and $B$, namely periodic and open. For these two cases, we derive the conformal field theory prediction for the leading terms in the $1/N$ expansion of $\mathcal F$, in a most general case that corresponds to the insertion of four and five fields, respectively. We provide lattice calculations of $\mathcal F$, both exact and numerical, for two free-fermionic lattice models: the XX spin chain and the model of critical dense polymers. We study the asymptotic behaviour of the lattice results for these two models and find an agreement with the predictions of conformal field theory.
\end{abstract}

%
%
\vspace{.5cm}
\noindent\textbf{Keywords:} Entanglement, bipartite fidelity, quantum spin chains, loop models, conformal field theory.
\newpage

\tableofcontents
\clearpage

\section{Introduction}

Understanding entanglement and correlations in quantum many-body systems is an important challenge in modern theoretical physics. The interest for these questions stems from the understanding that quantifying entanglement, an idea that originated in the information theory community \cite{VPRK97, VP98}, is useful to diagnose phase transitions and describe the critical behaviour of many-body systems \cite{OAFF02, ON02, VLRK03}. Entanglement now plays a prominent role in seemingly unrelated research areas, such as information theory, condensed matter, high energy physics and black holes physics.

Among the numerous existing entanglement measures, the so-called \textit{entanglement entropy} is the most broadly studied one, both in equilibrium situations \cite{HLW94, CC04, AFOV08, ECP08, CCD09, LR10} and out of equilibrium \cite{CC05,FC08,AC17, ABF19}. It is an efficient tool to measure bipartite entanglement in pure states. Consider a system in the pure state~$\ket{\psi}$, composed of two complementary subsystems $A$ and $B$. The entanglement entropy $S_A$ is defined as the \textit{von Neumann entropy} \cite{vN55} of the reduced density matrix of subsystem $A$,
\begin{equation}
S_A= - \text{tr}(\rho_A \log \rho_A), \qquad \rho_A = \text{tr}_B \ \rho, \qquad \rho = |\psi\rangle\langle \psi|,
\end{equation}
where $ \text{tr}_B$ indicates a trace over the degrees of freedom in $B$. The entanglement entropy does not depend on the subsystem that is traced over: $S_A=S_B$. For systems that are not critical, the entanglement entropy satisfies an area law \cite{S93, ECP08}: it is proportional to the area of the boundary between the two subsystems. In particular, for non-critical one-dimensional quantum systems, it saturates to a constant value in the limit of large system size $N \to \infty$. In contrast, for critical one-dimensional quantum systems, the entanglement entropy diverges logarithmically with the system size in the scaling limit: $S_A \propto \log N$. The prefactor is predicted by conformal field theory (CFT) \cite{HLW94, VLRK03, CC04} to be proportional to the central charge $c$, 
\begin{equation}
S_A = \frac{a c}{6} \log N + \mathcal{O}(1),
\end{equation}
where $a$ is the number of contact points between the two subsystems. If the whole system is defined on a periodic lattice, we have $a=2$. On the contrary, if one end of the subsystem $A$ is attached to a boundary, then $a=1$. 

Another observable that shares many features with the entanglement entropy is the \textit{fidelity} \cite{J94,ZP06, ZB08, S10, G10}. It is defined as the overlap between the groundstates of two Hamiltonians that differ by a small perturbation. The Hamiltonian of the system $H(\lambda)$, where $\lambda$ parameterises the perturbation, has the groundstate $\ket{\lambda}$. The fidelity is
\begin{equation}
f(\lambda_1, \lambda_2) = \bigg|\frac{\braket{\lambda_1 | \lambda_2}^2}{\braket{\lambda_1 | \lambda_1}\braket{\lambda_2 | \lambda_2}}\bigg|.
\end{equation}
As a particular example, we consider the situation where $\lambda$ parameterises the interaction between two complementary subsystems, 
\begin{equation}
H(\lambda) = H^A + H^B + \lambda H^{\text{int}},
\end{equation}
where $H^{A}$ and $H^{B}$ are the Hamiltonians of the subsystems $A$ and $B$, respectively. The term $H^{\text{int}}$ contains the interaction between the two subsystems. We denote by $\ket{\psi^A} $, $\ket{\psi^B} $ and $\ket{\psi^{AB}} $ the groundstates of $H^A$, $H^B$ and $H^{AB} = H(1)$, respectively. The \textit{logarithmic bipartite fidelity} \cite{DS11, SD13} is then defined as
\begin{equation}
\label{eq:FABGen}
\mathcal{F}_{A,B} = -\log \left | \frac{\braket{\psi^A \otimes \psi^B | \psi^{AB}}^2}{\braket{\psi^A \otimes \psi^B | \psi^A \otimes \psi^B}\braket{\psi^{AB}| \psi^{AB}}}\right |,
\end{equation}
with the notation $\ket{\psi^A \otimes \psi^B} \equiv \ket{\psi^A}\otimes \ket{\psi^B}$ for the groundstate of $H(0)$. Similarly to the entanglement entropy, this quantity vanishes when the involved groundstates are of the form $\ket{\psi} = \bigotimes_{j=1}^N\ket{\psi_j}$ with $\ket{\psi_j}$ independent of $j$ and $N$. Otherwise it is a positive real number. 

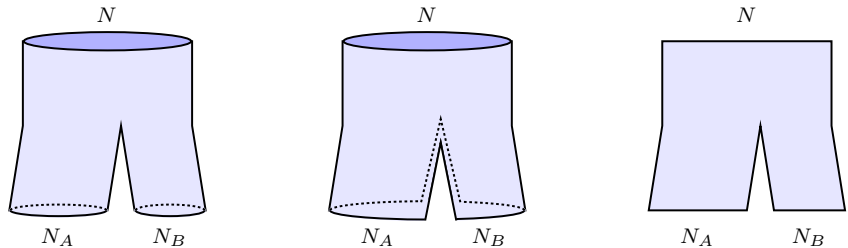
\begin{figure}
\begin{equation*}
\psset{unit=0.9}
\begin{pspicture}[shift=-0.525](0,-0.5)(2.5,3)
\pspolygon[fillstyle=solid,fillcolor=lightlightblue,linecolor=black,linewidth=0.75pt](-0.2,0)(0,1.25)(0,2.5)(2.5,2.5)(2.5,1.25)(2.7,0)(1.65,0)(1.45,1.25)(1.25,0)(0,0)
\psellipse[fillstyle=solid,fillcolor=midblue,linewidth=0.75pt](1.25,2.5)(1.25,0.15)
\psellipse[fillstyle=solid,fillcolor=lightlightblue,linewidth=0,linecolor=lightlightblue](0.525,0)(0.725,0.1)
\psellipse[fillstyle=solid,fillcolor=lightlightblue,linewidth=0,linecolor=lightlightblue](2.175,0)(0.525,0.1)
\psellipticarc[linestyle=dashed,dash=1pt 1pt,linewidth=0.75pt]{-}(0.525,0)(0.725,0.1){0}{180}\psellipticarc[linewidth=0.75pt]{-}(0.525,0)(0.725,0.1){180}{0}
\psellipticarc[linestyle=dashed,dash=1pt 1pt,linewidth=0.75pt]{-}(2.175,0)(0.525,0.1){0}{180}\psellipticarc[linewidth=0.75pt]{-}(2.175,0)(0.525,0.1){180}{0}
\rput(1.25,2.9){$_{N}$}
\rput(0.525,-0.4){$_{N_A}$}
\rput(2.175,-0.4){$_{N_B}$}
\end{pspicture}
\hspace{2cm}
\begin{pspicture}[shift=-0.525](0,-0.5)(2.5,3)
\pspolygon[fillstyle=solid,fillcolor=lightlightblue,linecolor=black,linewidth=0.75pt](-0.2,0)(0,1.25)(0,2.5)(2.5,2.5)(2.5,1.25)(2.7,0)(1.65,0)(1.45,1.0)(1.25,0)(0,0)
\psellipse[fillstyle=solid,fillcolor=midblue,linewidth=0.75pt](1.25,2.5)(1.25,0.15)
\psellipse[fillstyle=solid,fillcolor=lightlightblue,linewidth=0pt,linecolor=lightlightblue](1.25,0)(1.45,0.15)
\psellipticarc[linestyle=dashed,dash=1pt 1pt,linewidth=0.75pt]{-}(1.25,0)(1.45,0.15){0}{13}
\psellipticarc[linestyle=dashed,dash=1pt 1pt,linewidth=0.75pt]{-}(1.25,0)(1.45,0.15){135}{180}
\psellipticarc[linewidth=0.75pt]{-}(1.25,0)(1.45,0.15){180}{0}
\pspolygon[fillstyle=solid,fillcolor=white,linecolor=white,linewidth=0pt](1.2, -0.25)(1.45,1.0)(1.7,-0.25)
\psline[linewidth=0.75pt](1.22, -0.15)(1.45,1.0)(1.68,-0.15)
\psline[linestyle=dashed,dash=1pt 1pt,linewidth=0.75pt](1.16, 0.12)(1.45,1.35)(1.74,0.12)
\rput(1.25,2.9){$_{N}$}
\rput(0.525,-0.4){$_{N_A}$}
\rput(2.175,-0.4){$_{N_B}$}
\end{pspicture}
\hspace{2cm}
\begin{pspicture}[shift=-0.525](0,-0.5)(2.5,3)
\pspolygon[fillstyle=solid,fillcolor=lightlightblue,linecolor=black,linewidth=0.75pt](-0.2,0)(0,1.25)(0,2.5)(2.5,2.5)(2.5,1.25)(2.7,0)(1.65,0)(1.45,1.25)(1.25,0)(0,0)
\rput(1.25,2.9){$_{N}$}
\rput(0.525,-0.4){$_{N_A}$}
\rput(2.175,-0.4){$_{N_B}$}
\end{pspicture}
\end{equation*}
\caption{The periodic pants geometry, the skirt geometry and the flat pants geometry.}
\label{fig:geometries}
\end{figure}

An integrable quantum model in one dimension often underlies a statistical model in two dimensions \cite{Baxterbook}. The transfer matrices $T(u)$ for the latter commute at different values of the spectral parameter $u$, and the Hamiltonian of the former is obtained as a leading term in the expansion of $T(u)$ around $u = 0$. From its definition in terms of scalar products of groundstates, the bipartite fidelity then has an interpretation in terms of partition functions in the two-dimensional model. If in the one-dimensional model, the boundary conditions for the system and the two subsystems are open, the corresponding model in two dimensions is defined on a domain that resembles a pair of flat pants. We call it the {\it flat pants domain}. It is illustrated in the right panel of \cref{fig:geometries}.

Similarly to the entanglement entropy, the logarithmic bipartite fidelity is an efficient tool to detect quantum phase transitions. For non-critical models it satisfies an area law, whereas it diverges logarithmically with the system size for one-dimensional quantum critical systems.  On the flat pants geometry, Dubail and St\'ephan \cite{DS11,SD13} used CFT arguments to derive the $1/N$ expansion of the logarithmic bipartite fidelity up to order $N^{-1}\log N$. 
In particular, they found that the leading term is proportional to $\log N$, with a prefactor that
depends on the central charge $c$ of the theory. In the simple case where there is no change in boundary conditions between the two subsystems, this term is 
\begin{equation}
\label{eq:leading.F.flat}
\mathcal{F}_{\text{flat\,pants}} =\frac c8 \log N + \mathcal{O}(1). 
\end{equation}
In \cite{SD13}, the authors in fact considered a more general case where the Hamiltonians $H^A$, $H^B$ and $H^{AB}$ have different boundary conditions applied to the endpoints of the chains. In the conformal field theory, this corresponds to a situation where four primary fields are inserted on the flat pants domain. The three leading terms in the large-$N$ expansion of $\mathcal F$ are proportional to $\log N$, $1$ and $N^{-1}\log N$ and the explicit expressions for their coefficients are found to depend on the conformal data, namely the central charge, the conformal dimensions of the fields and the four-point function of these fields. 
In two previous papers, we checked these conformal predictions with analytical lattice computations for the XXZ spin chain at $\Delta = -\frac12$ \cite{HL17}, and for the model of critical dense polymers \cite{PMDR19}, a lattice model known to be described by a logarithmic CFT of central charge $c = -2$ \cite{PR07,PRV10}. In \cite{PMDR19}, we also extended the prediction of \cite{SD13} to the cases where one and two of the fields are logarithmic. We argued that the bipartite fidelity allows one to measure the central charge of the model, rather than the effective central charge. 

In \cite{DS11}, Dubail and St\'ephan also investigated the behaviour of the bipartite fidelity on a second geometry: the one drawn in the central panel of \cref{fig:geometries}. We call it the {\it skirt domain}. For the quantum chain, the state $|\psi^{AB}\rangle$ used to compute \eqref{eq:FABGen} in this case is the groundstate of the Hamiltonian with periodic boundary conditions, whereas $|\psi^{A}\rangle$ and $|\psi^{B}\rangle$ are the groundstates for the open chains of the subsystems $A$ and $B$.
The results of \cite{DS11,SD13} cover the simplest case where no boundary condition changing fields are inserted. The resulting $1/N$ expansion that they obtain for the bipartite fidelity reads 
\begin{equation}
\label{eq:FGenSkirt}
\mathcal{F}_{\textrm{skirt}} = \frac{c}{4} \log N + \frac c{24}(\tilde{G}(x)+\tilde{G}(1-x))+\tilde{C}+\mathcal{O}(N^{-1}\log N), 
\end{equation} 
where $x = N_A/N$ is the aspect ratio, $\tilde{C}$ is a non-universal constant, and
\begin{equation}
\tilde{G}(x) = \frac{3-6x+4x^2}{1-x}\log x.
\end{equation}

In this paper, we provide new CFT predictions and lattice computations of the bipartite fidelity on two domains with periodic boundary conditions. The first is the {\it periodic pants domain}, for which the full system and the two subsystems all have periodic boundary conditions. It is depicted in the left panel of \cref{fig:geometries}. To our knowledge, there are no previously known results for the bipartite fidelity on this geometry. The second is the skirt domain, where we will push further the investigation initiated in \cite{DS11}. We will consider the more complicated case where boundary condition changing fields are present in the CFT context, and will derive the leading terms in \eqref{eq:FGenSkirt} up to order $N^{-1}\log N$.

Accordingly, the bulk of the paper is divided into two large sections: \cref{sec:periodic.pants} covers the case of the periodic pants domain, and \cref{sec:skirt} focuses on the skirt domain. Each of these two sections is divided into four subsections. The first subsection gives the conformal predictions for the leading terms of $\mathcal F$ in its large-$N$ expansion, with the details of the calculations presented in \cref{app:CFT.derivations}. The second and third subsections give the exact calculations of the bipartite fidelity for the XX spin chain and the model of critical dense polymers, respectively. These are two free-fermionic models for which one can diagonalise the Hamiltonian explicitly and write $\mathcal F$ as a determinant, which can be evaluated in product form in certain favourable cases. The fourth subsection uses both exact asymptotic calculations and numerical evaluations of the determinants to compare the lattice results with the CFT prediction. Some of the technical details of the asymptotic calculations are relegated to \cref{app:Asympt}. We present final remarks in \cref{sec:conclusion}.

\section{Bipartite fidelity on the periodic pants geometry
}\label{sec:periodic.pants}

In this section, we consider the bipartite fidelity for physical systems $A$, $B$ and $AB$, of respective lengths $N_A$, $N_B$ and $N=N_A+N_B$, which are all endowed with periodic boundary conditions. For one-dimensional chains, the fidelity is defined as 
\be
\label{eq:fidelity.1D.general}
\mathcal F_p = - \log \big| \langle X^A \otimes X^B | X^{AB}\rangle\big|^2
\ee
where $\langle v \otimes w |$ is a short-hand notation for $\langle v| \otimes \langle w|$. The states $\langle X^S|$ and $| X^S \rangle$ are respectively the left and right groundstates of the Hamiltonian of the chain of the system $S$. These states are normalised in such a way that $\langle X^S| X^S \rangle = 1$.

For two-dimensional lattice models, the fidelity is defined as
\be
\label{eq:fidelity.2D.general}
\mathcal F_p = \lim_{M \to \infty} - \log \Bigg| \frac{\big(Z_p^{AB}\big)^2}{Z_c^{A}Z_c^{B}Z_c^{A\cup B}}\Bigg|.
\ee
Here $Z_p^{AB}$ is the partition function defined on the periodic pants geometry depicted in the left panel of \cref{fig:geometries}. The perimeter at the top is $N$, the perimeters of the legs $A$ and $B$ are $N_A$ and $N_B$, and the height is $2M$. Likewise, $Z_c^{A}$, $Z_c^{B}$ and $Z_c^{A\cup B}$ are partition functions on cylinders of height $2M$ and perimeters $N_A$, $N_B$ and $N$, respectively. Clearly, the bipartite fidelity depends on the choices of boundary conditions assigned to the top and bottom of these lattices. For suitable choices of these boundary conditions, the partition functions are all non-zero and the limit $M \to \infty$ in \eqref{eq:fidelity.2D.general} is well-defined.

As is well-known, certain families of integrable models in two dimensions are related to one-dimensional quantum spin chains \cite{Baxterbook}. In these cases, the two definitions \eqref{eq:fidelity.1D.general} and \eqref{eq:fidelity.2D.general} coincide. In \cref{sec:CFT.pants}, we give the conformal prediction for the leading terms in the large-$N$ expansion of the bipartite fidelity. In \cref{sec:XX.pants,sec:polymers.pants}, we compute $\mathcal F_p$ for the XX spin chain and for the model of critical dense polymers. Finally, in \cref{sec:asymptotics.pants}, we compare the asymptotical behaviour of the lattice results  with the predictions of conformal field theory.

\subsection{Predictions of conformal field theory}\label{sec:CFT.pants}

In this section, we give the conformal predictions for the bipartite fidelity on the periodic pants geometry. The details of the calculations are given in \cref{app:CFT.derivations}. Let us consider a one-dimensional quantum critical system of characteristic size $N$. For large $N$, its free energy $f$ behaves as 
\be
\label{label:energy.expansion}
f = f_{\text{bulk}}\, N^2 + f_{\text{surface}}\, N + f_{\text{shape}} \log N + f_{\text{cst}} + \dots
\ee
where the dots indicate that lower-order terms are omitted. In this expression, the first two terms are proportional to the area and the surface of the domain. They are non-universal in the sense that they depend on the details of the theory's short-range interactions. 
In contrast, the term $f_{\text{shape}}$ is universal. It depends on the geometry considered and on the data of the underlying CFT, namely the central charge $c$ and the dimensions $\Delta_i$, $\bar\Delta_i$ of the conformal fields. Finally, the term $f_{\text{cst}}$ also depends on the CFT data. Because the free energy can always be shifted by an overall non-universal constant, arguments of conformal field theory can only predict $f_{\text{cst}}$ up to a non-universal additive constant.

From the definition \eqref{eq:fidelity.2D.general}, the bipartite fidelity on the periodic pants domain corresponds to the following difference of free energies:
\be \label{eq:Fpdifference}
\mathcal F_p = 2 f_{p} - f_{c(N)} - f_{c(N_A)} - f_{c(N_B)},
\ee
where $f_p$ is the free energy on the periodic pants domain and $f_{c(P)}$ is the free energy on the cylinder of perimeter~$P$. The bipartite fidelity can then be seen as a renormalised free energy. It depends on two independent characteristic lengths, $N$ and $N_A$, with the third characteristic length given by $N_B = N-N_A$. 
We consider the asymptotic behaviour of $\mathcal F_p$ in the limit where $N$ and $N_A$ are sent to infinity with the aspect ratio $x = N_A/N$ kept fixed. It has the large-$N$ expansion
\be \label{eq:Fp.expansion}
\mathcal F_p = g_{0} \log N + g_1(x) + g_2(x) N^{-1}\log N + \dotsc .
\ee 

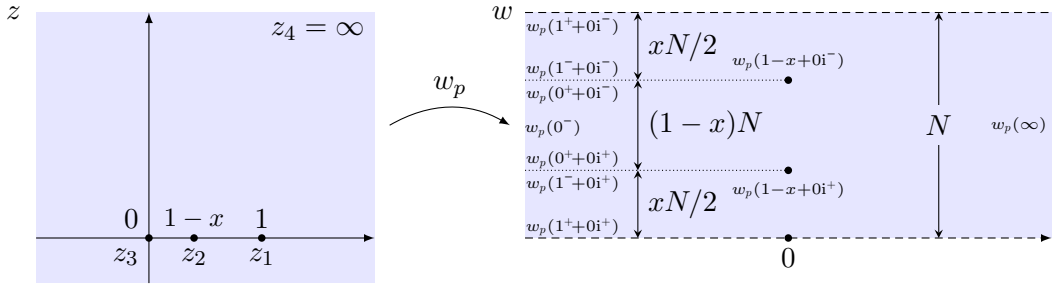
\begin{figure}
\centering
\begin{tikzpicture}
\def\a{3} 
\def\b{3.5} 
\def\L{3} 
\def\offset{0.3}
\def\ofsArr{1.5}
\def\x{0.6}
\def\shift{2} 
\def\un{1.5}
\newcommand{\mybullet}[3]{\fill #1 circle (.05) node#3 {#2}}
\newcommand{\Sminus}{\scalebox{.4}[.7]{$-$}}
\newcommand{\Splus}{\scalebox{.4}[.4]{$+$}}
\renewcommand{\wp}{w_{\scalebox{.4}{$p$}}}
%
\fill[blue!10] (-\a/2,-\a/5) rectangle (\a,\a);
\draw (-\a/2-\offset,\a) node{$z$};
\draw[->, >=latex] (-\a/2,0) to (\a,0);
\draw[->, >=latex] (0, -\a/5) to (0,\a);
\mybullet{(0,0)}{$0$}{[above left]};
\mybullet{(\un,0)}{$1$}{[above]};
\mybullet{(\un-\un*\x,0)}{{\small $1-x$}}{[above]};
\draw (0,0) node[below left] {$z_3$};
\draw (\un,0) node[below] {$z_1$};
\draw (\un-\un*\x,0) node[below] {$z_2$};
\draw (\a, \a) node[below left] {$z_4=\infty$};
\draw[->,>=latex] (\a+\shift/10,\a/2) to[bend left] node[midway, above] {$w_p$} (\a+9*\shift/10, \a/2);
%
\begin{scope}[shift={(\a+\b+\shift ,0)}]
\fill[blue!10] (-\b,0) rectangle (\b,\L);
\draw (-\b-\offset,\L) node{$w$};
\mybullet{(0,0)}{$0$}{[below]};
\mybullet{(0,{\L*\x/2})}{$\scriptscriptstyle \wp(1-x+0\ir^{\Splus})$}{[below]};
\mybullet{(0,{\L-\L*\x/2})}{$\scriptscriptstyle \wp(1-x+0\ir^{\Sminus})$}{[above]};
\draw[->, >=latex, densely dashed] (-\b,0) to (\b,0);
\draw[densely dashed] (-\b,\L) -- (\b,\L);
\draw[densely dotted] (-\b,{\L*\x/2}) -- (0,{\L*\x/2});
\draw[densely dotted] (-\b,{\L-\L*\x/2}) -- (0,{\L-\L*\x/2});
\draw[<->,>=stealth] ({\b-\ofsArr},0) -- ({\b-\ofsArr},\L )node[midway, fill=blue!10] {$N$};
\draw[<->,>=stealth] ({-\b+\ofsArr},0) -- ({-\b+\ofsArr},{\L*\x/2}) node[midway, right] {$xN/2$};
\draw[<->,>=stealth] ({-\b+\ofsArr},{\L*\x/2}) -- ({-\b+\ofsArr},{\L-\L*\x/2}) node[midway, right] {$(1-x)N$};
\draw[<->,>=stealth] ({-\b+\ofsArr},{\L-\L*\x/2})-- ({-\b+\ofsArr},{\L})  node[midway, right] {$xN/2$};
%
\draw (-\b,0) 	node[right=18pt,above=-3pt ]{$\scriptscriptstyle \wp(1^{\Splus}\!+0\ir^{\Splus})$};
\draw (-\b,{\L*\x/2}) node[right=18pt,below=-4pt ]{$\scriptscriptstyle \wp(1^{\Sminus}\!+0\ir^{\Splus})$};
\draw (-\b,{\L*\x/2}) node[right=18pt,above=-3pt ]{{$\scriptscriptstyle \wp(0^{\Splus}\!+0\ir^{\Splus})$}};
\draw (-\b,{\L/2}) 	node[right=-4pt]				 {$\scriptscriptstyle  \wp(0^{\Sminus})$}; 
\draw (-\b,\L) 			node[right=18pt,below=-4pt ]{$\scriptscriptstyle \wp(1^{\Splus}\!+0\ir^{\Sminus})$};
\draw (-\b,{\L-\L*\x/2}) 	node[right=18pt,above=-3pt ]{$\scriptscriptstyle \wp(1^{\Sminus}\!+0\ir^{\Sminus})$};
\draw (-\b,{\L-\L*\x/2}) 	node[right=18pt,below=-4pt ]{$\scriptscriptstyle \wp(0^{\Splus}\!+0\ir^{\Sminus})$};
\draw (\b,{\L/2}) 		node[left=-2 pt] 	{$\scriptscriptstyle \wp(\infty)$};
\end{scope}
%
%
\end{tikzpicture}
\caption{The function $w_{p}(z)$ maps the complex plane onto the periodic pants geometry. On the right part of the figure, the dashed and dotted lines are identified pairwise.}
\label{fig:MapPants}
\end{figure}

The coefficients $g_0$, $g_1(x)$ and $g_2(x)$ can be computed by the methods of conformal field theory. In this framework, the periodic pants domain is described as an infinite horizontal strip of width $N$ drawn in the complex plane.
The strip is decorated with two slits that divide its left half into two strips of width $Nx$ and $N(1-x)$. The boundary conditions are periodic along the strip's edges and along the slits so as to reproduce the geometry of the left panel of \cref{fig:geometries}. In particular, due to the periodicity, the two endpoints of the slits are identified as a unique point in the domain. We refer to this point as the {\it crotch point}. The mapping from the complex plane to the periodic pants geometry is
\be
\label{eq:w.pants}
w_p(z) = \frac{N}{2\pi} \big( x \log(z-1) + (1-x) \log z\big) + K_x, \qquad K_x = - \frac{N}{2\pi} \big( x \log x + (1-x) \log (1-x)\big).
\ee
This map is illustrated in \cref{fig:MapPants}.

In a general setting, we consider the situation where four fields are inserted on the periodic pants geometry. A field $\phi_1$ is inserted at $-\infty$ in the leg $A$, a field $\phi_2$ is inserted on the crotch point, a field $\phi_3$ is inserted at $-\infty$ in the leg $B$, and a field $\phi_4$ is inserted at $+\infty$. We denote by $w_i$, with $i = 1, \dots, 4$, these positions in the periodic pants domain, and by $z_i$ the corresponding positions in the complex plane obtained from the inverse map $w^{-1}_p(z)$. For simplicity, we assume that each field $\phi_i$ is spinless with conformal dimensions $\Delta_i = \bar \Delta_i$.

The first term $g_0$ in the expansion \eqref{eq:Fp.expansion} is obtained as a direct application of the Cardy-Peschel formula~\cite{CP88} for conical singularities. Indeed, the slit's endpoint in the pants domain corresponds to a single conical singularity of angle $4\pi$, as can be seen from the expansion of $w_p(z)$ about the point $z=z_2$. The resulting expression for $g_0$ depends on the dimension of the field $\phi_2$ and on the central charge:
\be
\label{eq:g0Pants}
g_0 = \frac c4 + 2 \Delta_2.
\ee

The next term $g_1(x)$ depends on the aspect ratio $x$ and on the dimensions $\Delta_i$ of the four fields. In the general setting where the four fields are non-trivial primary fields, $g_1(x)$ also depends on the four-point function of the fields $\phi_1$, $\phi_2$, $\phi_3$ and $\phi_4$. A similar dependence was found in \cite{SD13} for the flat pants domain. Here, we only state the resulting expressions, with their derivations given in \cref{app:CFT.derivations}. We give the result in two cases: (i) no field is inserted on the crotch point and the fields $\phi_1$, $\phi_3$ and $\phi_4$ are primary, and (ii) the four fields $\phi_i$ are vertex operators with charges $\alphaC_i = \bar\alphaC_i$ and conformal dimensions $\Delta_i = \bar \Delta_i = \alphaC_i^2/2$. For case (i), the known form of the three-point functions allows us to find
\be
\label{eq:g1.pants.3pts}
g_1(x) = 4 \Big(\Delta_4 - \frac{\Delta_1}x -\frac{\Delta_3}{1-x} \Big) \big(x \log x + (1-x)\log(1-x)\big) + \frac c{12} \big(G(x) + G(1-x)\big) + C
\ee
where 
\be
\label{eq:Gx}
G(x) = \frac{3 - 3x + 2 x^2}{1-x} \log x
\ee
and $C$ is a constant with respect to $x$. This constant is non-universal. For case (ii), the $n$-point function of the vertex fields is
\be
\label{eq:npoint.vertex}
\left\langle \prod_{i=1}^n \phi(z_i,\bar z_i)\right\rangle=\prod_{1\le i<j\le n} |z_i-z_j|^{2\alphaC_i \alphaC_j}, \qquad \sum_{i=1}^n \alphaC_i = 0,
\ee
and the function $g_1(x)$ reads
\begin{alignat}{2}\label{eq:g1.pants.4pts}
g_1(x) &=  
2\, \Big[\alphaC_1^2\Big(1-\frac1 x\Big) -\alphaC_3^2 - 2 \alphaC_3\alphaC_2 - \frac{\alphaC_2^2}2 +  \alphaC_4^2 (1-x) \Big] \log(1-x) + \{\alphaC_1 \leftrightarrow \alphaC_3, x\rightarrow 1-x\}
\nonumber \\[0.1cm] & + \frac c{12} \big(G(x) + G(1-x)\big) + C'
\end{alignat}
where $C'$ is a constant. On the first line of \eqref{eq:g1.pants.4pts}, the second contribution, marked by a bracket, indicates that the first term must be recopied but with the substitutions indicated inside this bracket.

Finally, we find, also in \cref{app:CFT.derivations}, that the prefactor $g_2(x)$ of the $N^{-1}\log N$ term is
\be
\label{eq:g2.pants}
g_2(x) = 2\,\Xi \times \bigg(\frac{c\,(1-x+x^2)}{24 x(1-x)} + \Delta_4 - \frac{\Delta_1}x - \frac {\Delta_3}{1-x}\bigg)
\ee
where the multiplicative constant $\Xi$, the {\it extrapolation length}, is a non-universal constant. This holds for cases (i) and (ii).

\subsection{Lattice calculation for the XX spin chain}\label{sec:XX.pants}

\subsubsection{Definition of the model}\label{sec:defXX}

The first lattice model for which we compute the bipartite fidelity is the spin-$\frac12$ quantum XX chain. It is a simple model of quantum magnetism where neighbouring spins interact isotropically in the $xy$ plane. We consider the XX spin chain on a ring of length $N$ with a diagonal twist. The Hamiltonian acts on the Hilbert space $(\mathbb C^2)^{\otimes N}$ with the local canonical basis 
\begin{equation}
|{\uparrow}\rangle = \begin{pmatrix}1\\0\end{pmatrix}, \qquad |{\downarrow}\rangle = \begin{pmatrix}0\\1\end{pmatrix}.
\end{equation}
It is convenient to introduce the Pauli matrices $\sigma^x = \left(\begin{smallmatrix}0&1\\1&0\end{smallmatrix}\right)$, $\sigma^y = \left(\begin{smallmatrix}0&-\ir\\\ir&0\end{smallmatrix}\right)$ and $\sigma^z = \left(\begin{smallmatrix}1&0\\0&-1\end{smallmatrix}\right)$, as well as the identity matrix $\mathbb I_2 = \left(\begin{smallmatrix}1&0\\0&1\end{smallmatrix}\right)$. We use the notation
\begin{equation}
\sigma_j^a = \underbrace{\mathbb I_2 \otimes \cdots \otimes \mathbb I_2}_{j-1} \otimes\, \sigma^a \otimes \underbrace{\mathbb I_2 \otimes \cdots \otimes \mathbb I_2}_{N-j}, \qquad a=x,y,z, \qquad j=1,\dots,N.
\end{equation}

The Hamiltonian of the model is 
\begin{equation}
\label{eq:Hxx}
H =- \sum_{j=1}^{N-1}  \Big(\sigma^+_j \sigma^-_{j+1} + \sigma^-_j \sigma^+_{j+1} \Big) - \eE^{\ir \phi} \sigma^+_N \sigma^-_1 - \eE^{-\ir \phi} \sigma^-_N \sigma^+_1
\end{equation}
where $\sigma^{\pm}=\frac12(\sigma^x \pm \ir \sigma^y)$ and $\phi$ is the \textit{twist angle}. We restrict our investigations to the case where $N$ is an even number. The Hamiltonian commutes with the total magnetisation $S^z = \frac12\sum_{j=1}^N \sigma^z_j$.

In order to study the bipartite fidelity, we consider the same model on two smaller systems $A$ and $B$. Their degrees of freedom are defined on smaller rings of lengths $N_A$ and $N_B$, and their respective twist angles are $\phi_A$ and $\phi_B$. The three twist angles $\phi$, $\phi_A$ and $\phi_B$ are three free parameters.  
We impose that both lengths are even numbers that satisfy $N_A+N_B = N$. The aspect ratio is 
$x = \frac{N_A}{N}.$

We denote by $H$, $H^A$ and $H^B$ the Hamiltonians of the whole system, of subsystem $A$ and of subsystem $B$, with respective groundstates $\ket{X_0^{AB}}$, $\ket{X_0^A}$ and $\ket{X_0^B}$. The subscript $0$ indicates that these states belong to the subspaces of zero magnetisation. Defining the dual states as
\be\langle v| = |v\rangle^\dagger,\ee 
the states $\langle{X_0^{AB}}|$, $\langle{X_0^A}|$ and $\langle{X_0^B}|$ are then the left-groundstates of the Hamiltonians $H$, $H^A$ and $H^B$. The logarithmic bipartite fidelity for the XX spin chain on the periodic pants domain is
\begin{equation}
\label{eq:LBFXXPants}
\mathcal F_p^{\tXX} = - \log \left| \braket{X_0^A \otimes X_0^B | X_0^{AB}}\right|^2,
\end{equation}
where $\ket{X_0^{AB}}$, $\ket{X_0^A}$ and $\ket{X_0^B}$ are assumed to have unit norms.

\subsubsection{Diagonalisation of the Hamiltonian}\label{sec:diagXX}

The diagonalisation of $H$ is standard. It uses a Jordan-Wigner transformation and a Fourier transform. The result is
\begin{equation}
H = - \sum_{k=1}^N2 \cos(\theta_k) \mu_k^\dagger \mu_k,
\end{equation}
where 
\begin{equation}
\label{eq:muk}
\mu_k = \frac1{\sqrt N} \sum_{j=1}^N \eE^{\ir j \theta_k} c_j, \qquad \mu^\dagger_k = \frac1{\sqrt N} \sum_{j=1}^N \eE^{-\ir j \theta_k} c_j^\dagger, \qquad 
\theta_k =
\left\{\begin{array}{cl}
\frac{2 \pi k - \phi}N & \frac N2 + S^z \textrm{ odd},\\[0.15cm]
\frac{2 \pi (k-\frac12) - \phi}N & \frac N2 + S^z \textrm{ even},
\end{array}\right.
\end{equation}
and
\begin{equation}  \label{eq:JW}
c_j = (-1)^{j-1}\bigg(\prod_{k=1}^{j-1} \sigma^z_k\bigg) \sigma^-_j, \qquad c_j^\dagger = (-1)^{j-1}\bigg(\prod_{k=1}^{j-1} \sigma^z_k\bigg) \sigma^+_j.
\end{equation}
The operators $\mu_k$ and $\mu^\dagger_\ell$ satisfy the usual fermionic anti-commutation rules
\begin{equation}
\label{eq:fermions.acomm}
\{\mu_k, \mu_\ell^\dagger\} = \delta_{k,\ell}, \qquad \{\mu_k, \mu_\ell\} = \{\mu^\dagger_k, \mu^\dagger_\ell\} = 0.
\end{equation}

With $k$ in the set $\{1, \dots, N\}$, we have a full set of fermionic operators. It is however useful to extend the range of $k$ to negative values using the periodicity properties $\mu_{k+N} = \mu_k$ and $\mu^\dagger_{k+N} = \mu^\dagger_k$. For $\phi \in (-\pi,\pi)$, the groundstate of $H$ is unique. It lies in the magnetisation sector $S^z = 0$ and is given by

\begin{equation}
\label{eq:X0}
|X_0\rangle = \left\{\begin{array}{cl}
\mu^\dagger_{(4-N)/4}\cdots\mu^\dagger_{N/4} |0\rangle & \frac N2 \textrm{ even,} \\[0.15cm]
\mu^\dagger_{(2-N)/4}\cdots\mu^\dagger_{(N-2)/4} |0\rangle & \frac N2 \textrm{ odd,}
\end{array}\right.
\qquad |0\rangle = |{\downarrow\cdots\downarrow}\rangle.
\end{equation}

Due to the anti-commutation relations of the operators $\mu_k$ and $\mu^\dagger_\ell$, the groundstate has norm one, namely $\braket{X_0 | X_0} = 1$,
where 
\begin{equation}
\label{eq:X0Left}
\langle X_0| = \left\{\begin{array}{cl}
 \langle 0 | \mu_{N/4}\cdots\mu_{(4-N)/4} & \frac N2 \textrm{ even,} \\[0.15cm]
 \langle 0 |  \mu_{(N-2)/4}\cdots\mu_{(2-N)/4} & \frac N2 \textrm{ odd,}
\end{array}\right.
\end{equation}
is the left groundstate of $H$.

\subsubsection{Bipartite fidelity}\label{sec:BP.XX}

In order to compute the bipartite fidelity, we introduce the fermion operators for the subsystems $A$ and $B$:
\begin{subequations}
\begin{alignat}{3}
	 \mu^A_k 			&= \frac1{\sqrt{N_A}} \sum_{j=1}^{N_A} \eE^{\ir j \theta^A_k} c_j, \qquad 
&&	(\mu^A_k)^\dagger 	  = \frac1{\sqrt{N_A}} \sum_{j=1}^{N_A} \eE^{-\ir j \theta^A_k} c_j^\dagger,
\\[0.2cm]
	 \mu^B_k 			&= \frac1{\sqrt{N_B}} \sum_{j=N_A+1}^{N} \eE^{\ir (j-N_A) \theta^B_k} c_j, \qquad 
&&	(\mu^B_k)^\dagger    = \frac1{\sqrt{N_B}} \sum_{j=N_A+1}^{N} \eE^{-\ir (j-N_A) \theta^B_k} c_j^\dagger, 
\end{alignat}
where
\begin{alignat}{2}
\theta^A_k &=
\left\{\begin{array}{cl}
\frac{2 \pi k - \phi_A}{N_A} & \frac {N_A}2 + (S^z)^A \textrm{ odd},\\[0.15cm]
\frac{2 \pi (k-\frac12) - \phi_A}{N_A} & \frac {N_A}2 + (S^z)^A \textrm{ even},
\end{array}\right.
\qquad 
&&(S^z)^A = \tfrac12 \sum_{j=1}^{N_A} \sigma_j^z,
\\[0.2cm]
\theta^B_k &=
\left\{\begin{array}{cl}
\frac{2 \pi k - \phi_B}{N_B} & \frac {N_B}2 + (S^z)^B \textrm{ odd},\\[0.15cm]
\frac{2 \pi (k-\frac12) - \phi_B}{N_B} & \frac {N_B}2 + (S^z)^B \textrm{ even},
\end{array}\right.
 \qquad
&&(S^z)^B = \tfrac12 \hspace{-0.2cm}\sum_{j=N_A+1}^{N} \sigma_j^z.
\end{alignat}
\end{subequations}
The three groundstates $\ket{X_0^{AB}}$, $\ket{X_0^{A}}$ and $\ket{X_0^{B}}$ have unit norms. Using Wick's theorem, we evaluate the overlap in \eqref{eq:LBFXXPants} as a determinant. The corresponding matrix contains the anti-commutators of the operators $\mu_{k'}^\dagger$ and $\mu^A_k$ or $\mu^B_k$. After simplification, the result is
\begin{equation}
\label{eq:exactOverlapPants}
\big|\!\braket{X_0^A \otimes X_0^B | X_0^{AB}}\!\big| =2^{-N/2}N^{-N/2}x^{-N x/4}(1-x)^{-N(1-x)/4} \big|\!\det \mathcal{A}\big|,
\end{equation}
with
\begin{equation}
\label{eq:Akk'}
\mathcal{A}_{k,k'} =  \left\{\begin{array}{cl}
\displaystyle\frac{1+\eE^{-\ir \pi x (2k'-1)}\eE^{-\ir(\phi_A-x \phi)}}{\sin \left[ \frac{\pi (2k-1)}{2Nx}-\frac{\pi (2k'-1)}{2N}-\frac{\phi_A-x \phi}{2Nx} \right]} &k=1,\dots,\frac{Nx}{2},  \\[0.6cm]
\displaystyle\frac{1+\eE^{-\ir (\phi-\phi_B)}\eE^{\ir \pi (1-x) (2k'-1)}\eE^{\ir x \phi}}{\sin \left[ \frac{\pi (2k-Nx-1)}{2N(1-x)}-\frac{\pi (2k'-1)}{2N}-\frac{\phi_B-(1-x) \phi}{2N(1-x)} \right]} &k=\frac{Nx}{2}+1,\dots, \frac N2,
\end{array}\right. \quad k' = 1, \dots, \tfrac N2.
\end{equation}
This holds for all parities of $N/2$, $N_A/2$ and $N_B/2$.
For generic twist angles, we do not know how to evaluate this determinant in closed form. This is due to the numerators in \eqref{eq:Akk'} which are different in the two parts of the matrix $\mathcal A$. 
We can however evaluate the determinant if the twist parameters $\phi$, $\phi_A$ and $\phi_B$ are such that
\begin{equation}
1+\eE^{-\ir \pi x (2k'-1)}\eE^{-\ir(\phi_A-x \phi)} = 1+\eE^{-\ir (\phi-\phi_B)}\eE^{\ir \pi (1-x) (2k'-1)}\eE^{\ir x \phi}.
\end{equation}
The solutions are
\begin{equation}
\label{eq:restrictions}
\phi = \phi_A+\phi_B-\pi + 2 \pi \ell, \qquad \ell \in \mathbb Z. 
\end{equation}
In the following, we focus on the case $\ell = 0$. Under this specialisation, we are able to simplify the determinant and obtain a closed-form formula. Indeed, we find
\begin{equation}
|\langle X_0^A \otimes X_0^B | X_0^{AB}\rangle| = N^{-N/2}x^{-N x/4}(1-x)^{-N(1-x)/4} \, \bigg|\prod \limits_{k'=1}^{N/2}\cos \Big[\pi x k'-\tfrac{\phi_A(x-1)+x \phi_B }{2}\Big] \bigg|\, |\!\det \mathcal{M}|
\end{equation}
where
\begin{equation}
\mathcal{M}_{k,k'} = \left\{\begin{array}{cl}
\sin \left[ \frac{\pi (2k-1)-\phi_A}{2Nx}-\frac{\pi (2k'-1)-\phi}{2N}\right]^{-1} &k=1,\dots,\frac{N x}{2},  \\[0.15cm]
\sin \left[ \frac{\pi (2k-Nx-1)-\phi_B}{2N(1-x)}-\frac{\pi (2k'-1)-\phi}{2N}\right]^{-1} &k=\frac{Nx}{2}+1,\dots, \frac N2,
\end{array}\right. \quad k'=1,\dots,\tfrac N2.
\end{equation}
We use the Cauchy identity 
\begin{equation}
\label{eq:Cauchy}
\det_{i,j} \frac1{\sin (x_i-y_j)} = \frac{\prod_{i<j} \sin(x_i-x_j) \sin(y_j-y_i)}{\prod_i \prod_j \sin (x_i-y_j)}
\end{equation}
and find
\begin{equation}
\begin{split}
\det \mathcal{M} &= \prod \limits_{1\leq k<k'\leq Nx/2} \sin \Big[\frac{\pi}{Nx}(k-k')\Big]\times \prod \limits_{1\leq k<k'\leq N(1-x)/2} \sin \Big[\frac{\pi}{N(1-x)}(k-k')\Big]\\[0.2cm]
&\times\prod \limits_{1\leq k<k'\leq N/2} \sin \Big[\frac{\pi}{N}(k'-k)\Big] \times \prod \limits_{k=1}^{Nx/2}\prod \limits_{k'=1}^{N(1-x)/2} \sin \Big[\frac{\pi (2k-1)-\phi_A}{2Nx}-\frac{\pi (2k'-1)-\phi_B}{2N(1-x)} \Big] \\
& \times \left( \prod \limits_{k=1}^{Nx/2}\prod \limits_{k'=1}^{N/2} \sin \Big[\frac{\pi (2k-1)-\phi_A}{2Nx}-\frac{\pi (2k'-1)-\phi}{2N} \Big] \right)^{-1} \\
&\times \left( \prod \limits_{k=1}^{N(1-x)/2}\prod \limits_{k'=1}^{N/2} \sin \Big[\frac{\pi (2k-1)-\phi_B}{2N(1-x)}-\frac{\pi (2k'-1)-\phi}{2N} \Big] \right)^{-1}.
\end{split}
\end{equation}

Many of these products have a similar structure. We define two generic products $P_1$ and $P_2$: 
\begin{subequations}
\label{eq:Ps}
\begin{alignat}{2}
P_1(N) &=\prod \limits_{1\leq k<k'\leq N/2}\left|  \sin \Big[\frac{\pi}{N}(k-k')\Big]\right|,
\\
P_2(N_1,N_2,\phi_1, \phi_2)&= \prod \limits_{k=1}^{N_1/2}\prod \limits_{k'=1}^{N_2/2}\left|  \sin \Big[\frac{\pi k}{N_1}-\frac{\pi k'}{N_2}-\frac{\pi+\phi_1}{2N_1}+\frac{\pi+\phi_2}{2N_2} \Big]\right|.
\end{alignat}
\end{subequations}
Their asymptotic large-$N$ behaviour are given in \cref{app:P1,app:P2}, respectively.
We recast the overlap $ |\langle X_0^A \otimes X_0^B | X_0^{AB}\rangle|$ in a simple way and find
\begin{alignat}{2}
\eE^{-\frac 12 \mathcal F_p^{\tXX} }\big|_{\phi = \phi_A+\phi_B-\pi} & = N^{-N/2}x^{-N x/4}(1-x)^{-N(1-x)/4} \bigg|\prod \limits_{k'=1}^{N/2}\cos \Big[\pi x k'-\tfrac{\phi_A(x-1)+x \phi_B }{2}\Big] \bigg|\nonumber\\
&\times \frac{P_1(N)\, P_1(Nx) \, P_1(N(1-x)) \, P_2(Nx,N(1-x),\phi_A,\phi_B)}{P_2(Nx,N,\phi_A,\phi_A+\phi_B-\pi)\, P_2(N(1-x),N,\phi_B,\phi_A+\phi_B-\pi)}.\label{eq:B}
\end{alignat}

\subsubsection{Other instances of the bipartite fidelity}\label{sec:modifiedLBFPants}

The CFT predictions of \cref{sec:CFT.pants} cover cases where fields are inserted in the legs and on the crotch point. In order to investigate these cases further, we consider modified instances of the bipartite fidelity on the periodic pants lattice. We do this by relaxing the condition tying $N$, $N_A$ and $N_B$, and define
\begin{equation} \label{eq:modifiedLBFPants}
\mathcal F_p^\tXX (n) = - \log \Big| \braket{X_0^A \otimes X_0^B \otimes 
\underset{n/2 \text{ times}}{\underbrace{{\uparrow\downarrow}\otimes \cdots  \otimes {\uparrow\downarrow}}}\, | X_0^{AB}}\Big|^2,
\end{equation}
where $n$ is an even integer. The new constraint on the lengths is $N=N_A+N_B+n$. In this scalar product, the magnetisation of the dual state vanishes, and the corresponding overlap is non-zero. Following the arguments of \cref{sec:BP.XX}, we obtain a determinant expression for this overlap similar to \eqref{eq:exactOverlapPants}, which we do not reproduce here. We are unable to evaluate it in product form. We analyse $\mathcal F_p^\tXX(n)$ for small values of $n$ in \cref{sec:asymptotics.pants} using the numerical evaluation of these determinants. The case $n=0$ corresponds to the standard definition \eqref{eq:LBFXXPants} of $\mathcal F_p^\tXX$.

\subsection{Lattice calculation for critical dense polymers}\label{sec:polymers.pants}

\subsubsection{Definition of the model}\label{sec:def}
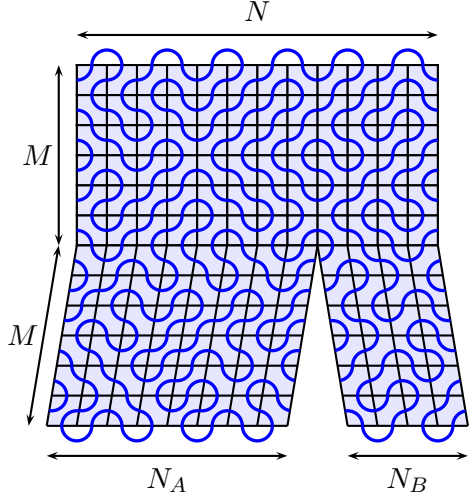
\begin{figure}
\begin{center}
\psset{unit=0.4cm}
\begin{pspicture}[shift=-9.4](-0.5,-3)(14.5,15.6)
\facegrid{(0,6)}{(12,12)}
\pspolygon[fillstyle=solid,fillcolor=lightlightblue,linecolor=white,linewidth=0pt](0,6)(-1,0)(7,0)(8,6)(0,6)
\pspolygon[fillstyle=solid,fillcolor=lightlightblue,linecolor=white,linewidth=0pt](8,6)(9,0)(13,0)(12,6)
\multiput(0,0)(1,0){9}{\psline{-}(0,6)(-1,0)}
\multiput(0,0)(1,0){5}{\psline{-}(8,6)(9,0)}
\multiput(0,0)(-0.166,-1){7}{\psline{-}(8,6)(0,6)}
\multiput(0,0)(0.166,-1){7}{\psline{-}(8,6)(12,6)}
\rput(0,11){\rput(0,0){\loopa}\rput(1,0){\loopa}\rput(2,0){\loopa}\rput(3,0){\loopb}\rput(4,0){\loopa}\rput(5,0){\loopa}\rput(6,0){\loopa}\rput(7,0){\loopb}\rput(8,0){\loopa}\rput(9,0){\loopa}\rput(10,0){\loopa}\rput(11,0){\loopb}}
\rput(0,10){\rput(0,0){\loopb}\rput(1,0){\loopb}\rput(2,0){\loopb}\rput(3,0){\loopb}\rput(4,0){\loopa}\rput(5,0){\loopa}\rput(6,0){\loopa}\rput(7,0){\loopa}\rput(8,0){\loopa}\rput(9,0){\loopa}\rput(10,0){\loopa}\rput(11,0){\loopb}}
\rput(0,09){\rput(0,0){\loopb}\rput(1,0){\loopb}\rput(2,0){\loopa}\rput(3,0){\loopb}\rput(4,0){\loopa}\rput(5,0){\loopa}\rput(6,0){\loopa}\rput(7,0){\loopa}\rput(8,0){\loopb}\rput(9,0){\loopb}\rput(10,0){\loopa}\rput(11,0){\loopa}}
\rput(0,08){\rput(0,0){\loopa}\rput(1,0){\loopb}\rput(2,0){\loopa}\rput(3,0){\loopa}\rput(4,0){\loopb}\rput(5,0){\loopb}\rput(6,0){\loopb}\rput(7,0){\loopb}\rput(8,0){\loopb}\rput(9,0){\loopa}\rput(10,0){\loopa}\rput(11,0){\loopa}}
\rput(0,07){\rput(0,0){\loopa}\rput(1,0){\loopa}\rput(2,0){\loopa}\rput(3,0){\loopa}\rput(4,0){\loopb}\rput(5,0){\loopb}\rput(6,0){\loopb}\rput(7,0){\loopb}\rput(8,0){\loopa}\rput(9,0){\loopa}\rput(10,0){\loopb}\rput(11,0){\loopb}}
\rput(0,06){\rput(0,0){\loopb}\rput(1,0){\loopa}\rput(2,0){\loopa}\rput(3,0){\loopa}\rput(4,0){\loopa}\rput(5,0){\loopb}\rput(6,0){\loopa}\rput(7,0){\loopa}\rput(8,0){\loopb}\rput(9,0){\loopb}\rput(10,0){\loopb}\rput(11,0){\loopa}}
\rput(-0.000,6){\rput(0,0){\bloopb}\rput(1,0){\bloopa}\rput(2,0){\bloopb}\rput(3,0){\bloopb}\rput(4,0){\bloopb}\rput(5,0){\bloopb}\rput(6,0){\bloopb}\rput(7,0){\bloopa}}
\rput(-0.166,5){\rput(0,0){\bloopa}\rput(1,0){\bloopa}\rput(2,0){\bloopb}\rput(3,0){\bloopa}\rput(4,0){\bloopb}\rput(5,0){\bloopa}\rput(6,0){\bloopb}\rput(7,0){\bloopa}}
\rput(-0.333,4){\rput(0,0){\bloopa}\rput(1,0){\bloopb}\rput(2,0){\bloopb}\rput(3,0){\bloopa}\rput(4,0){\bloopa}\rput(5,0){\bloopa}\rput(6,0){\bloopa}\rput(7,0){\bloopa}}
\rput(-0.500,3){\rput(0,0){\bloopb}\rput(1,0){\bloopb}\rput(2,0){\bloopa}\rput(3,0){\bloopa}\rput(4,0){\bloopa}\rput(5,0){\bloopb}\rput(6,0){\bloopb}\rput(7,0){\bloopa}}
\rput(-0.666,2){\rput(0,0){\bloopb}\rput(1,0){\bloopb}\rput(2,0){\bloopb}\rput(3,0){\bloopa}\rput(4,0){\bloopa}\rput(5,0){\bloopb}\rput(6,0){\bloopa}\rput(7,0){\bloopb}}
\rput(-0.833,1){\rput(0,0){\bloopa}\rput(1,0){\bloopa}\rput(2,0){\bloopb}\rput(3,0){\bloopb}\rput(4,0){\bloopa}\rput(5,0){\bloopa}\rput(6,0){\bloopb}\rput(7,0){\bloopb}}
\rput(0.000,6){\rput(8,0){\cloopa}\rput(9,0){\cloopb}\rput(10,0){\cloopb}\rput(11,0){\cloopa}}
\rput(0.166,5){\rput(8,0){\cloopa}\rput(9,0){\cloopa}\rput(10,0){\cloopa}\rput(11,0){\cloopb}}
\rput(0.333,4){\rput(8,0){\cloopb}\rput(9,0){\cloopb}\rput(10,0){\cloopa}\rput(11,0){\cloopb}}
\rput(0.500,3){\rput(8,0){\cloopa}\rput(9,0){\cloopa}\rput(10,0){\cloopb}\rput(11,0){\cloopb}}
\rput(0.666,2){\rput(8,0){\cloopa}\rput(9,0){\cloopa}\rput(10,0){\cloopb}\rput(11,0){\cloopa}}
\rput(0.833,1){\rput(8,0){\cloopb}\rput(9,0){\cloopa}\rput(10,0){\cloopa}\rput(11,0){\cloopa}}
\multiput(0,0)(2,0){6}{\psarc[linewidth=1.25pt,linecolor=blue](1,12){.5}{0}{180}}
\multiput(0,0)(2,0){4}{\psarc[linewidth=1.25pt,linecolor=blue](0,0){.5}{180}{0}}
\multiput(0,0)(2,0){2}{\psarc[linewidth=1.25pt,linecolor=blue](10,0){.5}{180}{0}}
\psline{<->}(-0.6,6)(-0.6,12)\rput(-1.3,9){$M$}
\psline{<->}(-0.6,6)(-1.6,0)\rput(-1.8,3){$M$}
\psline{<->}(0,13)(12,13)\rput(6,13.8){$N$}
\psline{<->}(-1,-1)(7,-1)\rput(3,-1.8){$N_A$}
\psline{<->}(9,-1)(13,-1)\rput(11,-1.8){$N_B$}
\end{pspicture}
\caption{A configuration of the model of critical dense polymers on the periodic pants geometry, with $M=6$, $N=12$, $N_A = 8$ and $N_B=4$. It contains four non-contractible loops and therefore has a weight~$\alpha^4$.}
\label{fig:loop.config}
\end{center}
\end{figure}

The second lattice model for which we compute the bipartite fidelity is the model of critical dense polymers. This is a two-dimensional model of interacting tiles on the square lattice, and here we define it on the periodic pants geometry. The lattice is divided into three parts, as illustrated in the example of \cref{fig:loop.config}. The top part is an $M \times N$ rectangular array of tiles, with $N$ even. The boundary conditions on this rectangle are periodic in the horizontal direction, so that the left and right segments of this rectangle are identified as the same segment. The bottom part is divided into two smaller arrays of tiles of respective sizes $M \times N_A$ and $M \times N_B$, with $N_A$ and $N_B$ both even and satisfying $N_A + N_B = N$. Each of these smaller arrays also has periodic horizontal boundary conditions. The tops of these two arrays are placed side-by-side and attached to the bottom segment of the $M \times N$ rectangle. 

A configuration of the model of critical dense polymers is a choice of the diagram
$
\psset{unit=0.3cm}
\, \begin{pspicture}[shift=-0.16](0,0)(1,1)
\pspolygon[fillstyle=solid,fillcolor=lightlightblue,linewidth=\mince](0,0)(0,1)(1,1)(1,0)
\psarc[linewidth=\moyen,linecolor=blue](1,0){.5}{90}{180}
\psarc[linewidth=\moyen,linecolor=blue](0,1){.5}{-90}{0}
\end{pspicture}\, $ or $\psset{unit=0.35cm}
\, \begin{pspicture}[shift=-0.16](0,0)(1,1)
\pspolygon[fillstyle=solid,fillcolor=lightlightblue,linewidth=\mince](0,0)(0,1)(1,1)(1,0)
\psarc[linewidth=\moyen,linecolor=blue](0,0){.5}{0}{90}
\psarc[linewidth=\moyen,linecolor=blue](1,1){.5}{180}{-90}
\end{pspicture}\, $ 
for each tile of this lattice. These are both assigned the weight $1$. The three free ends of the pants lattice, one at the top and two at the bottom of the lattice, are decorated by simple half-arcs that connect each tile to one of its neighbours. Put together, the loop segments drawn on the tiles and on the boundary half-arcs form a set of loops. The contractible loops, namely those that can be deformed to a point, are given a weight $\beta=0$. Non-contractible loops are divided into three families. They can (i) wrap around leg $A$, (ii) wrap around leg $B$, or (iii) wrap around the waist of the pants (or equivalently around both legs $A$ and $B$). We choose to assign a weight $\alpha \in \mathbb R$ to non-contractible loops in the families (i) and (iii), but a weight zero to those in the family (ii). The Boltzmann weight of a configuration $\sigma$ is then given by 
\be
W_\sigma = \alpha^{n_{\rm A} + n_{\rm AB}} \delta_{n_\beta,0} \delta_{n_{\rm B},0}
\ee
where $n_\beta$ is the number of contractible loops, and $n_{\rm A}$, $n_{\rm B}$ and $n_{\rm AB}$ are the numbers of non-contractible loops in families (i), (ii) and (iii), respectively. The partition function on the pants lattice, denoted $Z_p^{AB}$, is defined as
\begin{equation}
Z_p^{AB}=\sum_{\sigma} W_\sigma.
\end{equation}

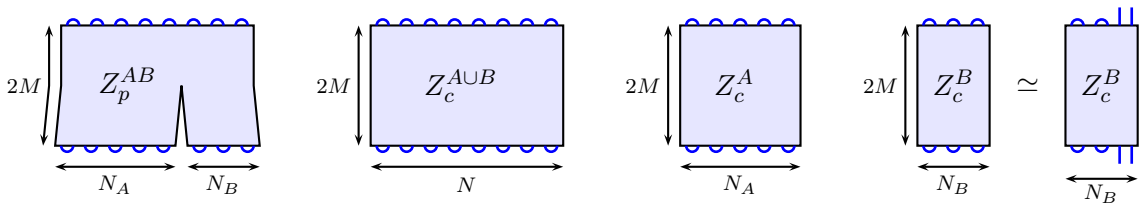
\begin{figure}
\begin{center}
\begin{equation*}
\psset{unit=0.4cm}
\begin{pspicture}[shift=-1.9](-1.4,-1.0)(6,4.6)
\pspolygon[fillstyle=solid,fillcolor=lightlightblue](-0.2,0)(0,2)(0,4)(6.4,4)(6.4,2)(6.6,0)(4.2,0)(4.0,2)(3.8,0)(-0.2,0)
\multiput(0,0)(0.8,0){8}{\psarc[linewidth=\moyen,linecolor=blue](0.4,4){0.2}{0}{180}}
\multiput(0,0)(0.8,0){5}{\psarc[linewidth=\moyen,linecolor=blue](0.2,0){0.2}{180}{0}}
\multiput(0,0)(0.8,0){3}{\psarc[linewidth=\moyen,linecolor=blue](4.6,0){0.2}{180}{0}}
\rput(2.2,2){$Z_p^{AB}$}
\psline{<->}(-0.2,-0.7)(3.8,-0.7)\rput(1.8,-1.3){$_{N_A}$}
\psline{<->}(4.2,-0.7)(6.6,-0.7)\rput(5.4,-1.3){$_{N_B}$}
\psline{<->}(-0.6,0)(-0.4,2)(-0.4,4)\rput(-1.2,2){$_{2M}$}
\end{pspicture}
\qquad\quad
\begin{pspicture}[shift=-1.9](-1.4,-1.0)(6,4.6)
\pspolygon[fillstyle=solid,fillcolor=lightlightblue](0,0)(6.4,0)(6.4,4)(0,4)
\multiput(0,0)(0.8,0){8}{\psarc[linewidth=\moyen,linecolor=blue](0.4,4){0.2}{0}{180}}
\multiput(0,0)(0.8,0){8}{\psarc[linewidth=\moyen,linecolor=blue](0.4,0){0.2}{180}{0}}
\rput(3,2){$Z_c^{A\cup B}$}
\psline{<->}(0,-0.7)(6.4,-0.7)\rput(3.2,-1.3){$_N$}
\psline{<->}(-0.4,0)(-0.4,4)\rput(-1.2,2){$_{2M}$}
\end{pspicture}
\qquad\quad
\begin{pspicture}[shift=-1.9](-1.4,-1.0)(3.6,4.6)
\pspolygon[fillstyle=solid,fillcolor=lightlightblue](0,0)(4.0,0)(4.0,4)(0,4)
\multiput(0,0)(0.8,0){5}{\psarc[linewidth=\moyen,linecolor=blue](0.4,4){0.2}{0}{180}}
\multiput(0,0)(0.8,0){5}{\psarc[linewidth=\moyen,linecolor=blue](0.4,0){0.2}{180}{0}}
\rput(1.8,2){$Z_c^{A}$}
\psline{<->}(0,-0.7)(4.0,-0.7)\rput(2,-1.3){$_{N_A}$}
\psline{<->}(-0.4,0)(-0.4,4)\rput(-1.2,2){$_{2M}$}
\end{pspicture}
\qquad \quad
\begin{pspicture}[shift=-1.9](-1.4,-1.0)(2.4,4.6)
\pspolygon[fillstyle=solid,fillcolor=lightlightblue](0,0)(2.4,0)(2.4,4)(0,4)
\multiput(0,0)(0.8,0){3}{\psarc[linewidth=\moyen,linecolor=blue](0.4,4){0.2}{0}{180}}
\multiput(0,0)(0.8,0){3}{\psarc[linewidth=\moyen,linecolor=blue](0.4,0){0.2}{180}{0}}
\rput(1.2,2){$Z_c^{B}$}
\psline{<->}(0,-0.7)(2.4,-0.7)\rput(1.2,-1.3){$_{N_B}$}
\psline{<->}(-0.4,0)(-0.4,4)\rput(-1.2,2){$_{2M}$}
\end{pspicture}
\quad 
\begin{pspicture}[shift=-1.9](0.2,-1.0)(0.8,4.6)
\rput(0.5,2){$\simeq$}
\end{pspicture}
\quad
\begin{pspicture}[shift=-1.9](0,-1.0)(2.4,4.6)
\pspolygon[fillstyle=solid,fillcolor=lightlightblue](0,0)(2.4,0)(2.4,4)(0,4)
\multiput(0,0)(0.4,0){2}{\psline[linewidth=\moyen,linecolor=blue](1.8,4)(1.8,4.6)}
\multiput(0,0)(0.4,0){2}{\psline[linewidth=\moyen,linecolor=blue](1.8,0)(1.8,-0.6)}
\multiput(0,0)(0.8,0){2}{\psarc[linewidth=\moyen,linecolor=blue](0.4,4){0.2}{0}{180}}
\multiput(0,0)(0.8,0){2}{\psarc[linewidth=\moyen,linecolor=blue](0.4,0){0.2}{180}{0}}
\rput(1.2,2){$Z_c^{B}$}
\psline{<->}(0,-1.1)(2.4,-1.1)\rput(1.2,-1.6){$_{N_B}$}
\end{pspicture}
\end{equation*}
\caption{The lattices and boundary conditions corresponding to $Z_p^{AB}$,  $Z_c^{A\cup B}$,  $Z_c^A$ and $Z_c^B$.}
\label{fig:PartitionFunctions}
\end{center}
\end{figure}

We also define $Z_c^{A\cup B}$ and $Z_c^{A}$, the partition functions for the model of dense polymers on cylinders of dimensions $2M \times N$ and $2M \times N_A$, respectively. On these cylinders, the bottom and top ends are also decorated with simple arcs. The contractible and non-contractible loops have the fugacities $0$ and $\alpha$, respectively. Finally, we define $Z_c^B$, a partition function defined on a cylinder of size $2M \times N_B$. It is different from the other partition functions, in that the only configurations with non-zero weights are those with exactly one loop. This loop can be contractible or non-contractible. The boundary conditions are as before, with both ends of the cylinder decorated with simple half-arcs. Each of the contributing configurations then has a weight $1$, and $Z_c^B$ is simply the number of such configurations. As argued in \cite{MDJ18}, one way to implement this without taking a limit on the fugacity of the loops is to replace one of the half-arcs at the top edge by two defects, and likewise at the bottom of the cylinder. One then imposes that the defects at the top connect with those at the bottom, with weight~$1$. For convenience, we choose that the last half-arc, on both the top and bottom segments, is the one replaced by a pair of defects. The lattices corresponding to the different partition functions are depicted in \cref{fig:PartitionFunctions}. 

The logarithmic bipartite fidelity is defined as
\begin{equation}\label{eq:Falpha}
\mathcal{F}_p^{\alpha} =- \lim_{M \to \infty} \log \Bigg( \frac{\big(Z_p^{AB}\big)^2}{Z_c^{A \cup B} Z_c^A Z_c^B}\Bigg).
\end{equation}
As we shall see, the ratio in the parenthesis is such that the limit is well-defined.

\subsubsection{The enlarged periodic Temperley-Lieb algebra}\label{sec:TL}

\paragraph{Definition of the algebra.}

The connectivity properties of loop segments are naturally described in the language of Temperley-Lieb algebras. Because of the periodic boundary conditions, our calculation of $\mathcal F_p^\alpha$ requires the periodic incarnation of the Temperley-Lieb algebra. This algebra was first introduced~\cite{L91} by D.~Levy in 1991 and was subsequently studied \cite{MS93,GL98,G98,EG99} by both physicists and mathematicians. Certain details in the definition of this algebra vary from one paper to the next, and here we work with the {\it enlarged periodic Temperley-Lieb algebra} $\eptl_N(\alpha, \beta)$, as defined in \cite{PRV10}.

This algebra is the linear span of connectivity diagrams drawn inside a rectangular box with $N$ marked nodes on its top segment and $N$ more on its bottom segment. Inside the box, the nodes are connected pairwise by non-intersecting loop segments. The boundary conditions are periodic in the horizontal direction. The loop segments can therefore cross this segment, in which case we say that they travel via the back of the cylinder. 

The product $a_1a_2$ of two connectivity diagrams in $\eptl_N(\alpha,\beta)$ is done by vertical concatenation: $a_1$ is drawn below $a_2$, the new connectivity is obtained from the connection of the top and bottom nodes, and each loop formed in the process is removed and replaced by a multiplicative weight of $\alpha$ or $\beta$ depending on its contractibility. Here are two examples of products of connectivity diagrams for $N=4$:

\begin{equation}
\psset{unit=0.9}
\begin{pspicture}[shift=-0.7](-0.0,0)(1.6,1.6)
\multiput(0,0)(0,0.8){2}{\pspolygon[fillstyle=solid,fillcolor=lightlightblue,linecolor=black,linewidth=0pt](0,0)(0,0.8)(1.6,0.8)(1.6,0)(0,0)}
\psarc[linecolor=blue,linewidth=\elegant]{-}(0.4,0.8){0.2}{180}{360}
\psarc[linecolor=blue,linewidth=\elegant]{-}(0,0){0.2}{0}{90}
\psarc[linecolor=blue,linewidth=\elegant]{-}(1.6,0){0.2}{90}{180}
\psbezier[linecolor=blue,linewidth=\elegant]{-}(0.6,0)(0.6,0.4)(1.0,0.4)(1.0,0.8)
\psbezier[linecolor=blue,linewidth=\elegant]{-}(1.0,0)(1.0,0.4)(1.4,0.4)(1.4,0.8)
\rput(0,0.8){
\psarc[linecolor=blue,linewidth=\elegant]{-}(0.4,0){0.2}{0}{180}
\psarc[linecolor=blue,linewidth=\elegant]{-}(1.2,0.8){0.2}{180}{0}
\psbezier[linecolor=blue,linewidth=\elegant]{-}(0.6,0.8)(0.6,0.4)(1.0,0.4)(1.0,0)
\psbezier[linecolor=blue,linewidth=\elegant]{-}(0.2,0.8)(0.2,0.45)(0.05,0.4)(0,0.4)
\psbezier[linecolor=blue,linewidth=\elegant]{-}(1.4,0)(1.4,0.35)(1.55,0.4)(1.6,0.4)
}
\end{pspicture} \ = \beta \ \ 
\begin{pspicture}[shift=-0.525](-0.0,-0.25)(1.6,0.8)
\pspolygon[fillstyle=solid,fillcolor=lightlightblue,linecolor=black,linewidth=0pt](0,0)(0,0.8)(1.6,0.8)(1.6,0)(0,0)
\psarc[linecolor=blue,linewidth=\elegant]{-}(0,0){0.2}{0}{90}
\psarc[linecolor=blue,linewidth=\elegant]{-}(1.6,0){0.2}{90}{180}
\psarc[linecolor=blue,linewidth=\elegant]{-}(1.2,0.8){0.2}{180}{0}
\psline[linecolor=blue,linewidth=\elegant]{-}(0.6,0)(0.6,0.8)
\psbezier[linecolor=blue,linewidth=\elegant]{-}(1.0,0)(1.0,0.3)(1.4,0.45)(1.6,0.45)
\psbezier[linecolor=blue,linewidth=\elegant]{-}(0.2,0.8)(0.2,0.55)(0.05,0.5)(0,0.5)
\end{pspicture} \ ,
\hspace{1.0cm}
\psset{unit=0.9}
\begin{pspicture}[shift=-0.7](-0.0,0)(1.6,1.6)
\pspolygon[fillstyle=solid,fillcolor=lightlightblue,linecolor=black,linewidth=0pt](0,-0.1)(0,0.8)(1.6,0.8)(1.6,-0.1)(0,-0.1)
\rput(0,0.8){\pspolygon[fillstyle=solid,fillcolor=lightlightblue,linecolor=black,linewidth=0pt](0,0)(0,0.8)(1.6,0.8)(1.6,0)(0,0)}
\psarc[linecolor=blue,linewidth=\elegant]{-}(0.8,0.8){0.2}{180}{360}
\psbezier[linecolor=blue,linewidth=\elegant]{-}(0.2,0.8)(0.2,0.3)(1.4,0.3)(1.4,0.8)
\rput(0,-0.1){
\psarc[linecolor=blue,linewidth=\elegant]{-}(0.4,0){0.2}{0}{180}
\psbezier[linecolor=blue,linewidth=\elegant]{-}(1.0,0)(1.0,0.45)(0.15,0.43)(-0.02,0.28)
\psbezier[linecolor=blue,linewidth=\elegant]{-}(1.4,0)(1.4,0.2)(1.56,0.26)(1.62,0.29)
\psframe[fillstyle=solid,linecolor=white,linewidth=0pt](-0.1,0)(-0.005,0.9)
\psframe[fillstyle=solid,linecolor=white,linewidth=0pt](1.605,0)(1.7,0.9)}
\rput(0,0.8){
\psarc[linecolor=blue,linewidth=\elegant]{-}(0.4,0.8){0.2}{180}{0}
\psarc[linecolor=blue,linewidth=\elegant]{-}(1.6,0){0.2}{90}{180}
\psarc[linecolor=blue,linewidth=\elegant]{-}(0,0){0.2}{0}{90}
\psbezier[linecolor=blue,linewidth=\elegant]{-}(0.6,0)(0.6,0.4)(1.0,0.4)(1.0,0.8)
\psbezier[linecolor=blue,linewidth=\elegant]{-}(1.0,0)(1.0,0.4)(1.4,0.4)(1.4,0.8)
}
\end{pspicture} \ = \alpha \ \ 
\begin{pspicture}[shift=-0.525](-0.0,-0.25)(1.6,0.8)
\pspolygon[fillstyle=solid,fillcolor=lightlightblue,linecolor=black,linewidth=0pt](0,0)(0,0.8)(1.6,0.8)(1.6,0)(0,0)
\psarc[linecolor=blue,linewidth=\elegant]{-}(0.4,0){0.2}{0}{180}
\psarc[linecolor=blue,linewidth=\elegant]{-}(0.4,0.8){0.2}{180}{0}
\psarc[linecolor=blue,linewidth=\elegant]{-}(1.2,0.8){0.2}{180}{0}
\psbezier[linecolor=blue,linewidth=\elegant]{-}(1.0,0)(1.0,0.45)(0.15,0.43)(-0.02,0.28)
\psbezier[linecolor=blue,linewidth=\elegant]{-}(1.4,0)(1.4,0.2)(1.56,0.26)(1.62,0.29)
\psframe[fillstyle=solid,linecolor=white,linewidth=0pt](-0.1,0)(-0.005,0.8)
\psframe[fillstyle=solid,linecolor=white,linewidth=0pt](1.605,0)(1.7,0.8)
\end{pspicture} \ .
\end{equation}

This algebra is generated by $N+2$ connectivity diagrams: $\Omega$, $\Omega^{-1}$, and $e_j$ with $j = 1, \dots, N$. These are depicted as
\begin{equation}
\begin{pspicture}[shift=-0.5](-0.7,-0.55)(2.0,0.35)
\rput(0.2,-0.55){$_1$}\rput(0.6,-0.55){$_2$}\rput(1.0,-0.55){$_3$}\rput(1.4,-0.55){\small$...$}\rput(1.8,-0.55){$_N$}
\pspolygon[fillstyle=solid,fillcolor=lightlightblue,linewidth=0pt](0,-0.35)(2.0,-0.35)(2.0,0.35)(0,0.35)
\multiput(0,0)(0.4,0){6}{\psbezier[linecolor=blue,linewidth=1.5pt]{-}(-0.2,-0.35)(-0.2,-0.0)(0.2,0.0)(0.2,0.35)}
\psframe[fillstyle=solid,linecolor=white,linewidth=0pt](-0.3,-0.4)(-0.005,0.4)
\psframe[fillstyle=solid,linecolor=white,linewidth=0pt](2.005,-0.4)(2.4,0.4)
\rput(-0.55,0.042){$\Omega=$}
\end{pspicture} \ ,
\quad \ \ 
\begin{pspicture}[shift=-0.5](-1.1,-0.55)(2.0,0.35)
\rput(0.2,-0.55){$_1$}\rput(0.6,-0.55){$_2$}\rput(1.0,-0.55){$_3$}\rput(1.4,-0.55){\small$...$}\rput(1.8,-0.55){$_N$}
\pspolygon[fillstyle=solid,fillcolor=lightlightblue,linewidth=0pt](0,-0.35)(2.0,-0.35)(2.0,0.35)(0,0.35)
\multiput(0,0)(0.4,0){6}{\psbezier[linecolor=blue,linewidth=1.5pt]{-}(-0.2,0.35)(-0.2,-0.0)(0.2,0.0)(0.2,-0.35)}
\psframe[fillstyle=solid,linecolor=white,linewidth=0pt](-0.3,-0.4)(-0.005,0.4)
\psframe[fillstyle=solid,linecolor=white,linewidth=0pt](2.005,-0.4)(2.4,0.4)
\rput(-0.75,0.07){$\Omega^{-1}=$}
\end{pspicture} \ ,
\quad
e_j =  \
\begin{pspicture}[shift=-0.525](-0.0,-0.55)(3.2,0.35)
\pspolygon[fillstyle=solid,fillcolor=lightlightblue,linecolor=black,linewidth=0pt](0,-0.35)(0,0.35)(3.2,0.35)(3.2,-0.35)(0,-0.35)
\psline[linecolor=blue,linewidth=1.5pt]{-}(0.2,-0.35)(0.2,0.35)
\rput(0.6,0){$...$}
\psline[linecolor=blue,linewidth=1.5pt]{-}(1.0,-0.35)(1.0,0.35)
\psarc[linecolor=blue,linewidth=1.5pt]{-}(1.6,0.35){0.2}{180}{360}
\psarc[linecolor=blue,linewidth=1.5pt]{-}(1.6,-0.35){0.2}{0}{180}
\psline[linecolor=blue,linewidth=1.5pt]{-}(2.2,-0.35)(2.2,0.35)
\rput(2.6,0){$...$}
\psline[linecolor=blue,linewidth=1.5pt]{-}(3.0,-0.35)(3.0,0.35)
\rput(0.2,-0.6){$_1$}
\rput(1.4,-0.6){$_{\phantom{+}j\phantom{+}}$}
\rput(3.0,-0.6){$_N$}
\end{pspicture} \ ,  
\quad
e_N= \
\begin{pspicture}[shift=-0.45](0,-0.55)(2.4,0.35)
\pspolygon[fillstyle=solid,fillcolor=lightlightblue,linewidth=0pt](0,-0.35)(2.4,-0.35)(2.4,0.35)(0,0.35)
\rput(0.2,-0.55){$_1$}\rput(0.6,-0.55){$_2$}\rput(1.0,-0.55){$_3$}\rput(2.2,-0.55){$_N$}
\rput(1.4,0.0){\small$...$}
\psarc[linecolor=blue,linewidth=1.5pt]{-}(0.0,0.35){0.2}{-90}{0}
\psarc[linecolor=blue,linewidth=1.5pt]{-}(0.0,-0.35){0.2}{0}{90}
\psline[linecolor=blue,linewidth=1.5pt]{-}(0.6,0.35)(0.6,-0.35)
\psline[linecolor=blue,linewidth=1.5pt]{-}(1.0,0.35)(1.0,-0.35)
\psline[linecolor=blue,linewidth=1.5pt]{-}(1.8,0.35)(1.8,-0.35)
\psarc[linecolor=blue,linewidth=1.5pt]{-}(2.4,-0.35){0.2}{90}{180}
\psarc[linecolor=blue,linewidth=1.5pt]{-}(2.4,0.35){0.2}{180}{-90}
\psframe[fillstyle=solid,linecolor=white,linewidth=0pt](-0.1,-0.4)(-0.005,0.4)
\psframe[fillstyle=solid,linecolor=white,linewidth=0pt](2.405,-0.4)(2.5,0.4)
\end{pspicture}\ ,
\end{equation}
where $j = 1, \dots, N-1$.
The identity $\boldsymbol I$ for this algebra is the element
\begin{equation}
\psset{unit=0.9}
\boldsymbol I = \Omega^{-1}\Omega = \Omega\Omega^{-1} = \ 
\begin{pspicture}[shift=-0.525](0,-0.25)(2.4,0.8)
\pspolygon[fillstyle=solid,fillcolor=lightlightblue,linecolor=black,linewidth=0pt](0,0)(0,0.8)(2.4,0.8)(2.4,0)(0,0)
\psline[linecolor=blue,linewidth=1.5pt]{-}(0.2,0)(0.2,0.8)
\psline[linecolor=blue,linewidth=1.5pt]{-}(0.6,0)(0.6,0.8)
\psline[linecolor=blue,linewidth=1.5pt]{-}(1.0,0)(1.0,0.8)
\rput(1.4,0.4){$...$}
\psline[linecolor=blue,linewidth=1.5pt]{-}(1.8,0)(1.8,0.8)
\psline[linecolor=blue,linewidth=1.5pt]{-}(2.2,0)(2.2,0.8)
\rput(0.2,-0.25){$_1$}
\rput(0.6,-0.25){$_2$}
\rput(1.0,-0.25){$_3$}
\rput(2.2,-0.25){$_N$}
\end{pspicture}\ .
\end{equation}
The diagrammatic rule describing the product in this algebra can be translated into relations satisfied by the generators. These are given for instance in \cite{PRV10}. The value of $\beta$ pertaining to the model of critical dense polymers is $\beta = 0$. In our calculations below, $\alpha$ is kept as a free parameter and $N$ is set to an even integer.

\paragraph{The transfer tangle and the Hamiltonian.}

The transfer tangle for the model of critical dense polymers with periodic boundary conditions is an element of $\eptl_N(\alpha, 0)$ defined as
\begin{equation}
\Tb (u)= \ 
\psset{unit=0.8}
\begin{pspicture}[shift=-1.1](-0.2,-0.7)(5.2,1.0)
\facegrid{(0,0)}{(5,1)}
\psarc[linewidth=0.025]{-}(0,0){0.16}{0}{90}
\psarc[linewidth=0.025]{-}(1,0){0.16}{0}{90}
\psarc[linewidth=0.025]{-}(4,0){0.16}{0}{90}
\psline[linewidth=1.5pt,linecolor=blue]{-}(0,0.5)(-0.2,0.5)
\psline[linewidth=1.5pt,linecolor=blue]{-}(5,0.5)(5.2,0.5)
\rput(2.5,0.5){$\ldots$}
\rput(3.5,0.5){$\ldots$}
\rput(0.5,.5){$u$}
\rput(1.5,.5){$u$}
\rput(4.5,.5){$u$}
\rput(2.5,-0.5){$\underbrace{\ \hspace{3.8cm} \ }_N$}
\end{pspicture}\ ,
\qquad 
 \begin{pspicture}[shift=-.40](1,1)
\facegrid{(0,0)}{(1,1)}
\psarc[linewidth=0.025]{-}(0,0){0.16}{0}{90}
\rput(.5,.5){$u$}
\end{pspicture}
\ = \cos u\ \
\begin{pspicture}[shift=-.40](1,1)
\facegrid{(0,0)}{(1,1)}
\rput[bl](0,0){\loopa}
\end{pspicture}
\ + \sin u \ \
\begin{pspicture}[shift=-.40](1,1)
\facegrid{(0,0)}{(1,1)}
\rput[bl](0,0){\loopb}
\end{pspicture}\ \ . 
\label{eq:Tu}
\end{equation}
It is a linear combination of $2^N$ connectivity diagrams. The isotropic value is $u = \frac \pi 4$, and we use the short-hand notation $\Tb=\Tb(\frac\pi4) $. This transfer tangle commutes for different values of the spectral parameter: $[\Tb(u),\Tb(v)] = 0$. Expanding $\Tb(u)$ in a Taylor series in $u$, we obtain
\begin{equation}
\label{eq:TH}
\Tb(u) = \Omega (\Ib- u \boldsymbol{H}) + \mathcal O(u^2), \qquad \boldsymbol{H} = - \sum_{j=1}^{N} e_j,
\end{equation}
where $\boldsymbol H$ is the Hamiltonian. It commutes with the transfer tangle: $[\boldsymbol H, \Tb(u)] = 0$. 

\paragraph{The standard modules $\boldsymbol{\stanW_{N,0}}$ and $\boldsymbol{\stanW_{N,2}}$.}

The algebra $\eptl_N(\alpha,\beta)$ possesses a family of standard modules $\stanW_{N,d}$ labelled by an integer number $d$ of defects. Our calculation of $\mathcal F_p^\alpha$ requires two standard modules: $\stanW_{N,0}$ and $\stanW_{N,2}$. These are respectively defined on the vector spaces generated by link states with zero and two defects. Link states are diagrams drawn over a segment where $N$ marked nodes are connected pairwise by non-intersecting loop segments or occupied by vertical defects that cannot be overarched. 

The boundary conditions are periodic in the horizontal direction. For example, here are the link states for $N=4$:
\begin{subequations}
\label{eq:link.states.bases}
\begin{alignat}{2}
&
\stanW_{4,0}: \ \ 
\psset{unit=0.8cm}
\begin{pspicture}[shift=-0.0](-0.0,0)(1.6,0.5)
\psline{-}(0,0)(1.6,0)
\psarc[linecolor=blue,linewidth=1.5pt]{-}(0.4,0){0.2}{0}{180}
\psarc[linecolor=blue,linewidth=1.5pt]{-}(1.2,0){0.2}{0}{180}
\end{pspicture}\ , \ \ 
\begin{pspicture}[shift=-0.0](-0.0,0)(1.6,0.5)
\psline{-}(0,0)(1.6,0)
\psbezier[linecolor=blue,linewidth=1.5pt]{-}(0.2,0)(0.2,0.7)(1.4,0.7)(1.4,0)
\psarc[linecolor=blue,linewidth=1.5pt]{-}(0.8,0){0.2}{0}{180}
\end{pspicture}\ , \ \ 
\begin{pspicture}[shift=-0.0](-0.0,0)(1.6,0.5)
\psline{-}(0,0)(1.6,0)
\psbezier[linecolor=blue,linewidth=1.5pt]{-}(0.6,0)(0.6,0.7)(1.8,0.7)(1.8,0)
\psbezier[linecolor=blue,linewidth=1.5pt]{-}(-0.2,0.54)(-0.12,0.43)(0.2,0.4)(0.2,0)
\psarc[linecolor=blue,linewidth=1.5pt]{-}(1.2,0){0.2}{0}{180}
\psframe[fillstyle=solid,linecolor=white,linewidth=0pt](1.6,-0.1)(2.0,0.6)
\psframe[fillstyle=solid,linecolor=white,linewidth=0pt](0.0,-0.1)(-0.4,0.6)
\end{pspicture}\ , \ \ 
\begin{pspicture}[shift=-0.0](-0.0,0)(1.6,0.5)
\psline{-}(0,0)(1.6,0)
\psarc[linecolor=blue,linewidth=1.5pt]{-}(0,0){0.2}{0}{90}
\psarc[linecolor=blue,linewidth=1.5pt]{-}(0.8,0){0.2}{0}{180}
\psarc[linecolor=blue,linewidth=1.5pt]{-}(1.6,0){0.2}{90}{180}
\end{pspicture}\ , \ \ 
\begin{pspicture}[shift=-0.0](-0.0,0)(1.6,0.5)
\psline{-}(0,0)(1.6,0)
\psbezier[linecolor=blue,linewidth=1.5pt]{-}(1.8,0.54)(1.72,0.43)(1.4,0.4)(1.4,0)
\psbezier[linecolor=blue,linewidth=1.5pt]{-}(-0.2,0)(-0.2,0.7)(1.0,0.7)(1.0,0)
\psarc[linecolor=blue,linewidth=1.5pt]{-}(0.4,0){0.2}{0}{180}
\psframe[fillstyle=solid,linecolor=white,linewidth=0pt](1.6,-0.1)(2.0,0.6)
\psframe[fillstyle=solid,linecolor=white,linewidth=0pt](0.0,-0.1)(-0.4,0.6)
\end{pspicture}\ , \ \ 
\begin{pspicture}[shift=-0.0](-0.0,0)(1.6,0.5)
\psline{-}(0,0)(1.6,0)
\psbezier[linecolor=blue,linewidth=1.5pt]{-}(-0.12,0.53)(-0.07,0.545)(0.6,0.545)(0.6,0)
\psbezier[linecolor=blue,linewidth=1.5pt]{-}(1.72,0.53)(1.67,0.545)(1.0,0.545)(1.0,0)
\psarc[linecolor=blue,linewidth=1.5pt]{-}(0.0,0){0.2}{0}{90}
\psarc[linecolor=blue,linewidth=1.5pt]{-}(1.6,0){0.2}{90}{180}
\psframe[fillstyle=solid,linecolor=white,linewidth=0pt](1.6,-0.1)(2.0,0.6)
\psframe[fillstyle=solid,linecolor=white,linewidth=0pt](0.0,-0.1)(-0.4,0.6)
\end{pspicture}\ ,
\\[0.3cm]
\label{eq:link.states.2.4}
&
\stanW_{4,2}: \ \ 
\psset{unit=0.8cm}
\begin{pspicture}[shift=-0.0](-0.0,0)(1.6,0.5)
\psline{-}(0,0)(1.6,0)
\psline[linecolor=blue,linewidth=1.5pt]{-}(0.2,0)(0.2,0.6)
\psline[linecolor=blue,linewidth=1.5pt]{-}(0.6,0)(0.6,0.6)
\psarc[linecolor=blue,linewidth=1.5pt]{-}(1.2,0){0.2}{0}{180}
\end{pspicture}\ , \ \ 
\begin{pspicture}[shift=-0.0](-0.0,0)(1.6,0.5)
\psline{-}(0,0)(1.6,0)
\psline[linecolor=blue,linewidth=1.5pt]{-}(0.2,0)(0.2,0.6)
\psline[linecolor=blue,linewidth=1.5pt]{-}(1.4,0)(1.4,0.6)
\psarc[linecolor=blue,linewidth=1.5pt]{-}(0.8,0){0.2}{0}{180}
\end{pspicture}\ , \ \ 
\begin{pspicture}[shift=-0.0](-0.0,0)(1.6,0.5)
\psline{-}(0,0)(1.6,0)
\psline[linecolor=blue,linewidth=1.5pt]{-}(1.0,0)(1.0,0.6)
\psline[linecolor=blue,linewidth=1.5pt]{-}(1.4,0)(1.4,0.6)
\psarc[linecolor=blue,linewidth=1.5pt]{-}(0.4,0){0.2}{0}{180}
\end{pspicture}\ , \ \ 
\begin{pspicture}[shift=-0.0](-0.0,0)(1.6,0.5)
\psline{-}(0,0)(1.6,0)
\psarc[linecolor=blue,linewidth=1.5pt]{-}(0,0){0.2}{0}{90}
\psline[linecolor=blue,linewidth=1.5pt]{-}(0.6,0)(0.6,0.6)
\psline[linecolor=blue,linewidth=1.5pt]{-}(1.0,0)(1.0,0.6)
\psarc[linecolor=blue,linewidth=1.5pt]{-}(1.6,0){0.2}{90}{180}
\end{pspicture}\ .  
\end{alignat}
\end{subequations}
The action of a connectivity in $\eptl_N(\alpha,\beta)$ on a link state $w$ in $\mathsf W_{N,0}$ is similar to the action of $\eptl_N(\alpha,\beta)$ on itself. The link state $w$ is drawn on top of the connectivity, the resulting new link state is read from the bottom nodes, and multiplicative factors of $\alpha$ and $\beta$ are included for each non-contractible and contractible loop, respectively. Moreover, for $\mathsf W_{N,2}$, the result is set to zero if the two defects are connected. Here are examples of this standard action:
\begin{equation}\label{eq:standard.examples}
\psset{unit=0.9}
\begin{pspicture}[shift=-0.525](-0.0,-0.25)(1.6,1.3)
\pspolygon[fillstyle=solid,fillcolor=lightlightblue,linecolor=black,linewidth=0pt](0,0)(0,0.8)(1.6,0.8)(1.6,0)(0,0)
\psarc[linecolor=blue,linewidth=\elegant]{-}(0,0){0.2}{0}{90}
\psarc[linecolor=blue,linewidth=\elegant]{-}(1.6,0){0.2}{90}{180}
\psarc[linecolor=blue,linewidth=\elegant]{-}(1.2,0.8){0.2}{180}{0}
\psline[linecolor=blue,linewidth=\elegant]{-}(0.6,0)(0.6,0.8)
\psbezier[linecolor=blue,linewidth=\elegant]{-}(1.0,0)(1.0,0.3)(1.4,0.45)(1.6,0.45)
\psbezier[linecolor=blue,linewidth=\elegant]{-}(0.2,0.8)(0.2,0.55)(0.05,0.5)(0,0.5)
\rput(0,0.8){
\psbezier[linecolor=blue,linewidth=1.5pt]{-}(1.8,0.54)(1.72,0.43)(1.4,0.4)(1.4,0)
\psbezier[linecolor=blue,linewidth=1.5pt]{-}(-0.2,0)(-0.2,0.7)(1.0,0.7)(1.0,0)
\psarc[linecolor=blue,linewidth=1.5pt]{-}(0.4,0){0.2}{0}{180}
\psframe[fillstyle=solid,linecolor=white,linewidth=0pt](1.6,0)(2.0,0.6)
\psframe[fillstyle=solid,linecolor=white,linewidth=0pt](0.0,0)(-0.4,0.6)
}
\end{pspicture}
\ = \alpha \ \
\begin{pspicture}[shift=0](0.0,0)(1.6,0.5)
\psline[linewidth=\mince](0,0)(1.6,0)
\psbezier[linecolor=blue,linewidth=1.5pt]{-}(-0.12,0.53)(-0.07,0.545)(0.6,0.545)(0.6,0)
\psbezier[linecolor=blue,linewidth=1.5pt]{-}(1.72,0.53)(1.67,0.545)(1.0,0.545)(1.0,0)
\psarc[linecolor=blue,linewidth=1.5pt]{-}(0.0,0){0.2}{0}{90}
\psarc[linecolor=blue,linewidth=1.5pt]{-}(1.6,0){0.2}{90}{180}
\psframe[fillstyle=solid,linecolor=white,linewidth=0pt](1.6,-0.1)(2.0,0.6)
\psframe[fillstyle=solid,linecolor=white,linewidth=0pt](0.0,-0.1)(-0.15,0.6)
\end{pspicture} \ ,
\qquad
\begin{pspicture}[shift=-0.525](-0.0,-0.25)(1.6,1.3)
\pspolygon[fillstyle=solid,fillcolor=lightlightblue,linecolor=black,linewidth=0pt](0,0)(0,0.8)(1.6,0.8)(1.6,0)(0,0)
\psarc[linecolor=blue,linewidth=\elegant]{-}(0.4,0){0.2}{0}{180}
\psarc[linecolor=blue,linewidth=\elegant]{-}(1.2,0.8){0.2}{180}{0}
\psbezier[linecolor=blue,linewidth=\elegant]{-}(0.6,0.8)(0.6,0.4)(1.0,0.4)(1.0,0)
\psbezier[linecolor=blue,linewidth=\elegant]{-}(0.2,0.8)(0.2,0.45)(0.05,0.4)(0,0.4)
\psbezier[linecolor=blue,linewidth=\elegant]{-}(1.4,0)(1.4,0.35)(1.55,0.4)(1.6,0.4)
\rput(0,0.8)
{
\psline[linecolor=blue,linewidth=\elegant]{-}(0.2,0)(0.2,0.5)
\psline[linecolor=blue,linewidth=\elegant]{-}(0.6,0)(0.6,0.5)
\psarc[linecolor=blue,linewidth=\elegant]{-}(1.2,0){0.2}{0}{180}
}
\end{pspicture}
\ = \beta \ 
\begin{pspicture}[shift=0](0.0,0)(1.6,0.5)
\psline[linewidth=\mince](0,0)(1.6,0)
\psline[linecolor=blue,linewidth=\elegant]{-}(1.0,0)(1.0,0.5)
\psline[linecolor=blue,linewidth=\elegant]{-}(1.4,0)(1.4,0.5)
\psarc[linecolor=blue,linewidth=\elegant]{-}(0.4,0){0.2}{0}{180}\end{pspicture} \ ,
\qquad
\begin{pspicture}[shift=-0.525](-0.0,-0.25)(1.6,1.3)
\pspolygon[fillstyle=solid,fillcolor=lightlightblue,linecolor=black,linewidth=0pt](0,0)(0,0.8)(1.6,0.8)(1.6,0)(0,0)
\psarc[linecolor=blue,linewidth=\elegant]{-}(0.4,0.8){0.2}{180}{360}
\psarc[linecolor=blue,linewidth=\elegant]{-}(0,0){0.2}{0}{90}
\psarc[linecolor=blue,linewidth=\elegant]{-}(1.6,0){0.2}{90}{180}
\psbezier[linecolor=blue,linewidth=\elegant]{-}(0.6,0)(0.6,0.4)(1.0,0.4)(1.0,0.8)
\psbezier[linecolor=blue,linewidth=\elegant]{-}(1.0,0)(1.0,0.4)(1.4,0.4)(1.4,0.8)
\rput(0,0.8)
{
\psline[linecolor=blue,linewidth=\elegant]{-}(0.2,0)(0.2,0.5)
\psline[linecolor=blue,linewidth=\elegant]{-}(0.6,0)(0.6,0.5)
\psarc[linecolor=blue,linewidth=\elegant]{-}(1.2,0){0.2}{0}{180}
}
\end{pspicture}
\ = 0.
\end{equation}
We note that the standard module $\stanW_{N,2}$ can be defined more generally with a twist parameter that keeps track of how much the defects wrap around the cylinder.

\paragraph{A link state module with identified connectivities.}

In the special case where $\alpha = \beta$, there exists another representation defined on link states with zero defects, with so-called {\it identified connectivities}~\cite{PRV10}. We denote it by $\widehat \stanW_{N,0}$. Its vector space is spanned by the link states with zero defects in $\stanW_{N,0}$ that have no arcs travelling via the back of the cylinder. These are in fact the same link states that span the standard module with zero defects of the usual Temperley-Lieb algebra $\tl_N(\beta)$. For $N=4$, these link states are:
\be
\widehat\stanW_{4,0}: \ \ 
\psset{unit=0.8cm}
\begin{pspicture}[shift=-0.0](-0.0,0)(1.6,0.5)
\psline{-}(0,0)(1.6,0)
\psarc[linecolor=blue,linewidth=1.5pt]{-}(0.4,0){0.2}{0}{180}
\psarc[linecolor=blue,linewidth=1.5pt]{-}(1.2,0){0.2}{0}{180}
\end{pspicture}\ , \ \ 
\begin{pspicture}[shift=-0.0](-0.0,0)(1.6,0.5)
\psline{-}(0,0)(1.6,0)
\psbezier[linecolor=blue,linewidth=1.5pt]{-}(0.2,0)(0.2,0.7)(1.4,0.7)(1.4,0)
\psarc[linecolor=blue,linewidth=1.5pt]{-}(0.8,0){0.2}{0}{180}
\end{pspicture}\ .
\ee

The action of connectivity diagrams on link states in $\widehat \stanW_{N,0}$ is defined using the same construction as for standard modules. One draws the link state above the connectivity diagram, reads the new link state from the bottom edge of the diagram, and includes factors of $\beta$ for each closed loop, both contractible and non-contractible. Crucially, in reading off the resulting link state, there is no distinction between arcs travelling via the front or back of the cylinder. Any arc travelling around the back of the cylinder is transformed into an arc travelling in the front of the cylinder. Here are two examples of this action for $N=4$:
\begin{equation}\label{eq:standard.examples2}
\psset{unit=0.9}
\begin{pspicture}[shift=-0.525](-0.0,-0.25)(1.6,1.3)
\pspolygon[fillstyle=solid,fillcolor=lightlightblue,linecolor=black,linewidth=0pt](0,0)(0,0.8)(1.6,0.8)(1.6,0)(0,0)
\psarc[linecolor=blue,linewidth=\elegant]{-}(0,0){0.2}{0}{90}
\psarc[linecolor=blue,linewidth=\elegant]{-}(1.6,0){0.2}{90}{180}
\psarc[linecolor=blue,linewidth=\elegant]{-}(1.2,0.8){0.2}{180}{0}
\psline[linecolor=blue,linewidth=\elegant]{-}(0.6,0)(0.6,0.8)
\psbezier[linecolor=blue,linewidth=\elegant]{-}(1.0,0)(1.0,0.3)(1.4,0.45)(1.6,0.45)
\psbezier[linecolor=blue,linewidth=\elegant]{-}(0.2,0.8)(0.2,0.55)(0.05,0.5)(0,0.5)
\rput(0,0.8){
\psarc[linecolor=blue,linewidth=1.5pt]{-}(0.4,0){0.2}{0}{180}
\psarc[linecolor=blue,linewidth=1.5pt]{-}(1.2,0){0.2}{0}{180}
}
\end{pspicture}
\ = \beta \ 
\begin{pspicture}[shift=0](0.0,0)(1.6,0.5)
\psline[linewidth=\mince](0,0)(1.6,0)
\psarc[linecolor=blue,linewidth=1.5pt]{-}(0.8,0){0.2}{0}{180}
\psbezier[linecolor=blue,linewidth=\elegant]{-}(0.2,0)(0.2,0.5)(1.4,0.5)(1.4,0)
\end{pspicture} \ ,
\qquad
\begin{pspicture}[shift=-0.525](-0.0,-0.25)(1.6,1.3)
\pspolygon[fillstyle=solid,fillcolor=lightlightblue,linecolor=black,linewidth=0pt](0,0)(0,0.8)(1.6,0.8)(1.6,0)(0,0)
\psarc[linecolor=blue,linewidth=\elegant]{-}(0.4,0.8){0.2}{180}{0}
\psarc[linecolor=blue,linewidth=\elegant]{-}(0.4,0){0.2}{0}{180}
\psarc[linecolor=blue,linewidth=\elegant]{-}(1.2,0){0.2}{0}{180}
\psbezier[linecolor=blue,linewidth=\elegant]{-}(1.0,0.8)(1.0,0.35)(0.15,0.37)(-0.02,0.52)
\psbezier[linecolor=blue,linewidth=\elegant]{-}(1.4,0.8)(1.4,0.6)(1.56,0.54)(1.62,0.51)
\psframe[fillstyle=solid,linecolor=white,linewidth=0pt](-0.1,0)(-0.005,0.8)
\psframe[fillstyle=solid,linecolor=white,linewidth=0pt](1.605,0)(1.7,0.8)
\rput(0,0.8)
{
\psarc[linecolor=blue,linewidth=1.5pt]{-}(0.8,0){0.2}{0}{180}
\psbezier[linecolor=blue,linewidth=\elegant]{-}(0.2,0)(0.2,0.5)(1.4,0.5)(1.4,0)
}
\end{pspicture}
\ = \beta \ 
\begin{pspicture}[shift=0](0.0,0)(1.6,0.5)
\psline[linewidth=\mince](0,0)(1.6,0)
\psarc[linecolor=blue,linewidth=1.5pt]{-}(0.4,0){0.2}{0}{180}
\psarc[linecolor=blue,linewidth=1.5pt]{-}(1.2,0){0.2}{0}{180}
\end{pspicture} \ .
\end{equation}

\paragraph{The XX representation.}

The XX representation of $\eptl_N(\alpha,0)$, which we denote by $\mathsf X_N$, is defined on the vector space $(\mathbb C^2)^{\otimes N}$. We use the canonical basis defined in \cref{sec:defXX}. In this basis, the generators $e_j$ with $j = 1, \dots, N-1$ are represented by the following matrices \cite{PS90}:
\begin{equation}
\label{eq:X.rep}
\mathsf X_N(e_j) = \underbrace{\mathbb I_2 \otimes \dots \otimes \mathbb I_2}_{j-1} \otimes 
\begin{pmatrix}
0 & 0 & 0 & 0 \\
0 & \ir & 1 & 0 \\
0 & 1 & \ir^{-1} & 0 \\
0 & 0 & 0 & 0
\end{pmatrix}
\otimes \underbrace{\mathbb I_2 \otimes \dots \otimes \mathbb I_2}_{N-j-1}\, ,\qquad j = 1, \dots, N-1.
\end{equation}
Likewise, the generators $\Omega^{\pm 1}$ and $e_N$ are represented by 
\be
\mathsf X_N(\Omega^{\pm 1}) = t^{\pm 1} \eE^{\pm \ir \phi\sigma^z_1/2}, \qquad
\mathsf X_N(e_N) = \mathsf X_N(\Omega)\mathsf X_N(e_1)\mathsf X_N(\Omega^{-1}), 
\ee
where $\phi$ is the twist angle and $t$ is the translation operator:
\be
t \big(|v_1\rangle \otimes |v_2\rangle \otimes \cdots \otimes |v_N\rangle\big) = |v_2\rangle  \otimes \cdots \otimes |v_N\rangle \otimes |v_1\rangle.
\ee
The matrices $\mathsf X_N(e_j)$ and $\mathsf X_N(\Omega^{\pm 1})$ realise a representation of $\eptl_N(\alpha,0)$ for 
\begin{equation}
\label{eq:alpha.phi}
\alpha = 2 \cos \Big(\frac{\phi}{2} \Big).
\end{equation}
They also commute with the total magnetisation $S^z = \frac12\sum_{i=1}^N \sigma^z_i$. As a consequence, the representation $\mathsf X_N$ splits as a direct sum of smaller representations labelled by the eigenvalues $m$ of $S^z$, which take the values $m = -\frac N2$, $-\frac{N-2}2$, \dots, $\frac N2$. 

In this representation, the Hamiltonian $\boldsymbol H$ is the twisted XX Hamiltonian \eqref{eq:Hxx}:
\be
H = \mathsf X_N(\boldsymbol{H})  = - \sum_{j=1}^{N-1} \Big( \sigma^+_j \sigma^-_{j+1} + \sigma^-_j \sigma^+_{j+1} \Big) - \eE^{\ir \phi} \sigma^+_N \sigma^-_1 - \eE^{-\ir \phi} \sigma^-_N \sigma^+_1.
\ee
 Likewise, the transfer matrix is defined as $T(u) = \mathsf X_N(\Tb(u))$. It is the transfer matrix of the six-vertex model at the anisotropy $\Delta = 0$, with periodic boundary conditions and a diagonal twist. We use the notation $T=T(\frac \pi 4)$ for this transfer matrix at the isotropic point.

\paragraph{Homomorphisms.}

There exists a map from link states to spin states that intertwines the link state and XX representations. Indeed, for a given link state $w$ in $\stanW_{N,0}$, $\stanW_{N,2}$ or $\widehat \stanW_{N,0}$, we write its image in $(\mathbb C^2)^{\otimes N}$ under this map as $|w\rangle$. It is defined from the following local maps:
\begin{equation}
\label{eq:localmaps}
|\,
\psset{unit=0.74}
\begin{pspicture}[shift=-0.08](0.0,0)(0.8,0.5)
\psline[linewidth=\mince](0,0)(0.8,0)
\psarc[linecolor=blue,linewidth=\elegant]{-}(0.4,0){0.2}{0}{180}
\end{pspicture}
\,  \rangle = \omega\, |{\uparrow \downarrow}\rangle + \omega^{-1}\, |{\downarrow \uparrow}\rangle, \qquad
|\,
\begin{pspicture}[shift=-0.08](0.0,0)(0.4,0.5)
\psline[linewidth=\mince](0,0)(0.4,0)
\psline[linecolor=blue,linewidth=\elegant]{-}(0.2,0)(0.2,0.5)
\end{pspicture}
\, \rangle =  |{\downarrow}\rangle, \qquad 
|\,
\psset{unit=0.74}
\begin{pspicture}[shift=-0.08](0.0,0)(0.8,0.5)
\psline[linewidth=\mince](0,0)(0.8,0)
\psarc[linecolor=blue,linewidth=1.5pt]{-}(0,0){0.2}{0}{90}
\psarc[linecolor=blue,linewidth=1.5pt]{-}(0.8,0){0.2}{90}{180}
\end{pspicture}
\,  \rangle = \omega^{-1}\eE^{-\frac{\ir \phi}{2}}\, |{\uparrow \downarrow}\rangle + \omega \ \eE^{\frac{\ir \phi}{2}}\, |{\downarrow \uparrow}\rangle, \qquad
\omega = \eE^{\ir \pi/4}.
\end{equation}
For a given link state, these local rules are applied to each arc and each defect.
For $w \in \stanW_{N,0}$, $\stanW_{N,2}$ and $\widehat \stanW_{N,0}$, the resulting state $|w \rangle$ has the magnetisation $m=0,-1$ and $0$, respectively. The homomorphism property then reads
\be
\mathsf X_N(e_j) |w\rangle = |e_jw\rangle, \qquad \mathsf X_N(\Omega^{\pm 1}) |w\rangle = |\Omega^{\pm 1}w\rangle, \qquad j = 1, \dots, N,
\ee
and is satisfied for each link state $w$. This holds for $\stanW_{N,0}$, $\widehat \stanW_{N,0}$ and $\stanW_{N,2}$, with the action of the generators on the right sides of the equations adapted accordingly, for each case. For $\stanW_{N,0}$, the homomorphism with the spin-chain representation holds for $\alpha$ fixed as a function of $\phi$ as in \eqref{eq:alpha.phi}. For $\widehat \stanW_{N,0}$, non-contractible loops have the weight $\beta$ and the homomorphism holds provided that $\phi$ is fixed to $\pm\pi$. In that case,  
$
|\,
\psset{unit=0.74}
\begin{pspicture}[shift=-0.08](0.0,0)(0.8,0.5)
\psline[linewidth=\mince](0,0)(0.8,0)
\psarc[linecolor=blue,linewidth=\elegant]{-}(0.4,0){0.2}{0}{180}
\end{pspicture}
\,  \rangle 
$
and
$
|\,
\psset{unit=0.74}
\begin{pspicture}[shift=-0.08](0.0,0)(0.8,0.5)
\psline[linewidth=\mince](0,0)(0.8,0)
\psarc[linecolor=blue,linewidth=1.5pt]{-}(0,0){0.2}{0}{90}
\psarc[linecolor=blue,linewidth=1.5pt]{-}(0.8,0){0.2}{90}{180}
\end{pspicture}
\,  \rangle
$ 
are equal up to a sign.
For $\stanW_{N,2}$, the map \eqref{eq:localmaps} is a homomorphism for the special value $\phi=0$. This is consistent with the fact that our definition of this module was given without including a parameter that keeps track of the winding of the defects.

\subsubsection{Bilinear forms}\label{sec:forms}

\paragraph{Bilinear forms for link states.}

The Gram bilinear form is an invariant form on the standard modules. For critical dense polymers, it is defined as follows. Let $w,w'$ be two link states in $\stanW_{N,0}$ or $\stanW_{N,2}$. Performing a vertical flip of $w$ and connecting its nodes to those of $w'$, we obtain a diagram where the loop segments form loops, that can be contractible or non-contractible loops. For $\mathsf W_{N,0}$, the Gram product of $w$ and $w'$, denoted $w \cdot w'$, is defined as $\alpha^{n_{\alpha}}\delta_{n_\beta,0} $, where $n_\alpha$ and $n_\beta$ count the non-contractible and the contractible loops. It is therefore non-zero only if the number of contractible loops is zero. 

For $\stanW_{N,2}$, the same diagram constructed from $w$ and $w'$ involves loops as before, but also finds the defects connected pairwise. If both defects of $w$ are connected to defects of $w'$ and there are no loops, then $w \cdot w' = 1$. Otherwise, $w\cdot w' = 0$. To illustrate, for $N=4$, the matrices encoding the Gram products between the link states in the bases \eqref{eq:link.states.bases} are
\be
\left(
\begin{array}{cccccc}
 0 & 0 & 0 & \alpha  & 0 & 0 \\
 0 & 0 & \alpha  & 0 & \alpha  & \alpha ^2 \\
 0 & \alpha  & 0 & 0 & \alpha ^2 & \alpha  \\
 \alpha  & 0 & 0 & 0 & 0 & 0 \\
 0 & \alpha  & \alpha ^2 & 0 & 0 & \alpha  \\
 0 & \alpha ^2 & \alpha  & 0 & \alpha  & 0 \\
\end{array}
\right), 
\qquad
\left(
\begin{array}{cccc}
 0 & 1 & 0 & 1 \\
 1 & 0 & 1 & 0 \\
 0 & 1 & 0 & 1 \\
 1 & 0 & 1 & 0 \\
\end{array}
\right).
\ee

These bilinear forms allow us to express the partition functions $Z^{A\cup B}_c$, $Z^A_c$ and $Z^B_c$ in terms of Gram products:
\begin{subequations}
\label{eq:ZFpants}
\begin{alignat}{2}
Z_c^{A\cup B}&=2^{MN} v^{AB}_0 \cdot (\Tb^{AB})^{2M} v^{AB}_0\big|_{\stanW_{N,0}}\ ,\\[0.2cm] 
Z_c^{A} &=2^{MN_A} v^A_0 \cdot (\Tb^{A})^{2M} v^{A}_0\big|_{\stanW_{N_A,0}}\ ,\\[0.2cm]
Z_c^{B} &=2^{MN_B} v^B_2 \cdot (\Tb^{B})^{2M} v^B_2\big|_{\stanW_{N_B,2}}\ ,
\end{alignat}
\end{subequations}
where the boundary states are
\begin{equation}
\label{eq:v0v2}
\psset{unit=0.8}
v_0 = \
\begin{pspicture}[shift=0](0.0,0)(4.8,0.5)
\psline[linewidth=\mince](0,0)(4.8,0)
\psarc[linecolor=blue,linewidth=\elegant]{-}(0.4,0){0.2}{0}{180}
\psarc[linecolor=blue,linewidth=\elegant]{-}(1.2,0){0.2}{0}{180}
\psarc[linecolor=blue,linewidth=\elegant]{-}(2.0,0){0.2}{0}{180}
\rput(2.8,0.15){...}
\psarc[linecolor=blue,linewidth=\elegant]{-}(3.6,0){0.2}{0}{180}
\psarc[linecolor=blue,linewidth=\elegant]{-}(4.4,0){0.2}{0}{180}
\end{pspicture}\ ,
\qquad
v_2 = \
\begin{pspicture}[shift=0](0.0,0)(4.8,0.5)
\psline[linewidth=\mince](0,0)(4.8,0)
\psarc[linecolor=blue,linewidth=\elegant]{-}(0.4,0){0.2}{0}{180}
\psarc[linecolor=blue,linewidth=\elegant]{-}(1.2,0){0.2}{0}{180}
\psarc[linecolor=blue,linewidth=\elegant]{-}(2.8,0){0.2}{0}{180}
\psarc[linecolor=blue,linewidth=\elegant]{-}(3.6,0){0.2}{0}{180}
\psline[linecolor=blue,linewidth=\elegant]{-}(4.2,0)(4.2,0.5)
\psline[linecolor=blue,linewidth=\elegant]{-}(4.6,0)(4.6,0.5)
\rput(2.0,0.15){...}
\end{pspicture}
 \ . 
\end{equation}
The superscripts $AB$, $A$ and $B$ for $\Tb$ in \eqref{eq:ZFpants} serve as a reminder that the corresponding objects are elements of the enlarged periodic Temperley-Lieb algebra with $N$, $N_A$ and $N_B$ nodes, respectively. We have also indicated by subscripts on the right side which action is used. Finally, we note that the powers of $2$ ensure that each tile has a weight $1$ instead of $\frac1{\sqrt2}$ as it does in \eqref{eq:Tu} for $u = \frac \pi 4$. 

\paragraph{Gram products on the pants geometry.}

We define a new Gram product, denoted $(w_1 \times w_2) \cdot w_3$, that is needed to compute $Z^{AB}_p$. In this product, the states $w_1$, $w_2$ and $w_3$ belong to $\stanW_{N_A, 0}$, $\widehat\stanW_{N_B, 0}$ and $\stanW_{N, 0}$ respectively, with $N = N_A + N_B$. The result of this product is obtained as follows. One flips $w_1$ and $w_2$ vertically, draws them on the legs $A$ and $B$ of the pants lattice, and connects its nodes to those of $w_3$ drawn on the top part of the pants lattice. Then $(w_1 \times w_2) \cdot w_3$ equals $\alpha^{n_{\rm A} + n_{\rm AB}} \delta_{n_\beta,0} \delta_{n_{\rm B},0}$, where $n_\beta$ counts the number of contractible loops, and the numbers $n_{\rm A}$, $n_{\rm B}$ and $n_{\rm AB}$ count the non-contractible loops in the three families, as explained in \cref{sec:def}. Here are two examples to illustrate:
\begin{subequations}
\begin{alignat}{2}
\Big(\,
\begin{pspicture}[shift=-0.1](-0.0,0)(1.6,0.5)
\psline{-}(0,0)(1.6,0)
\psbezier[linecolor=blue,linewidth=1.5pt]{-}(1.8,0.54)(1.72,0.43)(1.4,0.4)(1.4,0)
\psbezier[linecolor=blue,linewidth=\elegant]{-}(1.0,0)(1.0,0.58)(0.15,0.53)(-0.02,0.37)
\psarc[linecolor=blue,linewidth=1.5pt]{-}(0.4,0){0.2}{0}{180}
\psframe[fillstyle=solid,linecolor=white,linewidth=0pt](1.6,-0.1)(1.9,0.6)
\psframe[fillstyle=solid,linecolor=white,linewidth=0pt](0.0,-0.1)(-0.05,0.6)
\end{pspicture}
\, \times \,
\begin{pspicture}[shift=-0.1](-0.0,0)(0.8,0.5)
\psline{-}(0,0)(0.8,0)
\psarc[linecolor=blue,linewidth=1.5pt]{-}(0.4,0){0.2}{0}{180}
\end{pspicture}\,
 \Big) \cdot \,
\begin{pspicture}[shift=-0.1](-0.0,0)(2.4,0.5)
\psline{-}(0,0)(2.4,0)
\psarc[linecolor=blue,linewidth=1.5pt]{-}(1.2,0){0.2}{0}{180}
\psbezier[linecolor=blue,linewidth=1.5pt]{-}(0.6,0.0)(0.6,0.61)(1.8,0.61)(1.8,0)
\psarc[linecolor=blue,linewidth=1.5pt]{-}(0,0){0.2}{0}{90}
\psarc[linecolor=blue,linewidth=1.5pt]{-}(2.4,0){0.2}{90}{180}
\end{pspicture}
\ &= \alpha^2,
\\[0.2cm]
\Big(\,
\begin{pspicture}[shift=-0.1](-0.0,0)(1.6,0.5)
\psline{-}(0,0)(1.6,0)
\psbezier[linecolor=blue,linewidth=\elegant]{-}(0.6,0)(0.6,0.58)(1.45,0.53)(1.62,0.37)
\psarc[linecolor=blue,linewidth=1.5pt]{-}(1.2,0){0.2}{0}{180}
\psbezier[linecolor=blue,linewidth=1.5pt]{-}(-0.05,0.40)(-0,0.34)(0.2,0.26)(0.2,0)
\psframe[fillstyle=solid,linecolor=white,linewidth=0pt](0,-0.1)(-0.07,0.6)
\psframe[fillstyle=solid,linecolor=white,linewidth=0pt](1.6,-0.1)(1.67,0.6)
\end{pspicture}
\, \times \,
\begin{pspicture}[shift=-0.1](-0.0,0)(0.8,0.5)
\psline{-}(0,0)(0.8,0)
\psarc[linecolor=blue,linewidth=1.5pt]{-}(0.4,0){0.2}{0}{180}
\end{pspicture}\,
 \Big) \cdot \,
\begin{pspicture}[shift=-0.1](-0.0,0)(2.4,0.5)
\psline{-}(0,0)(2.4,0)
\psarc[linecolor=blue,linewidth=1.5pt]{-}(0.4,0){0.2}{0}{180}
\psarc[linecolor=blue,linewidth=1.5pt]{-}(1.6,0){0.2}{0}{180}
\psbezier[linecolor=blue,linewidth=1.5pt]{-}(1.0,0.0)(1.0,0.63)(2.2,0.63)(2.2,0)
\end{pspicture}
\ &= 0.
\end{alignat}
\end{subequations}
In terms of this Gram product, the partition function $Z^{AB}_p$ reads
\be
\label{eq:ZAB.standard}
Z_p^{AB} = 2^{MN}\Big((\Tb^{A})^M v_0^{A} \times (\Tb^{B})^M v_0^{B} \Big)\cdot (\Tb^{AB})^M v_0^{AB}.
\ee
In this expression, the transfer tangles $\Tb^{A}$ and $\Tb^{AB}$ act on the boundary states $v_0^{A}$ and $v_0^{AB}$ under the standard actions of $W_{N_A,0}$ and $W_{N,0}$ respectively (where $\alpha$ is a free parameter). In contrast, $(\Tb^{B})^M$ acts on $v_0^{B}$ under the module action of $\widehat W_{N_B,0}$ (where $\alpha = 0$).

We note that one can define more generally a Gram product on the pants geometry where the result of the product is $\alpha_1^{n_{\rm A}} \alpha_2^{n_{\rm B}} \alpha_3^{n_{\rm AB}} \delta_{n_\beta,0}$ and thus depends on three free parameters, one for each family of non-contractible loops. Our choice to set $\alpha_2 = 0$ and $\alpha_1 = \alpha_3 = \alpha$ however allows for an important simplification. Indeed, in this case, the connectivity of link states in leg $B$ can be described using the module $\widehat\stanW_{N_B,0}$, and the Gram products on the pants geometry can be computed using the Gram product on the cylinder, for $\stanW_{N,0}$. Our construction above achieves this by using the natural embedding from $\stanW_{N_A,0}\times \widehat\stanW_{N_B,0}$ into $\stanW_{N,0}$, where the two link states $w_1$ and $w_2$ are simply drawn side by side, and the loop segments of $w_1$ that travel via the back of the cylinder (if any) are extended to travel via the back of the larger cylinder. For instance, 
\be
\label{eq:embedding.examples}
\begin{pspicture}[shift=0](-0.0,0)(1.6,0.5)
\psline{-}(0,0)(1.6,0)
\psbezier[linecolor=blue,linewidth=1.5pt]{-}(1.8,0.54)(1.72,0.43)(1.4,0.4)(1.4,0)
\psbezier[linecolor=blue,linewidth=\elegant]{-}(1.0,0)(1.0,0.58)(0.15,0.53)(-0.02,0.37)
\psarc[linecolor=blue,linewidth=1.5pt]{-}(0.4,0){0.2}{0}{180}
\psframe[fillstyle=solid,linecolor=white,linewidth=0pt](1.6,-0.1)(1.9,0.6)
\psframe[fillstyle=solid,linecolor=white,linewidth=0pt](0.0,-0.1)(-0.05,0.6)
\end{pspicture}
\, \times \,
\begin{pspicture}[shift=0](-0.0,0)(0.8,0.5)
\psline{-}(0,0)(0.8,0)
\psarc[linecolor=blue,linewidth=1.5pt]{-}(0.4,0){0.2}{0}{180}
\end{pspicture}
\ \ \mapsto \ \
\begin{pspicture}[shift=0](-0.0,0)(2.4,0.5)
\psline{-}(0,0)(2.4,0)
\psbezier[linecolor=blue,linewidth=1.5pt]{-}(1.4,0)(1.4,0.5)(2.3,0.53)(2.4,0.53)
\psbezier[linecolor=blue,linewidth=1.5pt]{-}(1.0,0)(1.0,0.5)(0.1,0.53)(0.0,0.53)
\psarc[linecolor=blue,linewidth=1.5pt]{-}(0.4,0){0.2}{0}{180}
\psarc[linecolor=blue,linewidth=1.5pt]{-}(2.0,0){0.2}{0}{180}
\end{pspicture}\ \ ,
\qquad
\begin{pspicture}[shift=0](-0.0,0)(1.6,0.5)
\psline{-}(0,0)(1.6,0)
\psbezier[linecolor=blue,linewidth=\elegant]{-}(0.6,0)(0.6,0.58)(1.45,0.53)(1.62,0.37)
\psarc[linecolor=blue,linewidth=1.5pt]{-}(1.2,0){0.2}{0}{180}
\psbezier[linecolor=blue,linewidth=1.5pt]{-}(-0.05,0.40)(-0,0.34)(0.2,0.26)(0.2,0)
\psframe[fillstyle=solid,linecolor=white,linewidth=0pt](0,-0.1)(-0.07,0.6)
\psframe[fillstyle=solid,linecolor=white,linewidth=0pt](1.6,-0.1)(1.67,0.6)
\end{pspicture}
\, \times \,
\begin{pspicture}[shift=0](-0.0,0)(0.8,0.5)
\psline{-}(0,0)(0.8,0)
\psarc[linecolor=blue,linewidth=1.5pt]{-}(0.4,0){0.2}{0}{180}
\end{pspicture}
\ \ \mapsto \ \
\begin{pspicture}[shift=0](-0.0,0)(2.4,0.5)
\psline{-}(0,0)(2.4,0)
\psarc[linecolor=blue,linewidth=1.5pt]{-}(1.2,0){0.2}{0}{180}
\psbezier[linecolor=blue,linewidth=1.5pt]{-}(-0.05,0.40)(-0,0.34)(0.2,0.26)(0.2,0)
\psarc[linecolor=blue,linewidth=1.5pt]{-}(2.0,0){0.2}{0}{180}
\psbezier[linecolor=blue,linewidth=\elegant]{-}(0.6,0)(0.6,0.58)(2.0,0.68)(2.42,0.31)
\psframe[fillstyle=solid,linecolor=white,linewidth=0pt](0,-0.1)(-0.07,0.6)
\psframe[fillstyle=solid,linecolor=white,linewidth=0pt](2.4,-0.1)(2.47,0.6)
\end{pspicture}\ \ .
\ee
The same embedding in $\stanW_{N,0}$ does not naturally extended to $\stanW_{N_A,0}\times \stanW_{N_B,0}$. For example, we cannot embed 
$
\psset{unit=0.5}
\begin{pspicture}[shift=0](-0.0,0)(1.6,0.5)
\psline{-}(0,0)(1.6,0)
\psbezier[linecolor=blue,linewidth=1.2pt]{-}(1.8,0.54)(1.72,0.43)(1.4,0.4)(1.4,0)
\psbezier[linecolor=blue,linewidth=1.2pt]{-}(1.0,0)(1.0,0.58)(0.15,0.53)(-0.02,0.37)
\psarc[linecolor=blue,linewidth=1.2pt]{-}(0.4,0){0.2}{0}{180}
\psframe[fillstyle=solid,linecolor=white,linewidth=0pt](1.6,-0.1)(1.9,0.6)
\psframe[fillstyle=solid,linecolor=white,linewidth=0pt](0.0,-0.1)(-0.05,0.6)
\end{pspicture}
\, \times \,
\begin{pspicture}[shift=0](-0.0,0)(0.8,0.5)
\psline{-}(0,0)(0.8,0)
\psarc[linecolor=blue,linewidth=1.2pt]{-}(0,0){0.2}{0}{90}
\psarc[linecolor=blue,linewidth=1.2pt]{-}(0.8,0){0.2}{90}{180}
\end{pspicture}
$ in $\stanW_{N,0}$ in the same way as in \eqref{eq:embedding.examples}.

Thus, for the case $\alpha_2=0$, any Gram product on the pants geometry can be computed using the same product on the cylinder. In the spin-chain language, the resulting embedding translates simply to
\be
|w_1 \times w_2 \rangle = |w_1\rangle \otimes |w_2\rangle.
\ee

\paragraph{Overlaps in the XX chain.}

At the end of \cref{sec:TL}, we defined a map $w \mapsto |w\rangle$ that intertwines the link state and spin-chain representations of $\eptl_N(\alpha, 0)$. For each link state $w$, we define the dual state $\llangle w|$ as
\be
\label{eq:dual.vectors} 
\llangle w| = |w\rangle^\Tt\Big|_{\phi \to -\phi}
\ee
where the superscript $^{\Tt}$ stands for real transposition. The Gram product between two link states $w_1$ and $w_2$ can then be computed from the spin-chain representation \cite{MDSA13}:
\be
w_1 \cdot w_2 = \llangle w_1 | w_2 \rangle.
\ee
In the spin-chain language, the partition functions read
\begin{subequations}
\label{eq:Z.spins}
\begin{alignat}{2}
Z_{c}^{A\cup B}&=2^{MN} \llangle v^{AB}_0| (T^{AB})^{2M} | v^{AB}_0\rangle,\\[0.2cm] 
Z_c^{A} &=2^{MN_A} \llangle v^A_0 | (T^{A})^{2M} | v^{A}_0\rangle,\\[0.2cm]
Z_c^{B} &=2^{MN_B} \llangle v^B_2 | (T^{B})^{2M} | v^B_2 \rangle,\\[0.2cm]
Z_{p}^{AB} &= 2^{MN} \llangle v_0^{A} \otimes v_0^{B} | \big( (T^{A})^M \otimes (\widehat T^{B})^M \big) (T^{AB})^M |v_0^{AB}\rangle.
\end{alignat}
\end{subequations}
We recall that $T^{AB}$ and $T^A$ are the transfer matrices of the six-vertex model with a twist $\phi$ for system sizes $N$ and $N_A$ respectively, whereas $T^B$ and $\widehat T^B$ are the transfer matrices for the system size $N_B$ with the twists $\phi = 0$ and $\phi = \pi$, respectively.

\subsubsection{Bipartite fidelity}\label{sec:BF}

\paragraph{Diagonalisation of the Hamiltonian.}

The diagonalisation of $H$ is given in Section \ref{sec:diagXX}. We recall that for $\phi \in (-\pi,\pi)$, the groundstate of $H$ is unique, belongs to the magnetisation sector $S^z = 0$ and is given by
\begin{equation}
|X_0\rangle = \left\{\begin{array}{cl}
\mu^\dagger_{(4-N)/4}\cdots\mu^\dagger_{N/4} |0\rangle & \frac N2 \textrm{ even,} \\[0.15cm]
\mu^\dagger_{(2-N)/4}\cdots\mu^\dagger_{(N-2)/4} |0\rangle & \frac N2 \textrm{ odd,}
\end{array}\right.
\qquad |0\rangle = |{\downarrow\cdots\downarrow}\rangle.
\end{equation}
Likewise, in the sector $S^z = -1$, the groundstate of $H$ for $\phi = 0$ is unique and given by
\begin{equation}
|X_{-1}\rangle = \left\{\begin{array}{cl}
\mu^\dagger_{(4-N)/4}\cdots\mu^\dagger_{(N-4)/4} |0\rangle\big|_{\phi = 0} \quad& \frac N2 \textrm{ even,} \\[0.20cm]
\mu^\dagger_{(6-N)/4}\cdots\mu^\dagger_{(N-2)/4} |0\rangle\big|_{\phi = 0} \quad& \frac N2 \textrm{ odd.}
\end{array}\right.
\end{equation}
The transfer matrix and Hamiltonian commute, and thus we have 
\begin{equation}
T |X_0\rangle = \Lambda_0 |X_0\rangle, \qquad (T|_{\phi = 0}) |X_{-1}\rangle = \Lambda_{-1} |X_{-1}\rangle.
\end{equation}
The eigenvalues are \cite{PRV10,MDPR13}
\begin{subequations}
\begin{alignat}{2}
\Lambda_0 &= \frac{\cos (\frac\phi2)}{2^{N-1}} \prod_{j=1}^{N/2}\ (1+\tan x_j) \prod_{j=(N+2)/2}^{N} (1-\tan x_j), \qquad x_j = \frac {\pi(j-\frac12)-\frac\phi2} N,
\\[0.2cm]
\Lambda_{-1} &= \frac{N}{2^{N-1}} \prod_{j=1}^{(N-2)/2} \big(1+\tan(\tfrac{\pi j}N)\big)^2.
\end{alignat}
\end{subequations}
and are the largest eigenvalues in the sectors $S^z = 0$ and $S^z = -1$, respectively. We note in particular that $\Lambda_{-1} = \lim_{\phi \to \pi}\Lambda_0$. 

In the representations $\stanW_{N,0}$ and $\stanW_{N,2}$, the transfer tangle $\Tb$ has the right eigenstates $X_0$ and $X_{-1}$ with respective eigenvalues $\Lambda_0$ and $\Lambda_{-1}$, and $|X_0\rangle$ and $|X_{-1}\rangle$ are their images under the homomorphism map. Their duals $\llangle X_0|$ and $\llangle X_{-1}|$ are obtained from the definition \eqref{eq:dual.vectors}. Because
\be
\mu_k^\Tt \big|_{\phi \to - \phi} = 
\left\{\begin{array}{cl}
\mu^\dagger_{-k} & \frac N2 + S^z\textrm{ odd,} \\[0.15cm]
\mu^\dagger_{1-k} & \frac N2 + S^z \textrm{ even,}
\end{array}\right.
\ee
we have
\be
 \llangle X_0 |= (-1)^{N(N-2)/8} \langle X_0 |, \qquad \llangle X_{-1}| = (-1)^{(N-2)(N-4)/8} \langle X_{-1} |
\ee
where $\langle X_0 |= | X_0 \rangle ^\dagger$ and $\langle X_{-1} |= | X_{-1} \rangle ^\dagger$. The state $\langle X_0 |$ is given in \eqref{eq:X0Left} and is the left groundstate of $H$ in the sector $S^z=0$. Likewise, $\langle X_{-1}|$ is the left groundstate in the sector $S^z = -1$. Because of the fermionic relations \eqref{eq:fermions.acomm}, the groundstates have unit norms, and therefore we have
\be
\label{eq:norms}
\llangle  X_0|X_0\rangle = (-1)^{N(N-2)/8}, \qquad \llangle  X_{-1} |X_{-1}\rangle = (-1)^{(N-2)(N-4)/8}.
\ee

\paragraph{Ratios of overlaps.}

We now extract the leading behaviours of the partition functions \eqref{eq:ZFpants} and \eqref{eq:ZAB.standard} as $M$ tends to infinity. We start with $Z^{AB}_p$ and assign the extra labels $AB$, $A$ and $B$ to the groundstates over $(\mathbb C^2)^{\otimes N}$, $(\mathbb C^2)^{\otimes N_A}$ and $(\mathbb C^2)^{\otimes N_B}$, respectively. From \eqref{eq:norms}, we deduce that the identity matrix over $(\mathbb C^2)^{\otimes N}$ in the sector of zero magnetisation has a contribution along the groundstate of the form
\begin{equation}
\mathbb I\big|_{S^z = 0} = (-1)^{N(N-2)/8}|X^{AB}_0\rangle \llangle X^{AB}_0| + \dots\ .
\end{equation}
The next terms involve states that are not groundstates. Likewise, for $(\mathbb C^2)^{\otimes N_A}$ and $(\mathbb C^2)^{\otimes N_B}$, we have
\begin{subequations}
\begin{alignat}{2}
\mathbb I\big|_{S^z = 0} &= (-1)^{N_A(N_A-2)/8}|X^A_0\rangle \llangle X^A_0| + \dots\ , \\[0.2cm]
\mathbb I\big|_{S^z = 0} &= (-1)^{N_B(N_B-2)/8}|X^B_0\rangle \llangle X^B_0| + \dots \ .
\end{alignat}
\end{subequations}
We then have 
\begin{subequations}
\begin{alignat}{2}
(T^{AB})^{M} |v_0^{AB}\rangle &\simeq (-1)^{N(N-2)/8}(\Lambda_0^{AB})^{M} |X_0^{AB}\rangle\llangle X_0^{AB}|v_0^{AB}\rangle,\\[0.2cm]
\llangle v_0^{A}|(T^{A})^M &\simeq (-1)^{N_A(N_A-2)/8}(\Lambda_0^{A})^M \llangle v_0^{A}|X_0^{A}\rangle \llangle X_0^{A}|\,,\\[0.2cm]
\llangle v_0^{B}|(\widehat T^{B})^M &\simeq (-1)^{N_B(N_B-2)/8}(\Lambda_0^{B})^M\llangle v_0^{B}|X_0^{B}\rangle \llangle X_0^{B}|,
\end{alignat}
\end{subequations}
where $\Lambda_0^{AB}$, $\Lambda_0^{A}$ and $\Lambda_0^{B}$ are the eigenvalues of $T^{AB}$, $T^A$ and $\widehat T^B$ in the zero magnetization sector. The symbol $\simeq$ indicates that the smaller contributions coming from excited states have been omitted. We thus find
\begin{equation}
Z_p^{AB} \simeq \epsilon\, 2^{MN}(\Lambda_0^{AB}\Lambda_0^{A}\Lambda_0^{B})^{M} \llangle v_0^{A}|X_0^{A}\rangle\llangle v_0^{B}|X_0^{B}\rangle \llangle X_0^{A} \otimes X_0^{B} |X_0^{AB}\rangle\llangle X_0^{AB}|v_0^{AB}\rangle,
\end{equation}
where $\epsilon$ is a sign that will be irrelevant for our computation of $\mathcal F_p^\alpha$. We repeat the same argument for the other partition functions and find
\begin{subequations}
\begin{alignat}{2}
Z_c^{A\cup B} &\simeq 2^{MN} (\Lambda_0^{AB})^{2M} \llangle v_0^{AB}|X_0^{AB}\rangle\llangle X_0^{AB}|X_0^{AB}\rangle\llangle X_0^{AB}|v_0^{AB}\rangle, \\[0.2cm]
Z_c^{A} &\simeq 2^{MN_A} (\Lambda_0^{A})^{2M}\llangle v_0^{A}|X_0^{A}\rangle\llangle X_0^{A}|X_0^{A}\rangle\llangle X_0^{A}|v_0^{A}\rangle, \\[0.2cm]
Z_c^{B} &\simeq 2^{MN_B} (\Lambda_{-1}^{B})^{2M} \llangle v_2^{B}|X_{-1}^{B}\rangle\llangle X_{-1}^{B}|X_{-1}^{B}\rangle\llangle X_{-1}^{B}|v_{2}^{B}\rangle,
\end{alignat}
\end{subequations}
where $\Lambda^B_{-1}$ is the groundstate eigenvalue of $T^B$ in the sector $S^z = -1$.
Using \eqref{eq:norms} and $\Lambda_0^{B} = \Lambda_{-1}^{B}$, we find 
\begin{subequations}
\begin{equation}\label{eq:Foverlaps}
\mathcal{F}_p^{\alpha} = -  \log \Bigg(\sigma\, \llangle X_0^{A} \otimes X_0^{B} |X_0^{AB}\rangle^2  \frac{\llangle X_0^{AB}|v_0^{AB}\rangle}{\llangle v_0^{AB}|X_0^{AB}\rangle} \frac{\llangle v_0^{A}|X_0^{A}\rangle}{\llangle X_0^{A}|v_0^{A}\rangle} \frac{\llangle v_0^{B}|X_0^{B}\rangle^2}{\llangle v_2^{B}|X_{-1}^{B}\rangle\llangle X_{-1}^{B}|v_2^{B}\rangle} \Bigg)
\end{equation}
with 
\be
\sigma = (-1)^{\frac{N(N-2)}8}(-1)^{\frac{N_A(N_A-2)}8}(-1)^{\frac{(N_B-2)(N_B-4)}8}.
\ee
\end{subequations}

\paragraph{Product expressions for the overlaps involving boundary states.}

We use Wick's theorem to evaluate the various overlaps in \eqref{eq:Foverlaps}. To start, we note that the boundary states $\llangle v_0|$ and $\llangle v_2|$ can be written in terms of the fermionic operators $c_j$ as
\be
\label{eq:v0v2ferm}
\llangle v_0| = \langle 0 | a_{N-1} a_{N-3} \cdots a_{5} a_{3} a_{1}, \qquad \llangle v_2| = \langle 0 | a_{N-3}a_{N-5} \cdots a_{5} a_{3} a_{1}, \qquad a_j = \omega\, c_j + \omega^{-1} c_{j+1}.
\ee
We have the anti-commutation relation
\be
\label{eq:acomm.mu.a}
\{a_{2\ell-1},\mu_k^\dagger\} = \frac{\eE^{-2\ir\ell\theta_k}}{\sqrt N} \big(\omega\, \eE^{\ir \theta_k}+ \omega^{-1}\big).
\ee
Using Wick's theorem, we write the overlaps $\llangle X_0|v_0\rangle$ and $\llangle X_{-1}|v_2\rangle$ as 
\be
\llangle v_0|X_0\rangle = \det_{\substack{k\,\in\,K_0\\\ell = 1, \dots, N/2}} \{a_{2\ell-1},\mu_k^\dagger\}, 
\qquad
\llangle v_2|X_{-1}\rangle = \det_{\substack{k\,\in\,K_{-1} \\ \ell = 1, \dots, (N-2)/2}} \{a_{2\ell-1},\mu_k^\dagger\}\big|_{\phi = 0}\,, 
\ee
where
\be
K_0 = 
\left\{\begin{array}{cl}
\big\{\frac{4-N}4,\cdots, \frac N4\big\} & \frac N2 \textrm{ even,} \\[0.2cm]
\big\{\frac{2-N}4,\cdots, \frac{N-2}4\big\} & \frac N2 \textrm{ odd,}
\end{array}\right.
\qquad
K_{-1} = 
\left\{\begin{array}{cl}
\big\{\frac{4-N}4,\cdots, \frac {N-4}4\big\} & \frac N2 \textrm{ even,} \\[0.2cm]
\big\{\frac{6-N}4,\cdots, \frac{N-2}4\big\} & \frac N2 \textrm{ odd.}
\end{array}\right.
\ee
The factors that depend only on $k$ in \eqref{eq:acomm.mu.a} can be factorised from the determinant, which can then be evaluated in terms of a product using the Vandermonde identity. We simplify the result using the identities
\be
\label{eq:product.identities}
\prod_{j=1}^{N/2} \sin\big(\tfrac{\pi j}N\big) = \frac{N^{1/2}}{2^{(N-1)/2}},
\qquad 
\prod_{j=1}^{N/2}\prod_{k=j+1}^{N/2} \sin \big(\tfrac {2\pi(k-j)} N \big) = \frac{N^{N/4}}{2^{N^2/8}},
\ee
and find
\begin{subequations}
\begin{alignat}{2}
&\llangle v_0|X_0\rangle = 2^{N/4} \eE^{\ir \phi(N+1)/4} \eE^{-\ir \pi N(N-2)/16}\, \prod_{j=1}^{N/2}\, \sin\big(\tfrac{(2j-1)\pi + \phi}{2N}\big),
\\[0.05cm]
&\llangle v_0|X_0\rangle\big|_{\phi = \pi} = \frac{N^{1/2}}{2^{(N-2)/4}}\eE^{-\ir \pi(N^2-6N-4)/16}, 
\\[0.25cm]
&
\llangle v_2|X_{-1}\rangle = \frac{\eE^{-\ir \pi(N-2)(N-4)/16}}{2^{(N-2)/4}}.
\end{alignat}
\end{subequations}
It also follows from the definition \eqref{eq:dual.vectors} of dual states that 
\be
\llangle X_0|v_0\rangle = \llangle v_0|X_0\rangle \big |_{\phi \to -\phi}, \qquad 
\llangle X_{-1}|v_2\rangle =\llangle v_2|X_{-1}\rangle.
\ee
Putting these results together, we obtain
\begin{subequations}
\label{eq:Qs}
\begin{alignat}{2}
&\frac{\llangle X^{AB}_0|v^{AB}_0\rangle}{\llangle v^{AB}_0|X^{AB}_0\rangle} = \eE^{-\ir \phi (N+1)/2} \prod_{j=1}^{N/2} \frac{\sin\big(\tfrac{(2j-1)\pi - \phi}{2N}\big)}{\sin\big(\tfrac{(2j-1)\pi + \phi}{2N}\big)},
\\[0.1cm]
&\frac{\llangle v^{A}_0|X^{A}_0\rangle}{\llangle X^{A}_0|v^{A}_0\rangle} = \eE^{\ir \phi (N_A+1)/2} \prod_{j=1}^{N_A/2} \frac{\sin\big(\tfrac{(2j-1)\pi + \phi}{2N_A}\big)}{\sin\big(\tfrac{(2j-1)\pi - \phi}{2N_A}\big)},
\\[0.1cm]
&\frac{\llangle v^{B}_0|X^{B}_0\rangle^2}{\llangle v^{B}_2|X^{B}_{-1}\rangle\llangle X^{B}_{-1}|v^{B}_2\rangle} = - \ir N_B.
\end{alignat}
\end{subequations}

\paragraph{Exact expression for the bipartite fidelity.}

It remains to compute the overlap $\llangle X_0^{A} \otimes X_0^{B} | X_0^{AB}\rangle$. Up to unimportant signs that come from \eqref{eq:norms}, the overlap is exactly the one computed in \eqref{eq:B} with the specialisations $\phi_A = \phi$ and $\phi_B = \pi$. We note that these specialisations are compatible with \eqref{eq:restrictions} for $\ell=0$, for which we were able to obtain product expressions from the determinant expressions. From \eqref{eq:Foverlaps} and \eqref{eq:Qs} we find
\begin{equation}
\label{eq:FaFXX}
\mathcal{F}_p^{\alpha} = \mathcal{F}_p^{\tXX}\big|_{\substack{\phi_A = \phi\\ \phi_B = \pi}} -Q(N,\phi)+Q(Nx,\phi)-\log\big(N(1-x)\big)
\end{equation}
where
\begin{equation}
\label{eq:Q}
Q (N,\phi)= \sum \limits_{j=1}^{N/2} \log \left| \frac{\sin\big(\frac{(2j-1)\pi-\phi}{2N}\big)}{\sin\big(\frac{(2j-1)\pi+\phi}{2N}\big)}\right|.
\end{equation}
The closed-form expression for $\mathcal{F}_p^{\tXX}\big|_{\substack{\phi_A = \phi\\ \phi_B = \pi}}$ is read from \eqref{eq:B}.

\subsection{Asymptotics}\label{sec:asymptotics.pants}

In this section, we study the large-$N$ asymptotics of the bipartite fidelity for the XX chain and the model of critical dense polymers on the periodic pants lattice and compare the results with the CFT predictions of \cref{sec:CFT.pants}.

\subsubsection{XX spin chain}\label{sec:asy.pants.XX}

\paragraph{Exact results for $\boldsymbol{\mathcal F_p^{\tXX}}$ with $\boldsymbol{\phi = \phi_A + \phi_B - \pi}$.}
The calculation of the first terms in the large-$N$ asymptotics of \eqref{eq:B} is a long yet straightforward calculation. The details are given in \cref{app:Asympt}. We find
\begin{equation}
\label{eq:LBFXXFinal}
\begin{split}
\mathcal{F}_p^{\tXX}&\big|_{\phi = \phi_A+\phi_B-\pi}  = \frac{1}{2}\log N \\
&- \frac{\pi^2(-1+x)(1+2x)-6\pi x\Big((-1+x)\phi_A+x\phi_B\Big)+3\Big((-1+x)\phi_A+x\phi_B\Big)^2}{6 \pi^2 x}\log(1-x) \\
&+\frac{\pi^2(3-2x)x+6\pi(-1+x)\Big((-1+x)\phi_A+x\phi_B\Big)-3\Big((-1+x)\phi_A+x\phi_B\Big)^2}{6 \pi^2 (1-x)}\log x\\
&-\frac 12 -\frac 12 \log \pi + 6 \log A +\mathcal{O}\big(N^{-1}\big)
\end{split}
\end{equation}
where $A \simeq 1.282427$ is the Glaisher-Kinkelin constant. As expected, this function is symmetric under the simultaneous transformations $x \rightarrow 1-x$ and  $\phi_A \leftrightarrow \phi_B$. We see that this result is identical to the CFT prediction \eqref{eq:Fp.expansion} for case (ii), where $g_0$ and $g_1(x)$ are given in \eqref{eq:g0Pants} and \eqref{eq:g1.pants.4pts}, the conformal data is 
\begin{equation}
\label{eq:charges.c=1.v0}
c = 1, \qquad 
\alphaC_1 = \frac{\phi_A}{2\pi}, \qquad 
\alphaC_2 = -\frac{1}{2},\qquad
\alphaC_3 = \frac{\phi_B}{2\pi}, \qquad 
\alphaC_4=- \frac{\phi_A+\phi_B-\pi}{2\pi}, 
\end{equation}
and the non-universal constant is given by
\begin{equation}
C' = -\frac 12 -\frac 12 \log \pi + 6 \log A \simeq 0.420162.
\end{equation}
We also remark that there is no term proportional to $N^{-1}\log N$ in \eqref{eq:LBFXXFinal}. Comparing with \eqref{eq:g2.pants}, we observe that the content of the large parenthesis in this equation does not vanish in general. We therefore deduce that the extrapolation length $\Xi$ vanishes in this case.

\begin{figure}
\begin{pspicture}(0,0)(17,5)
\rput(4,2.5){\includegraphics[height=4.5cm]{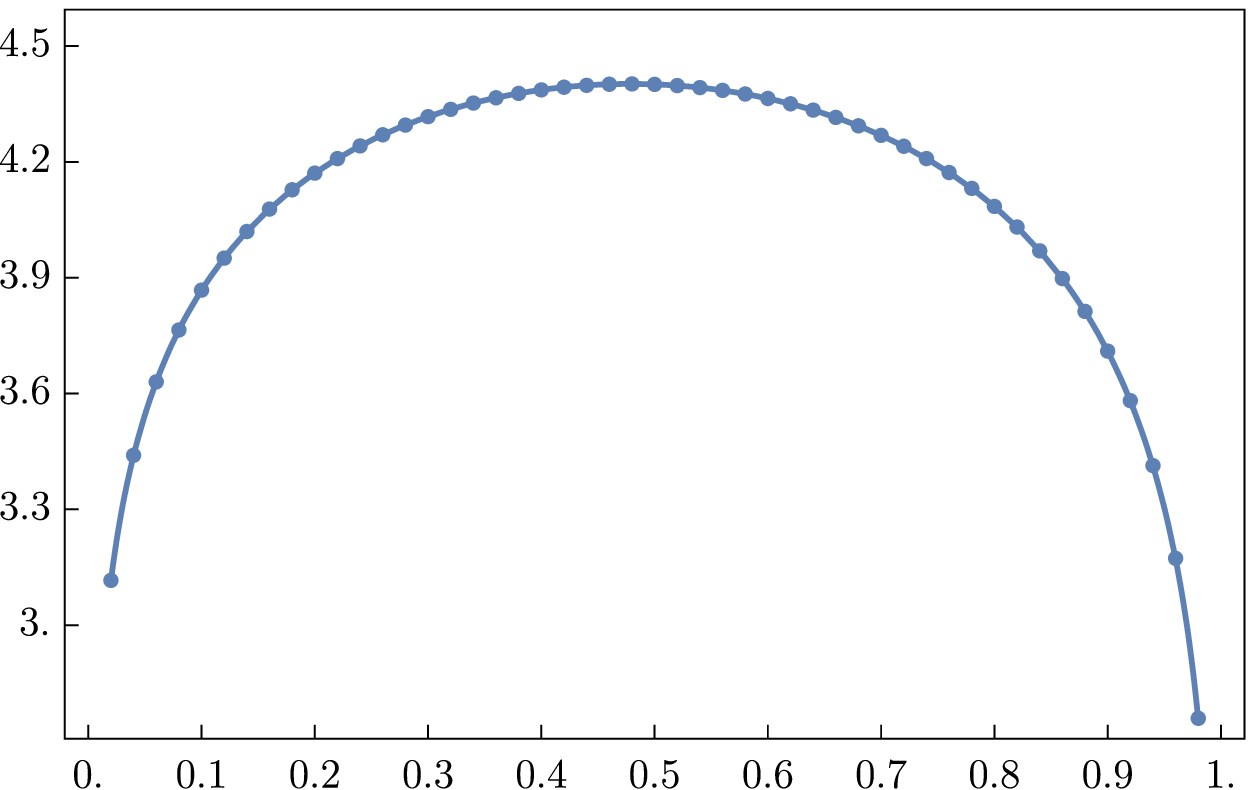}}
\rput(1.3,4.3){$\mathcal{F}_p^{\tXX}$}
\rput(7.1,0.8){$x$}
\rput(13,2.5){\includegraphics[height=4.5cm]{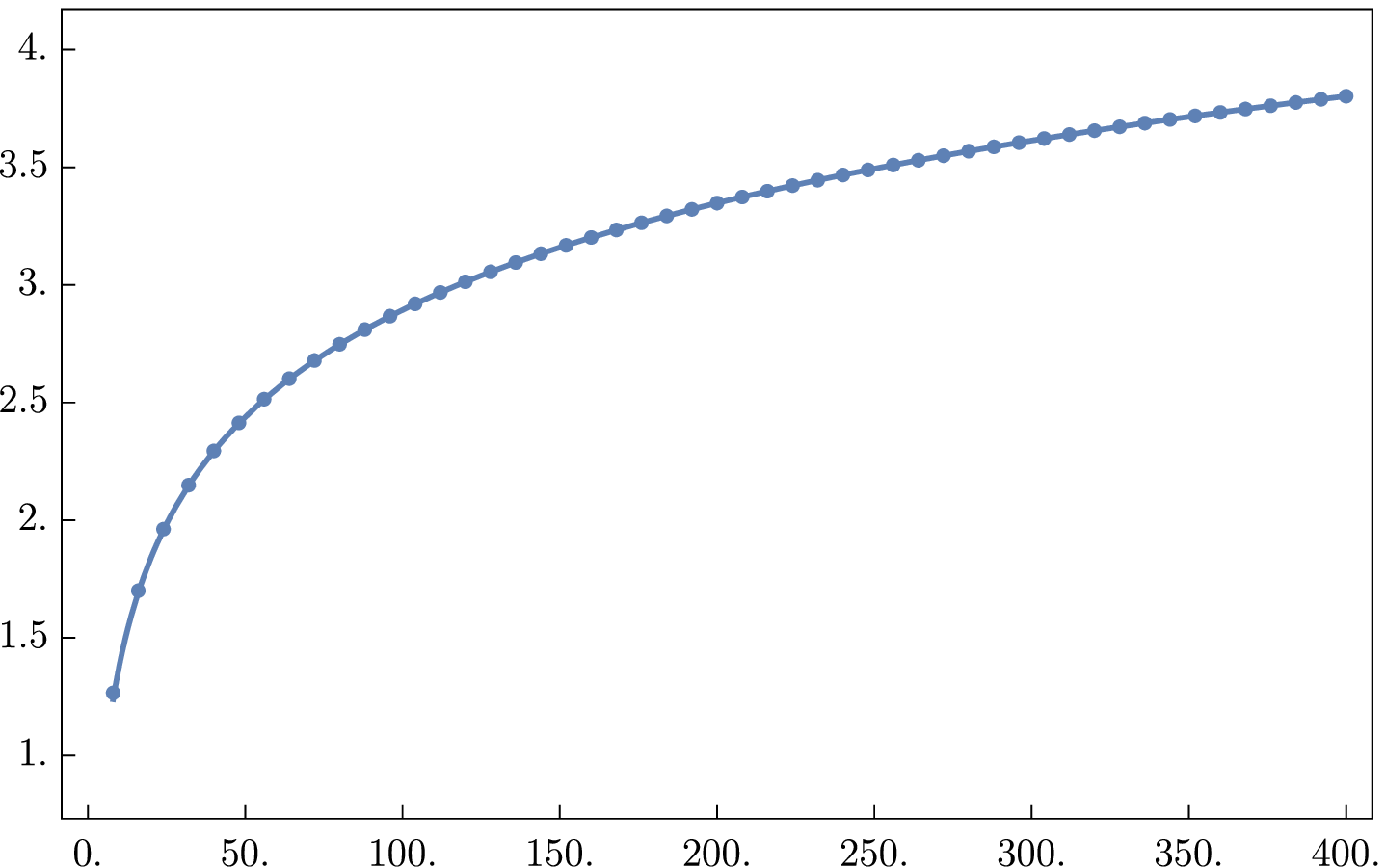}}
\rput(10.3,4.3){$\mathcal{F}_p^{\tXX}$}
\rput(16,0.8){$N$}
\end{pspicture}
\caption{The standard bipartite fidelity $\mathcal{F}_p^{\tXX}$ as a function of $x$ for $N=800$ (left panel) and as a function of $N$ for $x=\frac 14$ (right panel). In both cases, the twists are set to $\phi=1$, $\phi_A=2$ and $\phi_B=3$. The data points are computed from the determinant expression \eqref{eq:exactOverlapPants}, whereas the continuous line is \eqref{eq:conjPants} with \eqref{eq:charges.c=1.v1}, $\Xi=0$ and $C' = 0.447447$.}
\label{fig:FPantsCFT}
\end{figure}

\paragraph{Numerical results for $\boldsymbol{\mathcal F_p^{\tXX}(n)}$.}
We now study the large-$N$ behaviour of the bipartite fidelities in the case where the three twists $\phi$, $\phi_A$ and $\phi_B$ are arbitrary. We consider the standard bipartite fidelity \eqref{eq:LBFXXPants} as well as the modified instances $\mathcal{F}^\tXX_p(n)$ defined in \eqref{eq:modifiedLBFPants}. Since we do not have an exact product formula, we instead use the determinant expression \eqref{eq:exactOverlapPants}, and very similar formulas for the modified instances, to study the large-$N$ behaviour of these quantities numerically. To get precise numerical estimates, we follow the strategy of \cite{DS11} and fit $\mathcal{F}_p^{\tXX}(n)$ as
\begin{equation}
\label{eq:F.pants.fit}
 \mathcal{F}_p^{\tXX}(n) = g_0 \log N + g_1(x) + g_2 \frac{\log N}{N} +  \gamma_3 \frac{1}{N} + \gamma_4 \frac{\log N}{N^2} +  \gamma_5 \frac{1}{N^2} + \gamma_6 \frac{\log N}{N^3} +  \gamma_7 \frac{1}{N^3}.
\end{equation}
 From these numerical explorations, we find that the leading terms in the asymptotic expansion precisely reproduce the conformal prediction of \cref{sec:CFT.pants}. Namely, we have
\begin{equation}
\label{eq:conjPants}
\mathcal{F}_p^{\tXX}(n)  = \bigg(\frac 14 + \frac 14 \Big(\frac{\phi-\phi_A-\phi_B}{\pi}\Big)^2\bigg)\log N + g_1(x)+g_2(x)N^{-1}\log N +\mathcal{O}\big(N^{-1}\big),
\end{equation}
where $g_1(x)$ is given by the formula \eqref{eq:g1.pants.4pts} for vertex operators, with
\begin{equation}
\label{eq:charges.c=1.v1}
c=1, \qquad
\alphaC_1 = \frac{\phi_A}{2\pi}, \qquad 
\alphaC_2 = \frac{\phi-\phi_A-\phi_B}{2\pi}, \qquad
\alphaC_3 = \frac{\phi_B}{2\pi}, \qquad 
\alphaC_4=- \frac{\phi}{2\pi}.
\end{equation}
Likewise, $g_2(x)$ is given by \eqref{eq:g2.pants} with the conformal dimensions $\Delta_i = \alphaC_i^2/2$, the charges $q_i$ in \eqref{eq:charges.c=1.v1}, 
and the extrapolation length
\begin{equation}\label{eq:XiXXPants}
\Xi = -n.
\end{equation}
In particular, we have $\Xi=0$ for the standard bipartite fidelity. 

\begin{figure}
\begin{pspicture}(0,0)(17,5)
\rput(4,2.5){\includegraphics[height=4.5cm]{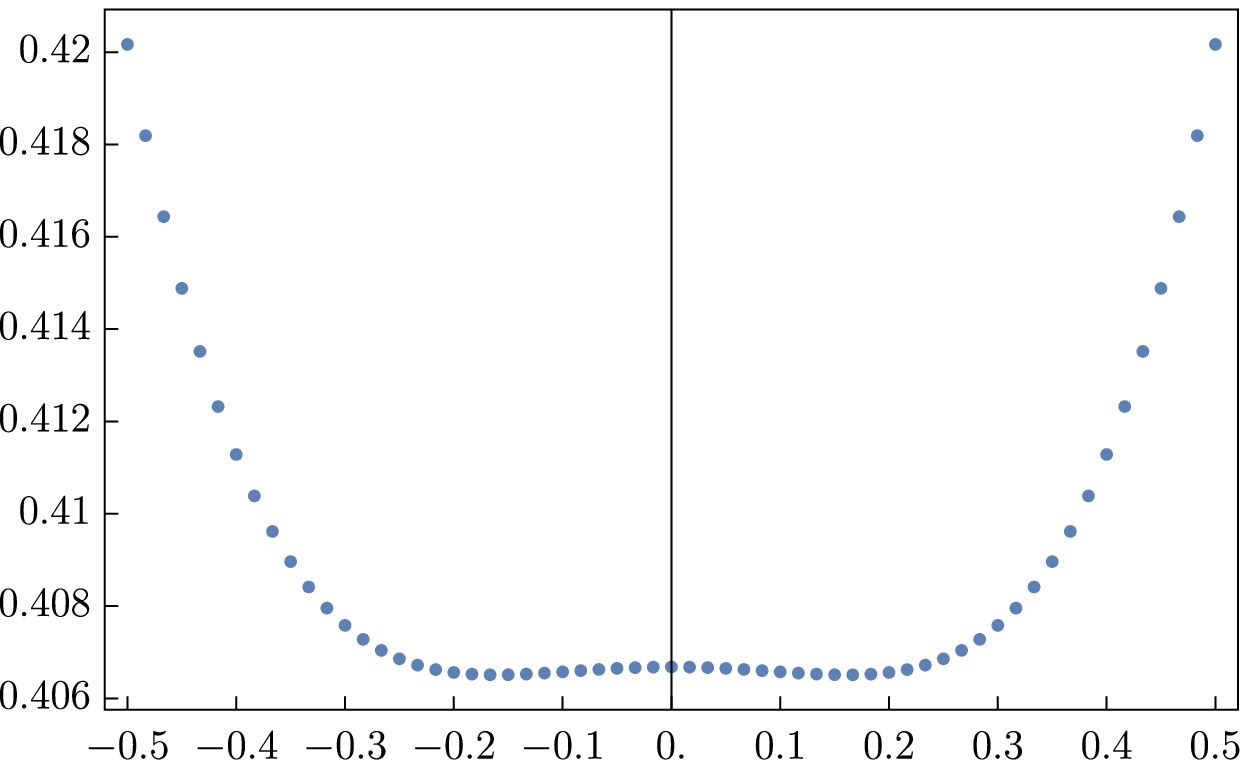}}
\rput(1.4,4.3){$C'$}
\rput(7.1,0.8){$\alphaC_2$}
\rput(13,2.5){\includegraphics[height=4.5cm]{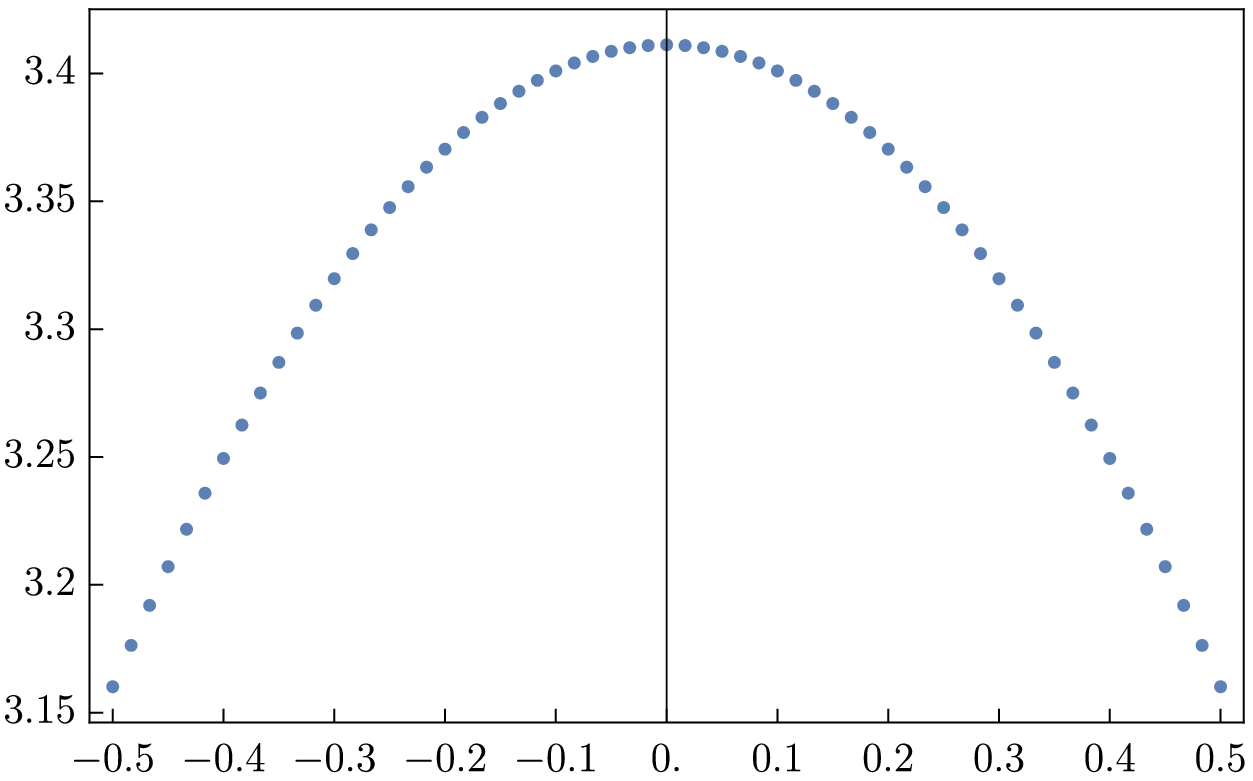}}
\rput(10.3,4.3){$C'$}
\rput(16,0.8){$\alphaC_2$}
\end{pspicture}
\caption{Numerical results for the constant $C'$ as a function of $\alphaC_2$ for the standard bipartite fidelity (left panel) and $n=4$ (right panel).}
\label{fig:Cprime}
\end{figure}

\begin{figure}
\begin{pspicture}(0,0)(17,5)
\rput(4,2.5){\includegraphics[height=4.5cm]{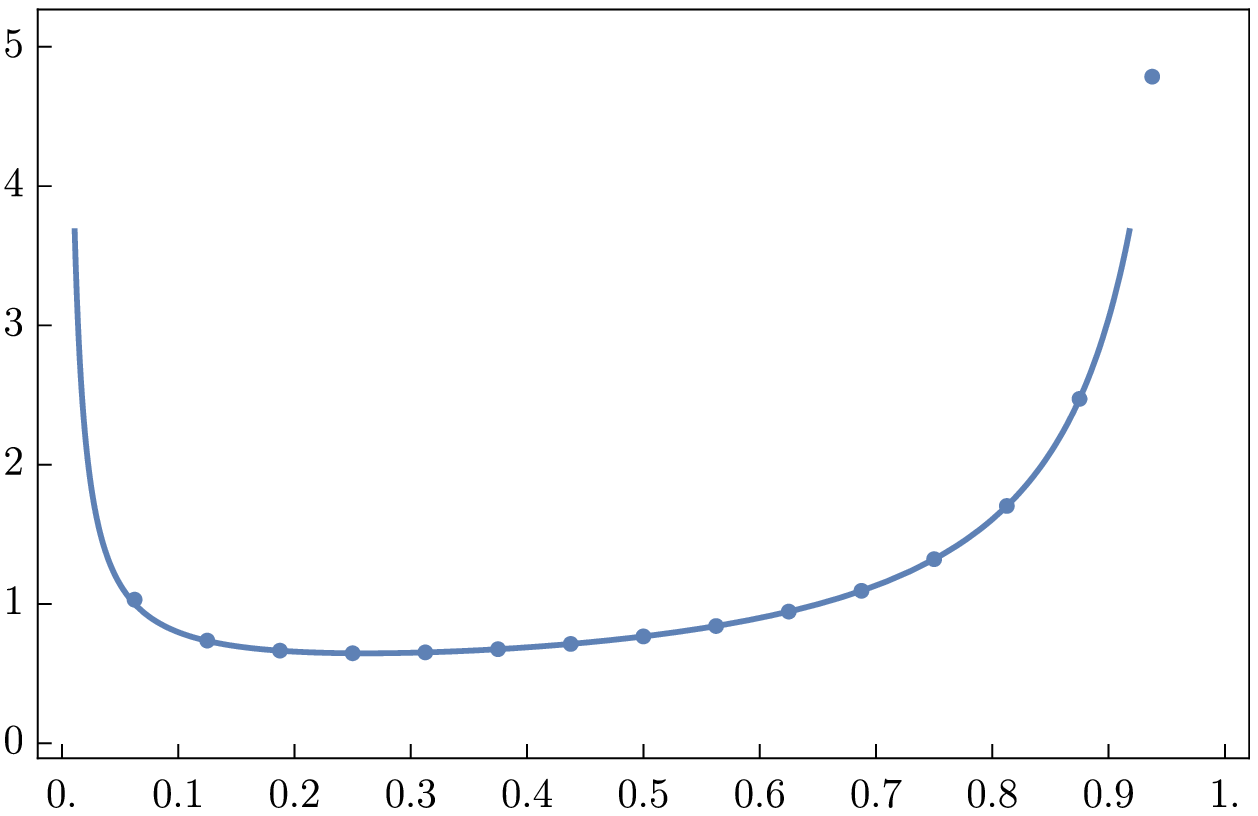}}
\rput(1.4,4.3){$g_2(x)$}
\rput(7.1,0.8){$x$}
\rput(4,2.8){$n=2$}
\rput(13,2.5){\includegraphics[height=4.5cm]{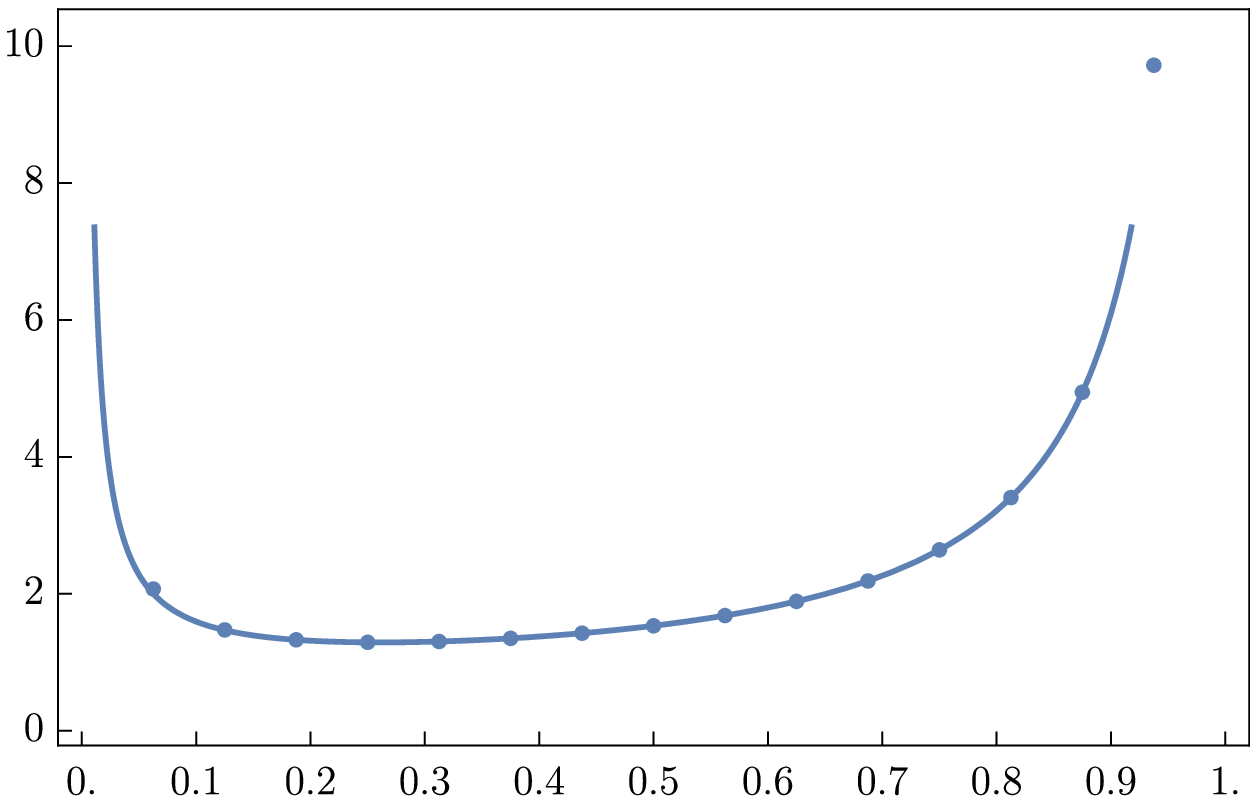}}
\rput(10.45,4.3){$g_2(x)$}
\rput(13.1,2.8){$n=4$}
\rput(16,0.8){$x$}
\end{pspicture}
\caption{The function $g_2(x)$ for the XX spin chain with $\phi=1$, $\phi_A=2$ and $\phi_B=3$ for $n=2$ and $n=4$. The blue dots are obtained by numerical evaluation of determinants similar to \eqref{eq:exactOverlapPants}. The blue curves plot the CFT predictions \eqref{eq:g2.pants} with the conformal data \eqref{eq:charges.c=1.v1} and $\Xi$ given in \eqref{eq:XiXXPants}. }
\label{fig:g2}
\end{figure}

The constant $C'$ is unknown and obtained from a fit in our analysis. 
In \cref{fig:FPantsCFT}, we plot \eqref{eq:conjPants} and compare it with numerical values obtained from the determinant formula \eqref{eq:exactOverlapPants}, as a function of~$x$ and~$N$. The agreement is remarkable. The same plots for $\mathcal F_p^{\tXX}(n)$ with $n=2,4,6$ are very similar, and the match is again convincing.
The numerical analysis reveals that $C'$ depends on the twists and on the spin state inserted at the crotch point. For a given choice of this spin state, we find that $C'$ depends on the twist parameters only through the difference $\phi - \phi_A - \phi_B$, or equivalently on the value of $\alphaC_2$. In \cref{fig:Cprime}, we plot $C'$ for various values of $\alphaC_2$ in the range $(-\frac12,\frac12)$, for $n=0$ and $n=4$. Finally, in \cref{fig:g2} we plot the function $g_2(x)$ for $n=2$ and $n=4$. This function depends on the spin state inserted at the crotch point only via the extrapolation length \eqref{eq:XiXXPants}.
The match between our numerical results and the CFT prediction is clear. 

\paragraph{Conformal interpretation.}
Let us now discuss how our results fit with previously known results about the conformal interpretation of the XX chain. It is well-known that the XX chain (and more generally the XXZ chain) is described by a CFT with central charge $c=1$. In this context, the presence of a twist line $\phi$ between two points $z_1$ and $z_2$ is accounted for by the insertion of two fields $\varphi(z_1)$ and $\varphi(z_2)$. The field $\varphi$ is a so-called {\it electric operator} \cite{dFHZ87}. Its multi-point functions are known to have the form \eqref{eq:npoint.vertex}, and our analysis in this subsection confirms this. It is also a spinless field. Its conformal dimensions $\Delta^{\tXX}_\phi = \bar \Delta^{\tXX}_\phi$ can be computed from the $\frac 1N$ finite-size correction term of the largest eigenvalue of the periodic transfer matrix $T$ with a diagonal twist. Indeed, this finite-size correction can be obtained either via the Bethe ansatz \cite{HQB87} or with the method of functional relations \cite{KBP91}, and it allows one to compute the difference $c-24 \Delta$ explicitly:
\be
\label{eq:diff.c.Delta}
c - 24 \Delta = 1 - \frac{3 \phi^2}{\pi^2}.
\ee
With $c=1$, the resulting conformal dimension is 
\be
\label{eq:Delta.XX}
\Delta^{\tXX}_\phi = \frac{\phi^2}{8\pi^2}.
\ee 
This is consistent with the known values $\alphaC = \bar \alphaC = \frac{\phi}{2\pi}$ for the charges of electric operators, and the values $\Delta_i = \alphaC_i^2/2$ obtained in \eqref{eq:charges.c=1.v0} and \eqref{eq:charges.c=1.v1}.
Comparing \eqref{eq:g0Pants} and \eqref{eq:conjPants}, we see that \eqref{eq:Delta.XX} also produces the correct value for $g_0$. We note that the field inserted on the crotch point is special, as it lies at the intersection of three twist lines. Accordingly, its conformal dimension is obtained from \eqref{eq:Delta.XX} with $\phi$ replaced by the difference $\phi-\phi_A - \phi_B$ of the twists at this intersection. The insertion of a N\'eel state of size $n$ on the crotch point does not modify this conformal dimension.

\subsubsection{Critical dense polymers}\label{sec:asy.pants.CDP}

The large-$N$ expansion for $\mathcal F_p^\alpha$ follows directly from the same expansions for $\mathcal{F}_p^{\tXX}\big|_{\phi = \phi_A+\phi_B-\pi}$ and the function $Q(N,\phi)$ defined in \eqref{eq:Q}. These are given in \eqref{eq:LBFXXFinal} and \eqref{eq:QAsympt}, respectively. This yields
\begin{alignat}{2}
\mathcal{F}_p^{\alpha} &= \mathcal{F}_p^{\tXX}\big|_{\substack{\phi_A = \phi\\ \phi_B = \pi}} -\log N - \frac{\pi \log(1-x)+\phi \log x}{\pi}
\nonumber\\[0.15cm]
&= -\frac{1}{2}\log N -\frac{1}{6}(G(x)+G(1-x))-\frac{1}{2}\Big(\frac{\phi^2}{\pi^2}-1\Big)\Big((1-x) \log x + \frac{(1-x)^2}{x} \log(1-x)\Big)\nonumber\\[0.15cm]
&\hspace{0.34cm} -\frac 12 -\frac 12 \log \pi + 6 \log A +\mathcal{O}\big(N^{-1}\big)
\label{eq:LBFaFinal}
\end{alignat}
where $G(x)$ is defined in \eqref{eq:Gx}. This exactly corresponds to the CFT prediction for case~(i) with
\begin{equation}
\label{eq:CDP.CFT.data}
c=-2, \qquad 
\Delta_1=\Delta_4= \frac{1}{8}\Big(\frac{\phi^2}{\pi^2}-1\Big), \qquad 
\Delta_3=0, \qquad
C=-\frac 12 -\frac 12 \log \pi + 6 \log A,
\qquad \Xi=0.
\end{equation}

This fits with previously known results for the conformal interpretation of the model of critical dense polymers. The central charge of this model is known to be $c=-2$ \cite{PR07,PRV10}.
In this context, for the model defined on an infinite cylinder, assigning a fugacity $\alpha \neq \beta$ to the non-contractible loops amounts to inserting two copies of a  conformal field $\rho(z)$ at the top and bottom of the cylinder. The properties of this field were investigated in detail in \cite{MDJ19}. The conformal dimensions $\Delta^\tCDP_\phi$ and $\bar \Delta^\tCDP_\phi$ of $\rho(z)$
are equal and can be obtained from the $\frac 1N$ finite-size correction term for the groundstate eigenvalue of the periodic transfer matrix. The spectrum of the transfer matrix for the six-vertex model and for the model of critical dense polymers are identical for $\alpha = 2 \cos (\frac\phi2)$. As a result, the difference $c-24 \Delta$ for critical dense polymers is also given by \eqref{eq:diff.c.Delta}. Setting $c=-2$ and solving for the conformal dimension yields
\be
\label{eq:Delta.phi.CDP}
\Delta^\tCDP_\phi = \frac{1}{8}\Big(\frac{\phi^2}{\pi^2}-1\Big).
\ee
This is precisely the conformal dimension that we obtained in \eqref{eq:CDP.CFT.data} for $\Delta_1$ and $\Delta_4$. It is also the value previously known from Coulomb gas calculations \cite{N84}.

\section{Bipartite fidelity on the skirt geometry
} \label{sec:skirt}

In this section, we consider the bipartite fidelity for physical systems $A$, $B$ and $AB$, of respective lengths $N_A$, $N_B$ and $N=N_A+N_B$, where the system $AB$ has periodic boundary conditions, whereas the systems $A$ and $B$ have open boundary conditions. For one-dimensional chains, the fidelity is defined as 
\be
\label{eq:fidelity.1D.general.skirt}
\mathcal F_s = - \log \big| \langle x^A \otimes x^B | X^{AB}\rangle\big|^2.
\ee
The states $\langle x^S|$ and $| x^S \rangle$ are respectively the left and right groundstates of the Hamiltonian of the system $S$ endowed with open boundary conditions. Moreover, $\langle X^S|$ and $| X^S \rangle$ are the groundstates of the Hamiltonian of the system $S$ with periodic boundary conditions. These states are assumed to be normalised in such a way that $\langle X^S| X^S \rangle = 1$ and $\langle x^S| x^S \rangle = 1$.

For two-dimensional lattice models, the fidelity is defined as
\be
\label{eq:fidelity.2D.general.skirt}
\mathcal F_s = \lim_{M \to \infty} - \log \Bigg| \frac{\big(Z_s^{AB}\big)^2}{Z_r^{A}Z_r^{B}Z_c^{A\cup B}}\Bigg|.
\ee
Here $Z_s^{AB}$ is the partition function defined on the skirt geometry depicted in the central panel of \cref{fig:geometries}. The height is $2M$, the perimeter at the top is $N$, the width of the legs $A$ and $B$ are $N_A$ and $N_B$. Likewise, $Z_r^{A}$ and $Z_r^{B}$ are partitions functions on rectangles of sizes $2M \times N_A$ and $2M \times N_B$, respectively. Finally, $Z_c^{A\cup B}$ is the model's partition function on the cylinder of perimeter~$N$ and height~$2M$. For suitable choices of these boundary conditions, the partition functions are all non-zero and the limit $M \to \infty$ in \eqref{eq:fidelity.2D.general.skirt} is well-defined.

For two-dimensional models that have one-dimensional quantum analogues \cite{Baxterbook}, the definitions \eqref{eq:fidelity.1D.general.skirt} and \eqref{eq:fidelity.2D.general.skirt} coincide. In \cref{sec:CFT.skirt}, we give the conformal prediction for the leading terms in the large-$N$ expansion of the bipartite fidelity. In \cref{sec:XX.skirt,sec:polymers.skirt}, we compute $\mathcal F_s$ for the XX spin chain and for the model of critical dense polymers. Finally, in \cref{sec:AsymptoticsSkirt}, we compare the asymptotical behaviour of the lattice results with the predictions of conformal field theory.

\subsection{Predictions of conformal field theory}\label{sec:CFT.skirt}

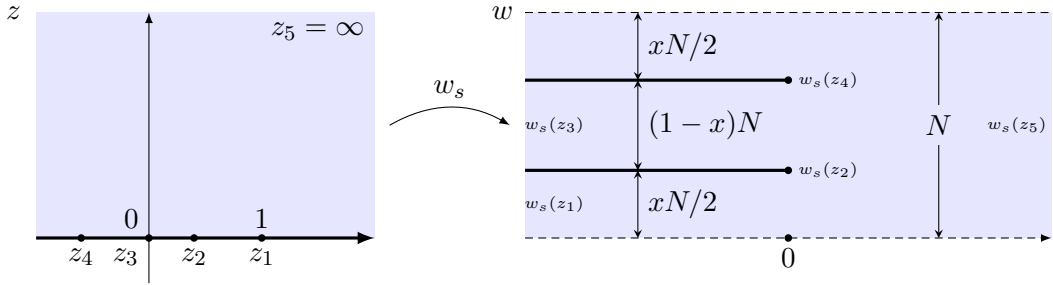
\begin{figure}
\centering
\begin{tikzpicture}
\def\a{3} 
\def\b{3.5} 
\def\L{3} 
\def\offset{0.3}
\def\ofsArr{1.5}
\def\x{0.6}
\def\shift{2} 
\def\un{1.5}
\newcommand{\mybullet}[3]{\fill #1 circle (.05) node#3 {#2}}
\newcommand{\Sminus}{\scalebox{.4}[.7]{$-$}}
\newcommand{\Splus}{\scalebox{.4}[.4]{$+$}}
\newcommand{\ws}{w_{\scalebox{.5}{$s$}}}
%
\fill[blue!10] (-\a/2,0) rectangle (\a,\a);
\draw (-\a/2-\offset,\a) node{$z$};
\draw[->, >=latex, line width=1.2pt] (-\a/2,0) to (\a,0);
\draw[->, >=latex] (0, -\a/5) to (0,\a);
\mybullet{(0,0)}{$0$}{[above left]};
\mybullet{(\un,0)}{$1$}{[above]};
\mybullet{(\un-\un*\x,0)}{{$z_2$}}{[below]};
\mybullet{(-\un*\x,0)}{{$z_4$}}{[below]};
\draw (0,0) node[below left] {$z_3$};
\draw (\un,0) node[below] {$z_1$};
\draw (\a, \a) node[below left] {$z_5=\infty$};
\draw[->,>=latex] (\a+\shift/10,\a/2) to[bend left] node[midway, above] {$w_s$} (\a+9*\shift/10, \a/2);
%
\begin{scope}[shift={(\a+\b+\shift ,0)}]
\fill[blue!10] (-\b,0) rectangle (\b,\L);
\draw (-\b-\offset,\L) node{$w$};
\mybullet{(0,0)}{$0$}{[below]};
\mybullet{(0,{\L*\x/2})}{{\tiny $w_s(z_2)$}}{[right]};
\mybullet{(0,{\L-\L*\x/2})}{{\tiny $w_s(z_4)$}}{[right]};
\draw[->, >=latex, densely dashed] (-\b,0) to (\b,0);
\draw[densely dashed] (-\b,\L) -- (\b,\L);
\draw[ line width=1.2pt] (-\b,{\L*\x/2}) -- (0,{\L*\x/2});
\draw[ line width=1.2pt] (-\b,{\L-\L*\x/2}) -- (0,{\L-\L*\x/2});
\draw[<->,>=stealth] ({\b-\ofsArr},0) -- ({\b-\ofsArr},\L )node[midway, fill=blue!10] {$N$};
\draw[<->,>=stealth] ({-\b+\ofsArr},0) -- ({-\b+\ofsArr},{\L*\x/2}) node[midway, right] {$xN/2$};
\draw[<->,>=stealth] ({-\b+\ofsArr},{\L*\x/2}) -- ({-\b+\ofsArr},{\L-\L*\x/2}) node[midway, right] {$(1-x)N$};
\draw[<->,>=stealth] ({-\b+\ofsArr},{\L-\L*\x/2})-- ({-\b+\ofsArr},{\L})  node[midway, right] {$xN/2$};
%
\draw (-\b,{\L*\x/4}) 	node[right=-4pt]	{{\tiny $w_s(z_1)$}}; 
\draw (-\b,{\L/2}) 	node[right=-4pt]	{{\tiny $w_s(z_3)$}}; 
\draw (\b,{\L/2}) 		node[left=-2 pt] 	{{\tiny $w_s(z_5)$}};
\end{scope}
\end{tikzpicture}
\caption{The function $w_{s}(z)$ maps the upper half-plane onto the skirt geometry. The positions $z_2$ and $z_4$ that are the preimages of the slits' endpoints are $z_2= (\sqrt{x-x^2}+x-1)/(2 x-1)$ and $z_4=(-\sqrt{x-x^2}+x-1)/(2 x-1)$. The bold lines are the boundaries of the domains. On the right part of the figure, the dashed lines are identified.}
\label{fig:MapSkirt}
\end{figure}

This section gives the conformal predictions for the bipartite fidelity on the skirt geometry. The details of the derivations are given in \cref{app:CFT.derivations}. From the definition \eqref{eq:fidelity.2D.general.skirt}, the bipartite fidelity on the skirt geometry corresponds to the following difference of free energies:
\be \label{eqn:Fsdifference}
\mathcal F_s = 2 f_{s} - f_{c(N)} - f_{r(N_A)} - f_{r(N_B)},
\ee
where $f_s$ is the free energy on the skirt geometry, $f_{c(N)}$ is the free energy on the cylinder of perimeter~$N$ and $f_{r(N)}$ is the free energy on the rectangle of width~$N$. The bipartite fidelity depends on two independent characteristic lengths, $N$ and $N_A$, with the third characteristic length given by $N_B = N-N_A$. We consider the asymptotic behaviour of $\mathcal F_s$ in the limit where $N$ and $N_A$ are sent to infinity with the aspect ratio $x = N_A/N$ kept fixed. It has the large-$N$ expansion
\be \label{eq:Fs.expansion}
\mathcal F_s = \tilde g_{0} \log N + \tilde g_1(x) + \tilde g_2(x) N^{-1}\log N + \dotsc.\ee 
The coefficients $\tilde g_0$, $\tilde g_1(x)$ and $\tilde g_2(x)$ can be computed by the methods of conformal field theory.
In this framework, the skirt domain is described as an infinite horizontal strip of width $N$ drawn in the complex plane,
 and decorated with two slits that divide the left half into two strips of width $Nx$ and $N(1-x)$. The boundary conditions are periodic along the strip's edges for Re$(w)>0$, but are open along the strips' edges for Re$(w)<0$. This indeed reproduces the geometry of the central panel of \cref{fig:geometries}. In contrast with the periodic pants domain, the skirt domain really has two distinct slit endpoints. The map from the upper half-plane to the skirt geometry is
\be
\label{eq:w_s}
w_s(z) = \frac{N}{2\pi} \big( x \log\big(\!-\!(z-1)^2\big) + (1-x) \log z^2 - \log(z^2 + (z-1)^2)\big) + K_x,
\ee
where $K_x$ is given in \eqref{eq:w.pants}. This map is illustrated in \cref{fig:MapSkirt}. It can be understood as the composition $w_s = w_p \circ v$ of the function $w_p(z)$ given in \eqref{eq:w.pants} and the function $v(z) = z^2/(2z^2- 2z +1)$. The latter maps the upper half-plane into the complex plane with a cut along the segment $[0, 1]$. Hence the boundary of the upper half-plane is mapped to the two slits of the skirt geometry. 

In a more general setting, one considers the situation where five fields are inserted in the skirt domain. A field $\phi_1$ is inserted at $-\infty$ in the leg $A$, a field $\phi_2$ is inserted on the first slit's endpoint, a field $\phi_3$ is inserted at $-\infty$ in the leg $B$, a field $\phi_4$ is inserted on the second slit's endpoint, and a field $\phi_5$ is inserted at $+\infty$.
The fields $\phi_i$ with $i=1,2,3,4$ are boundary fields that depend on a single variable and have the conformal dimension $\Delta_i$. We denote by $w_i$ these positions in the skirt domain, and by $z_i$ the corresponding positions in the complex plane obtained from the inverse map $w^{-1}_s(z)$. In contrast, $\phi_5$ is a bulk field that depends on two variables, which we write as $w_5,\bar w_5$ in the skirt domain and as $z_5,\bar z_5$ in the complex plane. For simplicity, we assume that $\phi_5$ is spinless. It has the conformal dimensions $\Delta_5 = \bar \Delta_5$. 

The first function $\tilde g_0$ in the expansion \eqref{eq:Fs.expansion} is obtained as a direct application of the Cardy-Peschel formula~\cite{CP88} for domains with corners. Indeed, the skirt domain has two corners of angles $2\pi$. The resulting expression for $\tilde g_0$ depends on the dimensions of the field $\phi_2$ and $\phi_4$ and on the central charge:
\be
\label{eq:g0Skirt}
\tilde g_0 = \frac c4 + \Delta_2 + \Delta_4.
\ee

The next-leading terms $\tilde g_1(x)$ and $\tilde g_2(x)$ depend non-trivially on the aspect ratio $x$. In this section, we write down the resulting expressions, with their derivations given in \cref{app:CFT.derivations}.
In a general setting where the five fields are primary fields, the function $\tilde g_1(x)$ also depends on the non-trivial functions of the cross-ratios that arise in the five-point function of these fields in the upper half-plane. Because $\phi_5$ is a bulk field, from Cardy's method of images, we know that this correlator is in fact a six-point function in the full complex plane. Here we restrict our focus to the following special cases: (i) only one non-trivial primary field is present, namely the bulk field $\phi_5$, and (ii) each of the fields $\phi_i$ is a vertex operator. In this case, the fields $\phi_i$ with $i = 1, \dots, 4$ are boundary fields with charges $\alphaC_i$ and dimension $\Delta_i=\alphaC_i^2/2$. In contrast, the field $\phi_5$ is a bulk field with the charges $\alphaC_5 =\bar\alphaC_5$ and dimensions $\Delta_5=\bar\Delta_5=\alphaC_5^2/2$. 

The results are as follows. For case~(i), we have
\be 
\label{eq:g1tilde.skirt.4pts}
\tilde g_1(x) =\frac{c}{24}\left( \tilde G(x)+ \tilde G(1-x)\right )+ 4\Delta_5 \left( x \log x+ (1-x)\log(1-x) +2 \log 2\right) + \tilde{C}
\ee
where 
\be
\label{eq:tildeGx}
\tilde G(x)=\frac{3-6x+4x^2}{1-x}\,\log x
\ee
and $\tilde{C}$ is a constant. For case~(ii), we have
\begin{multline}
 \label{eq:g1tilde.skirt.6pts}
\tilde g_1(x) = \frac{c}{24}\left( \tilde G(x)+ \tilde G(1-x)\right ) + 4\Delta_5 \left( x \log x+ (1-x)\log(1-x) +2 \log 2
\right)\\ 
 -\frac{\Delta_1}{x}\log(1-x) -\frac{\Delta_3}{1-x}\log x+(\Delta_2+\Delta_4) (\log x+\log(1-x)) + \tilde{C}'
\end{multline}
where $\tilde{C}'$ is a constant. We note that the functions $\tilde g_0$ and $\tilde g_1(x)$ were obtained in \cite{DS11} for the special case where all the conformal dimensions are zero.

Similarly, we compute the function $\tilde g_2(x)$ using the arguments of St\'ephan and Dubail. The details are given in \cref{app:CFT.derivations}. The authors argue that this function depends on a non-universal constant called the extrapolation length. As we shall see in \cref{sec:AsymptoticsSkirt}, a correct conformal interpretation of our lattice results requires a generalisation of their derivation to cases where each slit is assigned its own extrapolation length. We denote them by $\Xi_2$ and $\Xi_4$, and they correspond to the corners situated at $w_2$ and $w_4$, respectively. For case (i), we obtain
\begin{equation}
\label{eq:g2tilde.skirt.3pts}
\tilde g_2(x) = (\Xi_2+\Xi_4) \times \bigg(\frac{c\,(1-2x)^2}{48x(1-x)}+2 \Delta_5 \bigg).
\end{equation}
For the case (ii), the expression for $\tilde g_2(x)$ instead reads
\begin{equation}
\begin{split}
\label{eq:g2tilde.skirt.version.2}
\tilde{g}_2(x)
&= \Xi_2 \times \left[\frac{c (1-2 x)^2}{48 (1-x) x} + \frac{\alphaC _2^2-\alphaC _4^2}{16 x(1-x)}-\frac{\alphaC _1 \left(\alphaC _1-\alphaC _2+\alphaC _4\right)}{4 x} -\frac{ \alphaC _3 \left(\alphaC_3-\alphaC _2+\alphaC _4\right)}{4(1-x)}+\alphaC_5^2 \right]\\
& + \{ \Xi_2\to \Xi_4, \alphaC_2\leftrightarrow \alphaC_4\}. 
\end{split}
\end{equation}

We note that this expression greatly simplifies for $\Xi_2=\Xi_4$, and becomes independent of the charges $\alphaC_2$ and $\alphaC_4$ of the fields inserted at the endpoints of the slits. The function $\tilde{g}_2(x)$ also simplifies for $\alphaC_2=\alphaC_4$. In that case, it only depends on the sum $\Xi_2+\Xi_4$, as in \eqref{eq:g2tilde.skirt.3pts}, and can be written in terms of the dimensions $\Delta_i=\alphaC_i^2/2$:
\begin{equation}
\tilde g_2(x)\big|_{\alphaC_2=\alphaC_4} = (\Xi_2+\Xi_4) \times \bigg(\frac{c\,(1-2x)^2}{48 x(1-x)}+2\Delta_5 -\frac{\Delta_1}{2x}-\frac{\Delta_3}{2(1-x)} \bigg).
\end{equation}

\subsection{Lattice calculation for the XX spin chain}\label{sec:XX.skirt}

\subsubsection{Definition of the model}

The first model for which we compute the bipartite fidelity on the skirt geometry is the XX chain. In this case, the system $AB$ is a periodic chain of length $N$. Its Hamiltonian is \eqref{eq:Hxx} and depends on the twist $\phi$. The subsystems $A$ and $B$ are instead open chains of lengths $N_A$ and $N_B$. The Hamiltonian for the open chain of length $N$ with free boundary conditions, namely without fields applied to the endpoints, is 
\begin{equation}
\label{eq:Hf}
H_{\!f} = -\frac{1}{2}\sum_{j=1}^{N-1}\left(\sigma_j^x\sigma_{j+1}^x+\sigma_j^y\sigma_{j+1}^y \right).
\end{equation}
We denote by $\langle x_0^{A}|$ and $\langle x_0^{B}|$ the left groundstates of $H_{\!f}$ with $N_A$ and $N_B$ sites, respectively, in their zero-magnetisation sectors. As before, $N_A+N_B=N$ and all three lengths are even numbers.  The logarithmic bipartite is
\begin{equation}
\label{eq:LBFXXSkirt}
\mathcal{F}_s^{\tXX} = - \log \left| \braket{x_0^A \otimes x_0^B | X_0^{AB}}\right|^2
\end{equation}
where $\ket{X_0^{AB}}$ is the groundstate of $H$ given in \eqref{eq:X0}. The states $\ket{x_0^A}$ and $\ket{x_0^B}$, like $\ket{X_0^{AB}}$, are assumed to have unit norms.

\subsubsection{Bipartite fidelity}\label{sec:BFSkirt}

The diagonalisation of $H_f$ is standard. The tensor product of the left groundstates is
\begin{equation}
\bra{x_0^A \otimes x_0^B} = \bra{0}d^B_{N_B/2} \cdots d^B_1d^A_{N_A/2} \cdots d^A_1
\end{equation}
where 
\begin{subequations}
\label{eq:d}
\begin{alignat}{2}
d^A_k &=\Big(\frac{2}{N_A+1}\Big)^{1/2} \sum \limits_{j=1}^{N_A}\sin \Big(\frac{\pi k j}{N_A+1}\Big)c_j,\\ 
d^B_k &=\Big(\frac{2}{N_B+1}\Big)^{1/2} \sum \limits_{j=N_A+1}^{N}\sin \Big(\frac{\pi k (j-N_A)}{N_B+1}\Big)c_j.
\end{alignat}
\end{subequations}
The operators $c_j$ are the fermionic operators in \eqref{eq:JW}. With the choice of normalisation in \eqref{eq:d}, both $\bra{x_0^A}$ and $\bra{x_0^B}$ have unit norms. To compute \eqref{eq:LBFXXSkirt}, we use Wick's theorem and find
\begin{subequations}
\label{eq:overlappXXSkirt}
\begin{equation}
\big|\!\braket{x_0^A \otimes x_0^B | X_0^{AB}}\!\big| = 2^{-3N/4}N^{-N/4}(N_A+1)^{-(N_A-2)/4}(N_B+1)^{-(N_B-2)/4} \,|\!\det \mathcal{D}|
\end{equation}
where
\begin{equation}
\mathcal{D}_{k,k'} = \left\{\begin{array}{cl}
\displaystyle\frac{1-(-1)^k \eE^{-\ir (N_A+1)\vartheta_{k'}}}{\cos \big(\frac{\pi k}{N_A+1}\big)-\cos \vartheta_{k'}} &k=1,\dots,\frac{N_A}{2},  \\[0.6cm]
\displaystyle\eE^{-\ir N_A\vartheta_{k'}}  \frac{1-(-1)^{k-N_A/2} \ \eE^{-\ir (N_B+1)\vartheta_{k'}}}{\cos \big(\frac{\pi (k-N_A/2)}{N_B+1}\big)-\cos \vartheta_{k'}} &k=\frac{N_A}2+1,\dots, \frac N2,
\end{array}\right. \quad k'=1,\dots,\tfrac N2,
\end{equation}
and 
\begin{equation}
\label{eq:vartheta}
\vartheta_k =\frac{2 \pi (k -\frac12-\frac N4)- \phi}N.
\end{equation}
\end{subequations}
This holds for both parities of $N/2$. Sadly, we have been unable to push the calculation further and evaluate this determinant in product form using the methods employed in \cite{SD13,PMDR19}, even for special values of $x$ and $\phi$. Our analysis in \cref{sec:AsymptoticsSkirtXX} of the asymptotics of $\mathcal F_s^\tXX$ will thus rely on the numerical evaluations of determinants.

\subsubsection{Other instances of the bipartite fidelity}\label{sec:modifiedLBFSkirt}

The CFT predictions of \cref{sec:CFT.skirt} covers cases where fields are inserted in the legs and on the slits' endpoints. In order to investigate these cases on the lattice, we consider various modified instances of the bipartite fidelity on the skirt domain. In each case, the state $\ket{X_0^{AB}}$ is kept unchanged in the overlap, whereas the state $\bra{x_0^A \otimes x_0^B}$ is replaced by a state of the form
\begin{equation}
\label{eq:four.state}
\bra{x^{A}_{m_A}\otimes s_1 \otimes x^{B}_{m_B} \otimes s_2}.
\end{equation}
Here $\bra{x^S_m}$ is the left groundstate of the open XX chain of the system $S$ in the magnetisation sector $m$, and $s_1,s_2$ are selected from the set $\{\emptyset,\uparrow, \downarrow, \uparrow\downarrow, \uparrow\uparrow, \uparrow\uparrow\uparrow\}$. The condition on the even parity of $N_A$ and $N_B$ is relaxed and the relation tying them to $N$ is $N = N_A + n_1 + N_B + n_2$, where $n_1$ and $n_2$ are the lengths of the states $\ket{s_1}$ and $\ket{s_2}$. \cref{table:DefCasesSkirt} gives an overview of the thirteen cases that we consider. In particular, case 1 corresponds to the standard bipartite fidelity, as defined in \eqref{eq:LBFXXSkirt} with the overlap \eqref{eq:overlappXXSkirt}. In the other cases, $N_A$ is chosen even if $m_A$ is an integer and odd if $m_A$ is a half-integer, and likewise for $N_B$ in terms of $m_B$. In each case, the magnetisation of the state \eqref{eq:four.state} vanishes, and the corresponding overlap $\big|\!\braket{x^{A}_{m_A}\otimes s_1 \otimes x^{B}_{m_B} \otimes s_2| X_0^{AB}}\!\big|$ is non-zero. We obtain a determinant expression for each of these overlaps similar to \eqref{eq:overlappXXSkirt}, which we do not reproduce here. We are unable to evaluate these determinants in product form. Our analysis in \cref{sec:AsymptoticsSkirt} will instead rely on the numerical evaluation of these determinants.

\begin{table}
\begin{center}
\begin{tabular}{|c|c|c|c|c|}
\hline
 case & $m_A$ & $s_1$ & $m_B$ & $s_2$ \\
 \hline
 1
 & $0$ & $\emptyset$ & $0$& $\emptyset$ \\
 \hline
 2
 & $-1/2$ & $\emptyset$ & $+1/2$& $\emptyset$ \\
 \hline 
 3
 & $-1$ & $\emptyset$ & $1$& $\emptyset$ \\
 \hline
 4
 & $-3/2$ & $\emptyset$ & $+3/2$& $\emptyset$ \\
 \hline
 5
 & $-2$ & $\emptyset$ & $2$& $\emptyset$ \\
 \hline
 6
 & $0$ & $\uparrow$ &  $0$& $\downarrow$ \\ 

 \hline
 7
 & $-1/2$ & $\uparrow$ & $+1/2$& $\downarrow$ \\
 \hline
 8
 & $+1/2$ & $\downarrow$ & $+1/2$& $\downarrow$ \\
 \hline
 9
 & $0$ & $\emptyset$ & $0$& $\uparrow\downarrow$ \\
 \hline
 10
 & $-1/2$ & $\emptyset$ & $-1/2$& $\uparrow\uparrow$ \\ 
  \hline
 11
 & $-1/2$ & $\emptyset$ & $0$& $\uparrow$ \\
   \hline
 12
 & $-1$ & $\emptyset$ & $0$& $\uparrow\uparrow$ \\
   \hline
 13
 & $-3/2$ & $\emptyset$ & $0$& $\uparrow\uparrow\uparrow$ \\
 \hline
\end{tabular}
\end{center}
\caption{The thirteen cases considered for the XX chain on the skirt geometry.}
\label{table:DefCasesSkirt}
\end{table}

\subsection{Lattice calculation for critical dense polymers}\label{sec:polymers.skirt}

\subsubsection{Definition of the model}

We study the model of dense polymers on the skirt geometry. The lattice is a cylinder of height $4M$ and width $N$, which we choose to be even. There are two vertical slits that extend halfway across the cylinder. They divide the lower edge into two open segments of lengths $N_A$ and $N_B$, which are even numbers too. A configuration of the model of critical dense polymers on the skirt lattice is a tiling of the faces of this lattice by one of the two elementary tiles, 
$\psset{unit=0.3cm}
\,\begin{pspicture}[shift=-0.16](0,0)(1,1)
\pspolygon[fillstyle=solid,fillcolor=lightlightblue,linewidth=\mince](0,0)(0,1)(1,1)(1,0)
\psarc[linewidth=\moyen,linecolor=blue](1,0){.5}{90}{180}
\psarc[linewidth=\moyen,linecolor=blue](0,1){.5}{-90}{0}
\end{pspicture}$\, and $\,\psset{unit=0.3cm}
\begin{pspicture}[shift=-0.16](0,0)(1,1)
\pspolygon[fillstyle=solid,fillcolor=lightlightblue,linewidth=\mince](0,0)(0,1)(1,1)(1,0)
\psarc[linewidth=\moyen,linecolor=blue](0,0){.5}{0}{90}
\psarc[linewidth=\moyen,linecolor=blue](1,1){.5}{180}{-90}
\end{pspicture}\,$, with equal probability. 

The top circular edge, the bottom segments as well as the interior of the two slits are decorated with simple half-arcs. The situation is then similar to the example of \cref{fig:loop.config}, with the following differences: (i) the total height of the lattice is $4M$ instead of $2M$, and (ii) the four diagonal edges in the diagram's lower-half are decorated with simple half-arcs. The contractible loops are given a weight $\beta=0$, whereas the non-contractible loops have a fugacity $\alpha$. The Boltzmann weight of a configuration $\sigma$ is then given by $W_\sigma = \alpha^{n_\alpha} \delta_{n_\beta,0}$, where $n_\alpha$ and $n_\beta$ are the numbers of non-contractible and contractible loops in the configuration. The partition function on the skirt geometry, denoted $Z^{AB}_{s}$, is defined as
\begin{equation}
Z^{AB}_{s}=\sum_{\sigma} W_\sigma.
\end{equation}

The definition of the bipartite fidelity given below involves three more partition functions. The first, $Z_c^{A\cup B}$, is the partition function of the same model defined on a cylinder of height $4M$ and perimeter~$N$. The other two partition functions, $Z_{r}^A$ and $Z_{r}^B$, are defined on $4M \times N_A$ and $4M \times N_B$ rectangles, respectively. The boundary conditions consist of simple half-arcs on all four segments, and we restrict to configurations that have a single (contractible) loop. The partition functions $Z_{r}^A$ and $Z_{r}^B$ are then the numbers of these restricted configurations. 

Similarly to $Z_c^B$ discussed in \cref{sec:def}, the partition functions $Z_{r}^A$ and $Z_{r}^B$ can alternatively be defined on a lattice where the rightmost arcs on both the top and bottom edges are removed and replaced by a pair of defects. One then imposes 
 that both defects from the top segment connect to those of the bottom segment (with weight $1$). This produces exactly the same set of configurations and therefore the correct partition functions.

The logarithmic bipartite fidelity is then defined as
\begin{equation}\label{eq:Fsalpha}
\mathcal{F}_s^{\alpha} =- \lim_{M \to \infty} \log \Bigg( \frac{\big(Z_{s}^{AB}\big)^2}{\alpha \, Z_c^{A \cup B} Z_{r}^A Z_{r}^B}\Bigg).
\end{equation}
As we shall see, the limit $M \to \infty$ of this ratio is well defined. The factor $\alpha$ in the denominator ensures that $\mathcal{F}_s^{\alpha}$ is well-defined in the limit $\alpha \to 0$. Indeed, $Z^{AB}_{s}$ and $Z_c^{A\cup B}$ both vanish linearly in $\alpha$ as it tends to zero. With this convention for $\mathcal{F}_s^{\alpha}$, the leading powers of $\alpha$ coincide between the numerator and the denominator.

The choice to include $Z_{r}^A$ and $Z_{r}^B$ in the denominator is justified as follows. We note that for the other dense loop models for which the fugacity of contractible loops is non-zero, the natural choice would be to include in the denominator the partition functions with no defects. For the model of critical dense polymers, these partition functions are zero. As argued in \cite{MDJ18}, the reference partition functions in this case are instead those with pairs of defects on the top and bottom segments, as explained above.

\subsubsection{The Temperley-Lieb algebra}

Our goal is first to express the partition functions in  \eqref{eq:Fsalpha} in the language of the Temperley-Lieb algebra. The skirt geometry involves both periodic and open boundary conditions. For that reason, both the ordinary Temperley-Lieb algebra $\tl_N(\beta)$ and the enlarged periodic Temperley-Lieb algebra $\eptl_N(\alpha,\beta)$ are useful. Since $\tl_N(\beta)$ is a subalgebra of $\eptl_N(\alpha,\beta)$, all the tools needed to define $\tl_N(\beta)$ are given in \cref{sec:TL}, and below we only give a brief reminder. 

\paragraph{Definition of $\boldsymbol{\tl_N(\beta)}$.}
The Temperley-Lieb algebra \cite{TL71,J83,M91,GW93,W95,RSA14} is a unital, associative algebra generated by the linear span of connectivities. It is the subalgebra of $\eptl_N(\alpha,\beta)$ generated by the generators $\boldsymbol I$ and $e_j$, with $j = 1, \dots, N-1$. With products of these generators, one can produce all the connectivity diagrams in $\eptl_N(\alpha,\beta)$ that have no loop segments travelling via the back of the cylinder. The diagrammatic rules for computing products of connectivity diagrams are the same as those described in the periodic case. Non-contractible loops are never created in such products, and the algebra thus depends only on the weight $\beta$ of the contractible loops. We recall that the value of $\beta$ pertaining to the model of critical dense polymers is $\beta = 0$. 

\paragraph{The transfer tangle $\boldsymbol{\Db (u)}$.}

The double-row transfer tangle for the model of dense polymers is an element of $\tl_N(0)$ defined as \cite{PR07}
\begin{equation} 
\label{eq:Du}
\psset{unit=0.8cm}
\Db (u) = \frac{1}{\sin 2u}\ \ 
\begin{pspicture}[shift=-1.5](-0.5,-.6)(5.5,2)
\facegrid{(0,0)}{(5,2)}
\psarc[linewidth=0.025]{-}(0,0){0.16}{0}{90}
\psarc[linewidth=0.025]{-}(1,1){0.16}{90}{180}
\psarc[linewidth=0.025]{-}(1,0){0.16}{0}{90}
\psarc[linewidth=0.025]{-}(2,1){0.16}{90}{180}
\psarc[linewidth=0.025]{-}(4,0){0.16}{0}{90}
\psarc[linewidth=0.025]{-}(5,1){0.16}{90}{180}
\rput(2.5,0.5){$\ldots$}
\rput(2.5,1.5){$\ldots$}
\rput(3.5,0.5){$\ldots$}
\rput(3.5,1.5){$\ldots$}
\psarc[linewidth=1.5pt,linecolor=blue]{-}(0,1){0.5}{90}{-90}
\psarc[linewidth=1.5pt,linecolor=blue]{-}(5,1){0.5}{-90}{90}
\rput(0.5,.5){$u$}
\rput(0.5,1.5){$u$}
\rput(1.5,.5){$u$}
\rput(1.5,1.5){$u$}
\rput(4.5,.5){$u$}
\rput(4.5,1.5){$u$}
\rput(2.5,-0.5){$\underbrace{\qquad \hspace{2.5cm} \qquad}_N$}
\end{pspicture} \ ,
\qquad 
 \begin{pspicture}[shift=-.40](1,1)
\facegrid{(0,0)}{(1,1)}
\psarc[linewidth=0.025]{-}(0,0){0.16}{0}{90}
\rput(.5,.5){$u$}
\end{pspicture}
\ = \cos u\ \
\begin{pspicture}[shift=-.40](1,1)
\facegrid{(0,0)}{(1,1)}
\rput[bl](0,0){\loopa}
\end{pspicture}
\ + \sin u \ \
\begin{pspicture}[shift=-.40](1,1)
\facegrid{(0,0)}{(1,1)}
\rput[bl](0,0){\loopb}
\end{pspicture} \ , 
\end{equation}
where $u$ is the spectral parameter. We refer to the value $u= \frac \pi 4$, for which both tiles have equal weights, as the {\it isotropic point}, and use the short-hand notation $\Db = \Db(\frac \pi4)$. 

Two copies of the transfer tangle evaluated at different values of the spectral parameter commute: $[\Db(u),\Db(v)] = 0$. Furthermore, the Hamiltonian $\boldsymbol {H}_{\!o}$ of the model is related to the transfer tangle via the relation 
\begin{equation}
\label{eq:DH}
\Db(u) = \Ib -2 u \boldsymbol{H}_{\!o} + \mathcal \mathcal \mathcal O(u^2), \qquad \boldsymbol H_{\!o} = - \sum_{j=1}^{N-1}e_j,\\
\end{equation}
and is also in the commuting family. 

\paragraph{The standard modules $\boldsymbol{\stanV_{N,0}}$ and $\boldsymbol{\stanV_{N,2}}$.} 
The standard modules $\stanV_{N,0}$ and $\stanV_{N,2}$ for $\tl_N(\beta)$ are built on the vector space generated by link states on $N$ nodes with zero and two defects, respectively. The standard module $\stanV_{N,0}$ is defined on the same vector space as the module $\widehat\stanW_{N,0}$ over $\eptl_N(\beta,\beta)$. It is in fact the restriction of this module to the action of elements of $\tl_N(\beta) \subset \eptl_N(\beta,\beta)$. Likewise, the module $\stanV_{N,2}$ is defined on the subspace of $\stanW_{N,2}$ spanned by link states with no arcs travelling via the back of the cylinder. The diagrammatic action of the connectivities in $\tl_N(\beta)$ on the link state of $\stanV_{N,0}$ and $\stanV_{N,2}$  is the same as in the periodic case, with the difference that non-contractible loops are never formed. 

\paragraph{Partition functions.} The Gram products for the standard modules of $\tl_N(\beta)$ are defined in the same way as those introduced in \cref{sec:forms} for the periodic case. We express the partition functions in \eqref{eq:Fsalpha} as
\begin{subequations}
\label{eq:ZF}
\begin{alignat}{2}
Z_s^{AB} &= 2^{2MN} (v^A_0 \otimes v^B_0)\cdot \big(\Db^{A}\otimes \Db^{B}\big)^M \big(\Tb^{AB}\big)^{2M} v^{AB}_0\big|_{\mathsf W_{N,0}}\, ,\label{eq:ZFAB}\\[0.2cm]
Z_c^{A\cup B}&=2^{2MN} v^{AB}_0 \cdot (\Tb^{AB})^{4M} v^{AB}_0\big|_{\mathsf W_{N,0}}\, ,\\[0.2cm] 
Z_r^{A} &=2^{2MN_A} v^A_2 \cdot (\Db^{A})^{2M} v^{A}_2\big|_{\mathsf V_{N_A,2}}\, ,\\[0.2cm]
Z_r^{B}&=2^{2MN_B} v^B_2 \cdot (\Db^{B})^{2M} v^B_2\big|_{\mathsf V_{N_B,2}}\, ,
\end{alignat}
\end{subequations}
where $v_0$ and $v_2$ are defined in \eqref{eq:v0v2}. The conventions are the same as those used in \cref{sec:forms}.

\paragraph{The XX representation and spin-chain overlaps.} The XX representation of $\tl_N(\beta =0)$ is given in \eqref{eq:X.rep}. In this representation, the Hamiltonian with open boundary conditions $H_{o}$ is the XX Hamiltonian with the $U_q(s\ell_2)$-invariant boundary magnetic fields of Pasquier and Saleur \cite{PS90},
\begin{equation}
\label{eq:Ho}
H_{\!o} = \mathsf X_N (\boldsymbol H_{\!o}) = -\frac{1}{2}\sum_{j=1}^{N-1}\big(\sigma_j^x\sigma_{j+1}^x+\sigma_j^y\sigma_{j+1}^y \big)-\frac{\ir}{2}(\sigma_1^z-\sigma_N^z).
\end{equation}

The representative of $\Db(u)$ in the XX representation is the double-row transfer matrix $D(u)$. We use the notation $D = D(\frac \pi 4)$ for the transfer matrix at the isotropic point. The map from link states to spin states given in \eqref{eq:localmaps} is a homomorphism between link state and spin representations for the algebra $\tl_N(\beta)$ as well. The partition functions can then be written in terms of spin-chain overlaps as
\begin{subequations}\label{eq:ZS}
\begin{alignat}{2}
Z_s^{AB} &= 2^{2MN}\llangle v^A_0 \otimes v^B_0| \big(D^{A}\otimes D^{B}\big)^M \big(T^{AB}\big)^{2M} |v^{AB}_0\rangle,\label{eq:ZSAB}\\[0.2cm]
Z_c^{A\cup B} &= 2^{2MN}\llangle v^{AB}_0| (T^{AB})^{4M} |v^{AB}_0\rangle,\\[0.2cm] 
Z_r^{A} &=2^{2MN_A} \llangle v^A_2 | (D^{A})^{2M} |v^{A}_2\rangle,\\[0.2cm]
Z_r^{B} &= 2^{2MN_B}\llangle v^B_2| (D^{B})^{2M} |v^B_2\rangle.
\end{alignat}
\end{subequations}
We note that all the states that appear as dual states do not depend on the twist parameter $\phi$, so in this case \eqref{eq:dual.vectors} simply becomes $\llangle w| = | w \rangle^\Tt$.

\subsubsection{Bipartite fidelity}\label{sec:polymers.skirt.BF}

\paragraph{Diagonalisation of $\boldsymbol{H_{\!o}}$.}
The first step is to use the Jordan-Wigner transformation to write $H_{\!o}$ as
\begin{equation}
H_{\!o} = - \sum_{j=1}^{N-1} \big(c^\dagger_{j+1}c_{j} + c^\dagger_j c_{j+1}  \big) - \ir \left(c^\dagger_1c_1 -c^\dagger_Nc_N\right)
\end{equation}
where the fermionic operators $c_j$ and $c_j^\dagger$ are defined in \eqref{eq:JW}. Recalling that $\omega = \eE^{\ir \pi/4}$, the second step is to perform a Fourier transform of these operators, by defining
\begin{equation}
\eta_k = \frac{1}{\kappa_k} \sum_{j=1}^{N-1} \sin(\tfrac {\pi k j}{N})\, a_j, \qquad \eta_k^{\text t} = \frac{1}{\kappa_k} \sum_{j=1}^{N-1} \sin(\tfrac {\pi k j}{N})\, a_j^{\text t}, \qquad \kappa_k = \sqrt{N \cos(\tfrac{\pi k}N)},
\end{equation}
where
\begin{equation}
a_j = \omega\, c_j + \omega^{-1} c_{j+1}, \qquad a_j^{\text t} = \omega\, c_j^\dagger + \omega^{-1} c_{j+1}^\dagger.
\end{equation}
These operators satisfy the fermionic relations
\begin{equation}
\{a_j, a_k^\Tt\} = \delta_{j,k-1}+\delta_{j,k+1}, \quad \{\eta_k, \eta_\ell^\Tt\} = \delta_{k,\ell}, \quad  \{a_j, a_k\} = \{a_j^\Tt, a_k^\Tt\} = \{\eta_k, \eta_\ell\} = \{\eta_k^\Tt, \eta_\ell^\Tt\} = 0.
\end{equation}
The Hamiltonian can be expressed in Jordan-normal form using these operators. For $N$ even, the set of operators $\eta_k$ and $\eta^\Tt_k$, with $k \in \{1, \dots, \frac{N-2}2\} \cup \{\frac{N+2}2, \dots, N-1\}$ is complemented with the operators 
\begin{equation}
\chit = -\omega\sqrt{\frac 2N} \sum_{j=1}^N \ir^{-(j-1)} \big(\big\lfloor \tfrac j2 \big\rfloor- \tfrac N4\big) \, c_j, \qquad \chit^\Tt =-\omega\sqrt{\frac 2N} \sum_{j=1}^N \ir^{-(j-1)} \big(\big\lfloor \tfrac j2 \big\rfloor- \tfrac N4\big)  \,c_j^\dagger,
\end{equation}
as well as with the operators $\varphi$ and $\varphi^\Tt$,
\begin{equation}
\label{eq:phi}
\varphi = \omega^{-1}\sqrt{\frac2{N}} \sum_{j=1}^{N}\ir^{-(j-1)} c_j, \qquad \varphi^\Tt = \omega^{-1}\sqrt{\frac2{N}}\sum_{j=1}^{N}\ir^{-(j-1)} c^\dagger_j.
\end{equation}
The anti-commutation relations are
\begin{equation}
\{\varphi, \chit^\Tt\} = \{\varphi^\Tt, \chit\} =1, \qquad \{\varphi^\Tt, \varphi\} = \{\chit^\Tt, \chit\} = \{\varphi, \chit\} = \{\varphi^\Tt, \chit^\Tt\} =  0.
\end{equation}
All the anti-commutators involving the operators $\eta_k$ and $\eta^\Tt_k$ and one of $\varphi, \varphi^\Tt, \chit$ and $\chit^\Tt$ also vanish.
In terms of these operators, the Hamiltonian takes the form
\begin{equation}
\label{eq:Heven}
H_{\!o} = \varphi^\Tt \varphi-2  \sum_{\substack{k=1\\ k\neq N/2}}^{N-1}\cos(\tfrac{\pi k}N) \eta_k^\Tt \eta_k.
\end{equation}
The groundstate eigenspace in the zero-magnetisation sector is two-dimensional and is spanned by the states
\begin{equation}
\label{eq:w0}
|w_0 \rangle = \varphi^\Tt \eta_1^\Tt \eta_2^\Tt \dots \eta_{N/2-1}^\Tt |0\rangle, \qquad
|\hat w_0 \rangle = \chit^\Tt \eta_1^\Tt \eta_2^\Tt \dots \eta_{N/2-1}^\Tt  |0\rangle.
\end{equation}
These form a rank-two Jordan cell:
\begin{equation}
\label{eq:HJordan}
H_{\!o}|w_0\rangle = h_0 |w_0\rangle, \qquad H_{\!o}|\hat w_0\rangle = h_0 |\hat w_0\rangle + |w_0 \rangle,
\qquad h_0 = 1- \textrm{cot}(\tfrac\pi{2N}).
\end{equation}
Likewise, in the sectors of magnetisation $m=-1$, the state with the lowest energy is 
\begin{equation}
|w_{-1}\rangle = \eta_1^\Tt \eta_2^\Tt \dots \eta^\Tt_{N/2-1} |0 \rangle
\end{equation}
and its eigenvalue is also $h_0$.
Because $D(u)$ and $H_{\!o}$ commute, we have
\begin{equation}
D(u) \ket{w_0} = \lambda_0(u) \ket{w_0}\!,\qquad  D(u) \ket{\hat w_0}= \lambda_0(u)\ket{\hat w_0} + f(u) \ket{w_0}, \qquad
D(u)  \ket{w_{-1}} = \lambda_{-1}(u) \ket{w_{-1}}\!.
\end{equation}
where $\lambda_0(u) = \lambda_{-1}(u)$ are the eigenvalues of $D(u)$ and $f(u)$ is a non-zero function.
At the isotropic point, these states generate the eigenspace of $D$ of maximal eigenvalue in the sectors $S^z = 0, -1$.

\paragraph{Ratios of partition functions in the limit $\boldsymbol{M \to \infty}$.}

Following the same steps and logic as in \cref{sec:BF} and in \cite{PMDR19}, we extract the leading behaviour of the overlaps \eqref{eq:ZS} in the limit $M \to \infty$. The result is 

\begin{equation}\label{eq:Fsoverlaps}
\mathcal{F}_s^{\alpha} = -  \log \Bigg( \alpha^{-1}\llangle w_0^{A} \otimes w_0^{B} |X_0^{AB}\rangle^2 \frac{ \llangle v_0^{A}|\hat w_0^{A}\rangle^2}{\llangle v_2^{A}|w_{-1}^{A}\rangle^2} \frac{\llangle v_0^{B}|\hat w_0^{B}\rangle^2}{\llangle v_2^{B}|w_{-1}^{B}\rangle^2} \frac{\llangle X_0^{AB}|v_0^{AB}\rangle}{\llangle v_0^{AB}|X_0^{AB}\rangle}\Bigg)
\end{equation}
where
$\llangle v_0|$ and $\llangle v_2|$ are defined in \eqref{eq:v0v2ferm}.

\paragraph{Determinant forms for the overlaps.}
We express the overlaps in \eqref{eq:Fsoverlaps} in determinant form with Wick's theorem. Two of the ratios involving boundary states were already computed in \cite{PMDR19},
\begin{equation}
\label{eq:simple.overlap}
\frac{\llangle  v_0^{A}|\hat w_0^{A}\rangle}{\llangle v_2^{A}|w_{-1}^{A}\rangle} =  \omega^{2}\sqrt{\frac{N_{A}}{2}},\qquad  \frac{\llangle  v_0^{B}|\hat w_0^{B}\rangle}{\llangle v_2^{B}|w_{-1}^{B}\rangle} =  \omega^{2}\sqrt{\frac{N_{B}}{2}}.
\end{equation}
To compute the last ratio in \eqref{eq:Fsoverlaps}, we need the explicit form of the anti-commutators \eqref{eq:acomm.mu.a}. After some algebra, we obtain
\begin{equation}
\frac{\llangle X_0^{AB}|v_0^{AB}\rangle}{\llangle v_0^{AB}|X_0^{AB}\rangle} = \Bigg( \prod_{k=1}^{N/2} \frac{\cos(\frac{\pi}{4}-\frac{\vartheta_k}{2})}{\cos(\frac{\pi}{4}+\frac{\vartheta_k}{2})} \Bigg)\eE^{-\ir \frac{\phi}{2}(N+1)} (-1)^{\frac{N}{2}+1 + \lfloor \frac{N-2}{4} \rfloor}
\end{equation}
where $\vartheta_k$ is defined in \eqref{eq:vartheta}.

It only remains to compute the overlap $\llangle w_0^{A} \otimes w_0^{B} |X_0^{AB}\rangle$. It involves the state
\begin{equation}
\llangle w_0^{A} \otimes w_0^{B} | = \langle 0 |\eta^{B}_{\frac{N_B-2}{2}} \dots \eta^{B}_{1} \varphi^B \eta^{A}_{\frac{N_A-2}{2}} \dots \eta^{A}_{1} \varphi^A
\end{equation}
where
\begin{subequations}
\label{eq:OperatorsSubsystems}
\begin{alignat}{3}
\eta_k^{A} &= \frac{1}{\kappa^A_k}\sum_{j=1}^{N_A-1} \sin ( \tfrac{\pi k j}{N_A} )a_j, 
\qquad 
&&\varphi^{A} = \omega^{-1}\sqrt{\frac{2}{N_A}} \sum_{j=1}^{N_A} \ir^{-(j-1)} \, c_j, 
\\[0.15cm]
\eta_k^{B} &= \frac{1}{\kappa^B_k}\sum_{j=N_A+1}^{N-1} \sin ( \tfrac{\pi k (j-N_A)}{N_B})a_j,
\qquad
&&\varphi^{B} = \omega^{-1}\sqrt{\frac{2}{N_B}} \sum_{j=N_A+1}^{N} \ir^{-(j-N_A-1)} \, c_j,
\\[0.15cm]
\kappa^{A}_k &= \big(N_{A} \cos ( \tfrac{\pi k}{N_{A}})\big)^{1/2}, \qquad &&
\kappa^{B}_k = \big(N_{B} \cos ( \tfrac{\pi k}{N_{B}})\big)^{1/2}.
\end{alignat}
\end{subequations}
To compute the overlap, we introduce the rescaled operators
\begin{equation}
\label{eq:twiddle.eta}
\tilde{\eta}_k = \sum_{j=1}^{N-1} \sin ( \tfrac{\pi k j}{N} )\,a_j, \qquad k = 1, \dots, N-1.
\end{equation}
We similarly define rescaled operators for $\tilde{\eta}^A_k$ and $\tilde{\eta}^B_k$ from \eqref{eq:OperatorsSubsystems}, by removing the prefactors $\kappa_k^A$ and $\kappa_k^B$. These operators have the advantage of being well defined for $k=N/2$. All the fermionic operators appearing in $\llangle w_0^{A} \otimes w_0^{B} |$ can in fact be written in terms of the $\tilde{\eta}_k$, as indeed we have 
$\tilde{\eta}_k=\kappa_k\eta_k$ and $\tilde{\eta}_{\frac{N}{2}} = \omega^{2}\sqrt{N/2}\,\varphi$. 

The anti-commutators between the operators $\tilde{\eta}^A_k$, $\tilde{\eta}^B_k$ and $\mu_j^{\dagger}$ are
\begin{subequations}
\begin{alignat}{2}
&\{ \tilde{\eta}^A_k, \mu_j^{\dagger} \} = \frac{1}{\sqrt{N}}(\omega + \omega^{-1}\eE^{- \ir \theta_j})\sin\Big(\frac{\pi k}{N_A}\Big)\frac{1-(-1)^k\eE^{- \ir N_A \theta_j}}{2 \cos \theta_j-2 \cos (\frac{\pi k}{N_A})} , \\[0.2cm]
&\{ \tilde{\eta}^B_k, \mu_j^{\dagger} \} = \frac{\eE^{- \ir N_A \theta_j}}{\sqrt{N}} (\omega + \omega^{-1}\eE^{- \ir \theta_j})\sin\Big(\frac{\pi k}{N_B}\Big)\frac{1-(-1)^k\eE^{- \ir N_B \theta_j}}{2 \cos \theta_j-2 \cos (\frac{\pi k}{N_B})} .
\end{alignat}
\end{subequations}
The final overlap $\llangle w_0^{A} \otimes w_0^{B} |X_0^{AB}\rangle$ is then obtained as the determinant of these commutators. 
Using the identities
\begin{subequations}
\begin{alignat}{2}
& \prod_{k=1}^{\frac N2-1}\sin \Big (\frac{\pi k}{N} \Big) = \prod_{k=1}^{\frac N2-1}\cos \Big (\frac{\pi k}{N} \Big) = N^{1/2}2^{-\frac{N-1}{2}},
\\[0.2cm]
& \prod_{k=1}^{\frac N2} \cos\Big (\frac{\pi}{4}+\frac{\theta_k}{2}\Big ) \cos\Big (\frac{\pi}{4}-\frac{\theta_k}{2}\Big) = 2^{-N+1} \cos \Big(\frac{\phi}{2}\Big) = 2^{-N} \alpha,
\end{alignat}
\end{subequations}
we find that the result simplifies to
\begin{equation}\label{eq:Fclosed}
\eE^{-\mathcal{F}_s^{\alpha} } = N^{-\frac N2} (N x)^{-\frac{N x}{2}+\frac{3}{2}}(N (1-x))^{-\frac{N (1-x)}{2}+\frac{3}{2}} 2^{-\frac{N}{2}+1} (\det \mathcal{C})^2
\end{equation}
where
\begin{equation}
\mathcal{C}_{k,k'} = \left\{\begin{array}{cl}
\sin\big(\frac{N x \vartheta_{k'}}{2}- \frac{\pi k}{2}\big)\Big( \cos \vartheta_{k'}- \cos \big(\frac{\pi k}{N x}\big)\Big)^{-1} &k=1,\dots,\frac{N x}{2},  \\[0.15cm]
(-1)^{k'}\sin\big(\frac{N (1-x) \vartheta_{k'}}{2}- \frac{\pi (k-\frac{N x}{2})}{2}\big)\Big( \cos \vartheta_{k'}- \cos \big(\frac{\pi (k-\frac{N x}{2})}{N(1-x)}\big)\Big)^{-1} &k=\frac{N x}{2}+1,\dots, \frac N2,
\end{array}\right.
\end{equation}
with $k'=1,\dots,\frac N2$. For arbitrary values of $x$ and $\phi$, we are unable to evaluate the determinant in \eqref{eq:Fclosed} in closed form using the Cauchy determinant formula \eqref{eq:Cauchy}. We are however able to compute the determinant for two specialisations of $x$ and $\phi$: (i) $x=1/2$ with $\phi$ arbitrary and (ii) $\phi=0$ with $x$ arbitrary. For simplicity, our results for these two cases are given below for $N \equiv 0 \textrm{ mod } 4$.
 
\paragraph{Specialisation (i): $\boldsymbol{x=1/2}$ and arbitrary $\boldsymbol{\phi}$. }
For $x=1/2$, $N\equiv 0\textrm{ mod }4$ and arbitrary values of $\phi$, after some simplifications, we find
 \begin{equation}
 \label{eq:FSkirti}
| \det \mathcal{C}| = \cos(\tfrac{\phi}{2})^{N/4} \bigg | \det_{k,k' =1}^{N/4} \Big( \cos \vartheta_{2k'}- \cos \big(\tfrac{\pi k}{N }\big)\Big)^{-1} \det_{k,k' =1}^{N/4} \Big( \cos \vartheta_{2k'-1}- \cos \big(\tfrac{\pi k}{N }\big)\Big)^{-1}  \bigg |. 
 \end{equation}
 In this case, we can apply \eqref{eq:Cauchy} to evaluate both determinants.
 
\paragraph{Specialisation (ii): $\boldsymbol{\phi=0}$ and arbitrary $\boldsymbol{x}$. }
 For $\phi=0$, $N\equiv 0\textrm{ mod }4$ and arbitrary values of $x$, after some simplifications, we find
\begin{subequations}
\label{eq:FSkirtii}
\begin{equation}
|\det \mathcal{C}| = 2^{N/4} \Big|\det \mathcal{C}^1\det \mathcal{C}^2\prod _{k=1}^{N/4} \cos \Big( \tfrac{Nx \vartheta_{k}}{2}\Big)\prod _{k=N/4+1}^{N/2} \sin \Big( \tfrac{Nx \vartheta_{k}}{2}\Big)\Big|
\end{equation} 
with
 \begin{alignat}{2}
&\mathcal{C}^1_{k,k'} = \left\{\begin{array}{cl}
\Big( \cos \vartheta_{k'}- \cos \big(\frac{\pi (2k-1)}{N x}\big)\Big)^{-1} \hphantom{^{-1}}
&k=1,\dots,\frac{N x}{4},  \\[0.15cm]
\Big( \cos \vartheta_{k'}- \cos \big(\frac{\pi (2k-\frac{N x}{2})}{N(1- x)}\big)\Big)^{-1} \hphantom{^{-1}}
 &k=\frac{N x}{4}+1,\dots, \frac N4,
\end{array}\right. 
\quad &&{k'}=1,\dots ,\tfrac{N}{4}, \\[0.2cm]
&\mathcal{C}^2_{k,k'} = \left\{\begin{array}{cl}
\Big( \cos \vartheta_{k'}- \cos \big(\frac{\pi 2k}{N x}\big)\Big)^{-1} &k=1,\dots,\frac{N x}{4},  \\[0.15cm]
\Big( \cos \vartheta_{k'}- \cos \big(\frac{\pi (2k-\frac{N x}{2}-1)}{N(1- x)}\big)\Big)^{-1} &k=\frac{N x}{4}+1,\dots, \frac N4,
\end{array}\right. 
\quad &&{k'}=\tfrac N4 +1,\dots , \tfrac{N}{2}.
 \end{alignat}
  \end{subequations}
In this case as well, both determinants can be evaluated with \eqref{eq:Cauchy}.

\subsection{Asymptotics}\label{sec:AsymptoticsSkirt}

In this subsection, we study the large-$N$ asymptotics of the bipartite fidelity for the XX chain and the model of critical dense polymers on the skirt geometry. We compare these results with the conformal predictions of \cref{sec:CFT.skirt}.

\subsubsection{XX spin chain}\label{sec:AsymptoticsSkirtXX}

\begin{table}
\begin{center}
\begin{tabular}{|c|c|c|c|c|c|c|c|c|}
 \hline\xrowht{10pt}
case & $\Delta_1$ &  $\Delta_2$ &  $\Delta_3$ &  $\Delta_4$ &  $\Delta_5$&  $\tilde{C}'$&  $\Xi_2$& $\Xi_4$ \\
\hline 
1 &$0$ &$0$ & $0$& $0$ &$\Delta^\tXX_\phi$ &$0.03669$ & $1$ & $1$\\
\hline 
2 &$1/8$ &$0$ & $1/8$ &$0$ & $\Delta^\tXX_\phi$& $0.03669$ & $1$ & $1$\\
\hline 
3 &$1/2$ &$0$ & $1/2$ &$0$ & $\Delta^\tXX_\phi$&$0.03669$ & $1$ & $1$ \\
\hline 
4 &$9/8$ &$0$ & $9/8$ &$0$ & $\Delta^\tXX_\phi$&$0.03669$ & $1$ & $1$ \\
\hline 
5 &$2$ &$0$ & $2$ &$0$ & $\Delta^\tXX_\phi$&$0.03669$ & $1$ & $1$ \\
\hline 
6 &$0$ &$1/8$ &$0$ &$1/8$ &$\Delta^\tXX_\phi$ & $1.11331$ & $0$ & $0$\\
\hline 
7 &$1/8$  &$1/8$  &$1/8$  & $1/8$ &$\Delta^\tXX_\phi$ &$1.11331$ & $0$ & $0$ \\
\hline 
8 &$1/8$  &$1/8$  &$1/8$  & $1/8$ &$\Delta^\tXX_\phi$ &$0.42016$ & $0$ &$0$ \\
\hline 
9 &$0$ &$0$ & $0$& $0$ &$\Delta^\tXX_\phi$ &$0.96639$ & $1$ & $-1$ \\
\hline 
10 &$1/8$  & $0$& $1/8$ &$1/2$ &$\Delta^\tXX_\phi$  &$0.78562$& $1$ & $-1$ \\
\hline 
11 &$1/8$  & $0$& $0$ &$1/8$ &$\Delta^\tXX_\phi$  &$0.40172$& $1$ & $0$\\
\hline 
12 &$1/2$  & $0$& $0$ &$1/2$ &$\Delta^\tXX_\phi$  &$0.78562$ & $1$ & $-1$\\
\hline 
13 &$9/8$  & $0$& $0$ &$9/8$ &$\Delta^\tXX_\phi$  & $1.06794$& $1$& $-2$\\
\hline 
\end{tabular}
\end{center}
\caption{The conformal dimensions, the constant $\tilde{C}'$ and the extrapolation lengths for each of the thirteen cases, with $\Delta^\tXX_\phi$ defined in \eqref{eq:Delta.XX}.}
\label{table:ConfDataCasesSkirt}
\end{table}

We study the asymptotic behaviour of the bipartite fidelity for the XX chain numerically, for each case defined in \cref{table:DefCasesSkirt}. The data points are obtained by evaluating the determinants numerically, namely the expression \eqref{eq:overlappXXSkirt} for case 1 and similar determinant expressions for the other cases. We compare these numerical values with a fit of the form 
\begin{equation}
\label{eq:F.skirt.fit}
 \mathcal{F}_s^{\tXX} = \tilde g_0 \log N + \tilde g_1(x) + \tilde g_2 \frac{\log N}{N} +  \gamma_3 \frac{1}{N} + \gamma_4 \frac{\log N}{N^2} +  \gamma_5 \frac{1}{N^2} + \gamma_6 \frac{\log N}{N^3} +  \gamma_7 \frac{1}{N^3}.
\end{equation}

\begin{figure}[!t]
\begin{center}
\begin{pspicture}(0,0.7)(17,11)
\rput(4,8.5){\includegraphics[height=4.5cm]{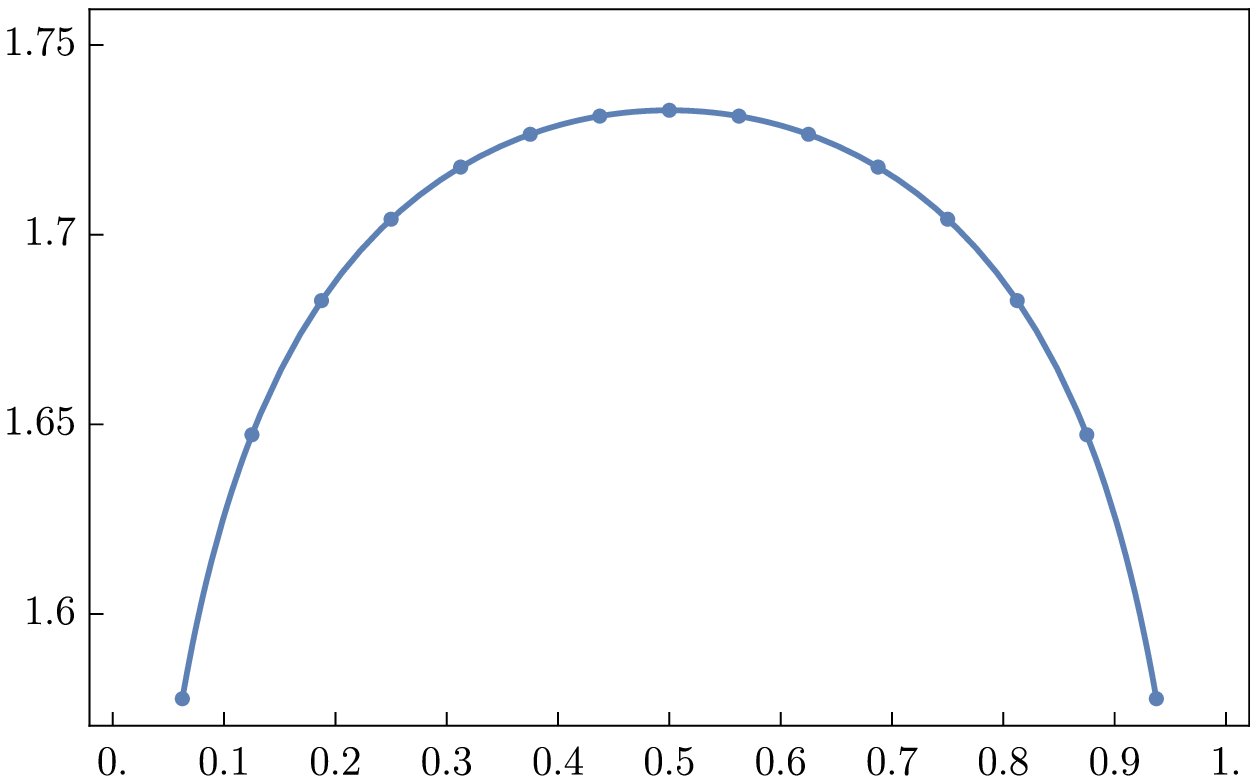}}
\rput(1.7,11){ $\tilde g_0 \log N +\tilde g_1(x)$}
\rput(7.4,6.8){$x$}
\rput(4.2,8.5){case 1}
\rput(13,8.5){\includegraphics[height=4.5cm]{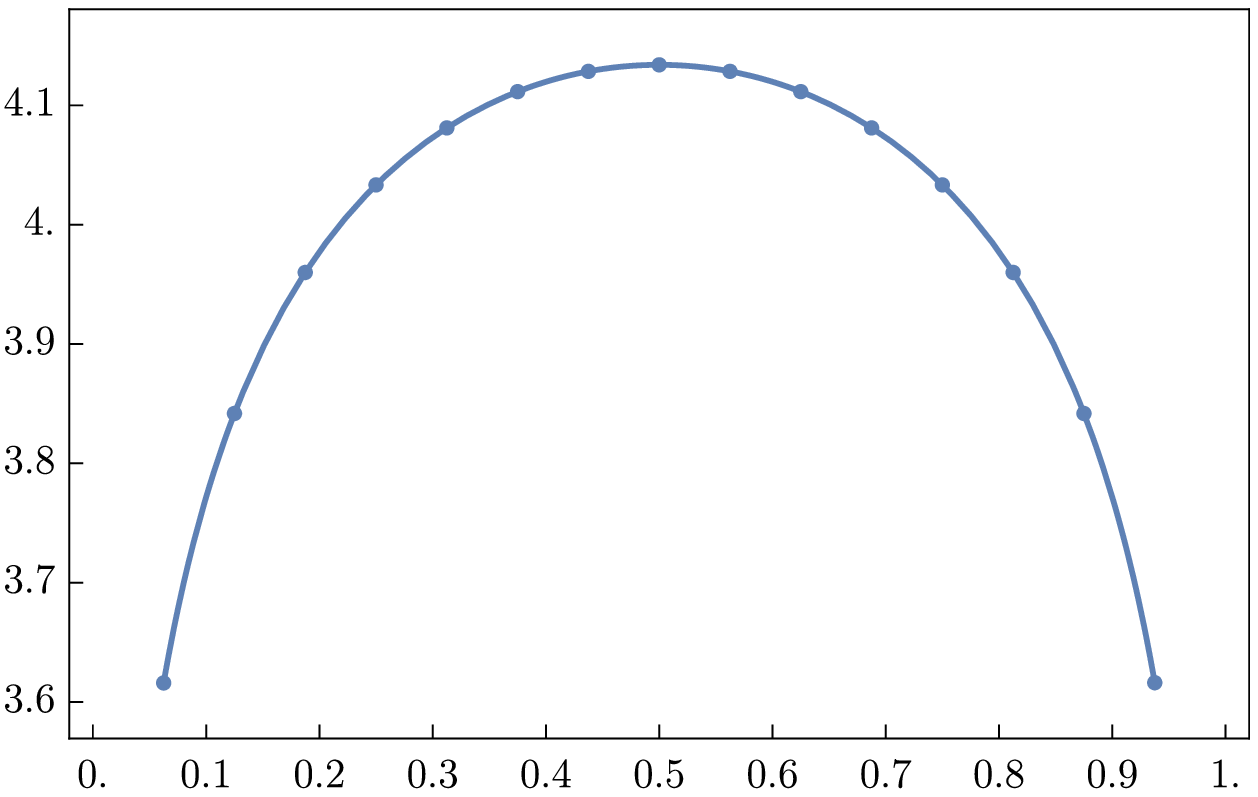}}
\rput(10.6,11){ $\tilde g_0 \log N +\tilde g_1(x)$}
\rput(16.4,6.8){$x$}
\rput(13.2,8.5){case 6}

\rput(4.1,3){\includegraphics[height=4.5cm]{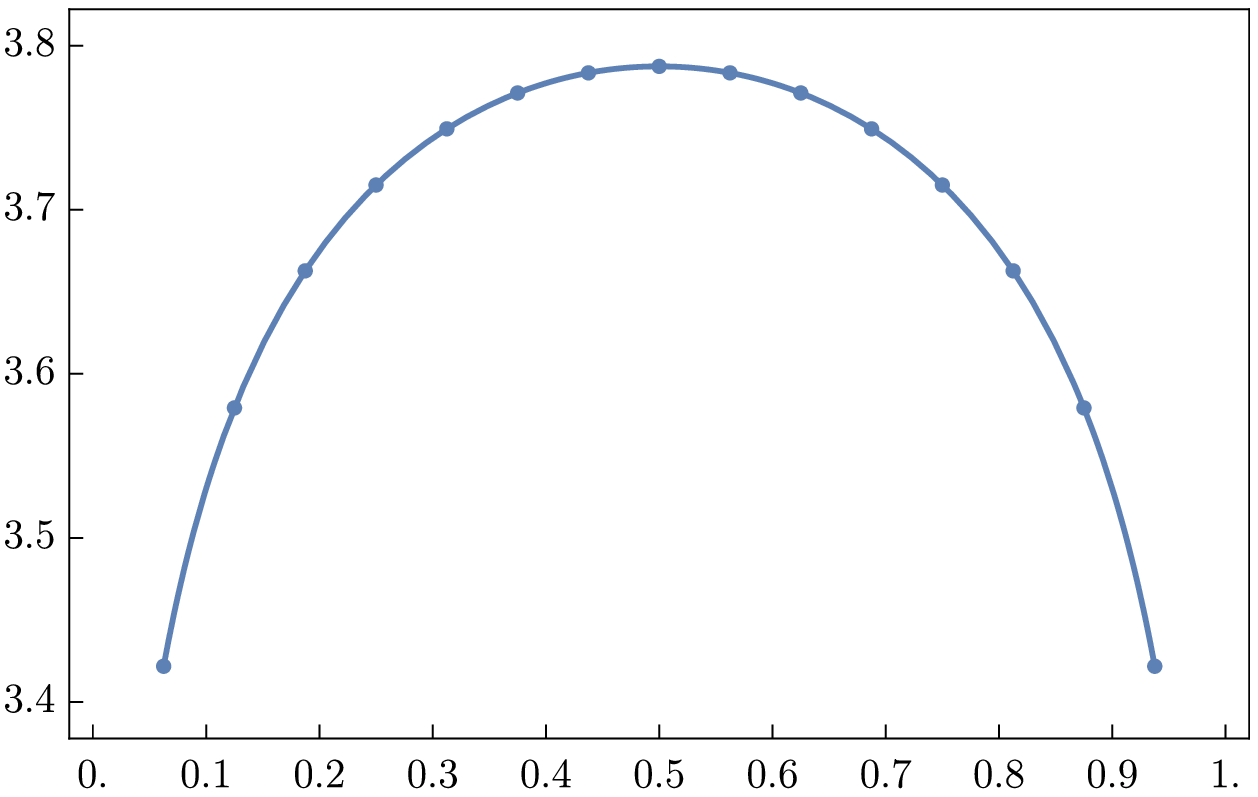}}
\rput(1.7,5.5){ $\tilde g_0 \log N +\tilde g_1(x)$}
\rput(7.5,1.3){$x$}
\rput(4.2,3){case 8}

\rput(13,3){\includegraphics[height=4.5cm]{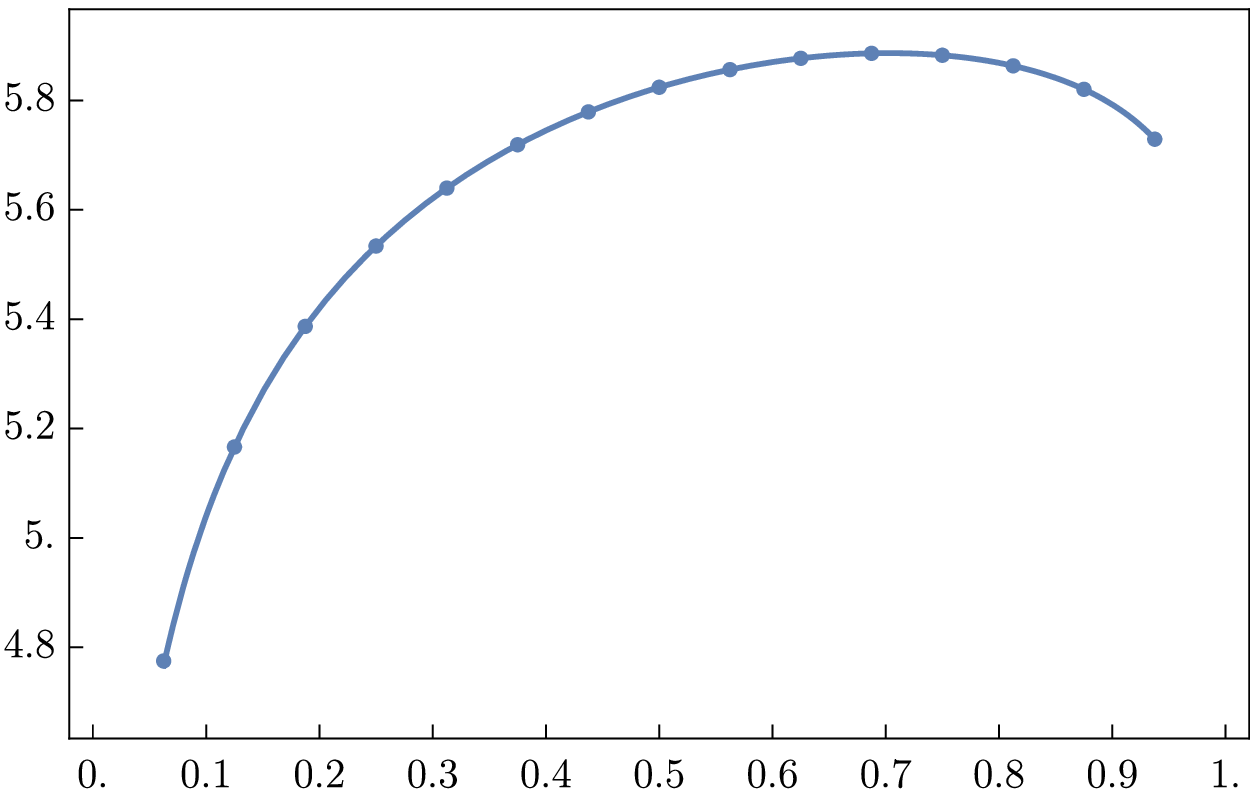}} 
\rput(10.6,5.5){ $\tilde g_0 \log N +\tilde g_1(x)$}
\rput(16.4,1.3){$x$}
\rput(13.2,3){case 12} 
\end{pspicture}
\end{center}
\caption{The function $\tilde g_0\log N +\tilde g_1(x)$ as a function of $x$ for $N=800$ and $\phi=2$ for the cases 1, 6, 8 and 12. 
For the data points, $\tilde g_0$ and $\tilde g_1(x)$
are obtained by fitting \eqref{eq:F.skirt.fit}, whereas the solid curve is the CFT prediction 
for these two functions with the data of \cref{table:ConfDataCasesSkirt}.
}
\label{fig:FSkirt}
\end{figure}

From this analysis, we find that the asymptotic behaviour of $\mathcal{F}_s^{\tXX}$ is correctly predicted by the CFT formula \eqref{eq:Fs.expansion} for case (ii), with the functions $\tilde g_{0}$, $\tilde g_{1}(x)$ and $\tilde g_2(x)$ given in \eqref{eq:g0Skirt}, \eqref{eq:g1tilde.skirt.6pts} and \eqref{eq:g2tilde.skirt.version.2}. These are specialised to the value $c=1$ of the central charge and to values of the conformal dimensions $\Delta$ and of the constant $\tilde{C}'$
that depend on the case considered. These values are given in \cref{table:ConfDataCasesSkirt}. In \cref{fig:FSkirt}, we plot the numerical data and the curve $\tilde g_0 \log N+ \tilde g_1(x)$, for the cases 1, 6, 8 and 12. We find a perfect agreement.

We observe that the constant $\tilde C'$ is independent of $\phi$. In fact, one should note the presence of a factor of $8\Delta_5 \log 2$ in \eqref{eq:g1tilde.skirt.6pts} which could a priori be included in the constant $C'$. In general, $C'$ is a constant with respect to $x$, but can depend on the dimensions of the fields. In \cref{sec:constant.skirt}, we argue that the overall additive term in \eqref{eq:g1tilde.skirt.6pts} should indeed include a factor $8\Delta_5 \log 2$, which we have chosen to write separately from $\tilde C'$. With this choice, we find that the values of $\tilde C'$ depend only on the fields inserted on the endpoints of the slits, corresponding to the states $s_1$ and $s_2$ in \eqref{eq:four.state}. 

The conformal dimensions in \cref{table:ConfDataCasesSkirt} are consistent with known results for the CFT description of the XX spin chain and the six-vertex model. The field $\varphi$ inserted at $z=z_5$ accounts for the presence of the twist~$\phi$ in the model. As discussed in \cref{sec:asy.pants.XX}, this electric operator has the dimensions $\Delta = \bar\Delta = \Delta^{\tXX}_\phi$. In the current setting, the twist line connects the point $z_5$ with the endpoint of one of the two slits, which corresponds to the boundary of the domain. In the six-vertex model, the point on the boundary where this twist line is chosen to terminate can be chosen arbitrary, and the resulting partition function is independent of this choice. In the conformal interpretation, the method of images tells us that we can replace the field $\varphi(z_5,\bar z_5)$ by the product $\varphi(z_5)\varphi(\bar z_5)$ of two chiral fields. These two fields have charges $\alphaC_5$ and $-\alphaC_5$, respectively, and the conformal dimensions are $\Delta^{\tXX}_\phi = \bar \Delta^{\tXX}_\phi = \alphaC_5^2/2$. This is precisely the assumption we make in \cref{app:CFT.derivations} to derive the CFT prediction for the asymptotic expansion of~$\mathcal F_s$.

The other fields in the positions $z_i$ with $i = 1, \dots, 4$ are so-called {\it magnetic operators}. These are primary fields that account for the presence of spin states of fixed magnetisation $m$. Here they live on the boundary and thus depend on a single variable. Their conformal dimension $\Delta_m$ depends only on the magnetisation~$m$ of the state inserted at $z=z_i$. This dimension can be computed from the finite-size correction of the groundstate of the Hamiltonian $H_f$ with free boundary conditions, defined in \eqref{eq:Hf}. The scaling limit of this Hamiltonian was for instance studied in \cite{MDRR16}. In this case, from the correction term proportional to $1/N$, one obtains the difference $c-24 \Delta_m$ as
\be
c - 24 \Delta_m = 1 - 12 m^2.
\ee
Setting $c=1$ and solving for $\Delta_m$ yields
\be
\label{eq:Delta.m}
\Delta_m = \frac {m^2}2.
\ee
The values of $\Delta_1$, $\Delta_2$, $\Delta_3$ and $\Delta_4$ in \cref{table:ConfDataCasesSkirt} are precisely given by \eqref{eq:Delta.m} with the corresponding values for $m \in \{0, \pm\frac12, \pm1, \pm \frac32, \pm 2\}$.

\begin{figure}[!t]
\begin{center}
\begin{pspicture}(0,0.7)(17,11)
\rput(4,8.5){\includegraphics[height=4.5cm]{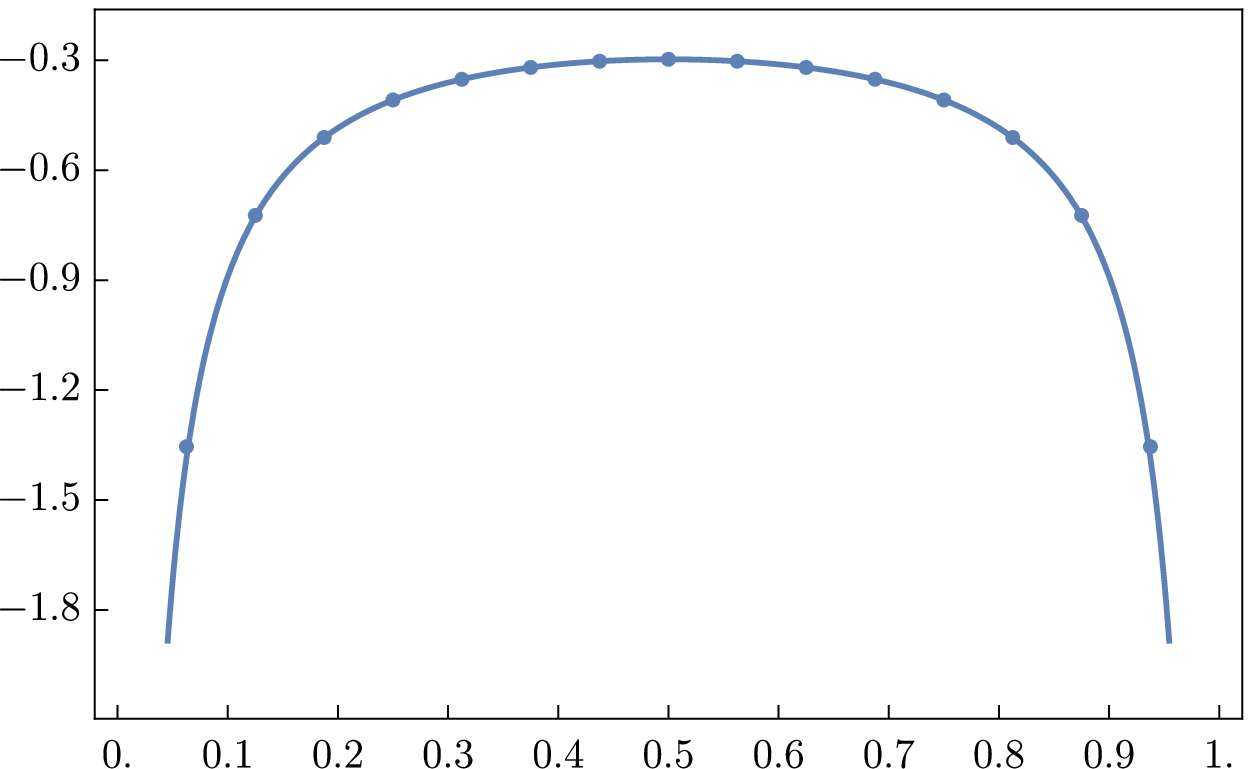}}
\rput(1.5,10.4){$\tilde g_2(x)$}
\rput(6.9,6.8){$x$}
\rput(4.2,8.5){case 2}

\rput(13,8.5){\includegraphics[height=4.5cm]{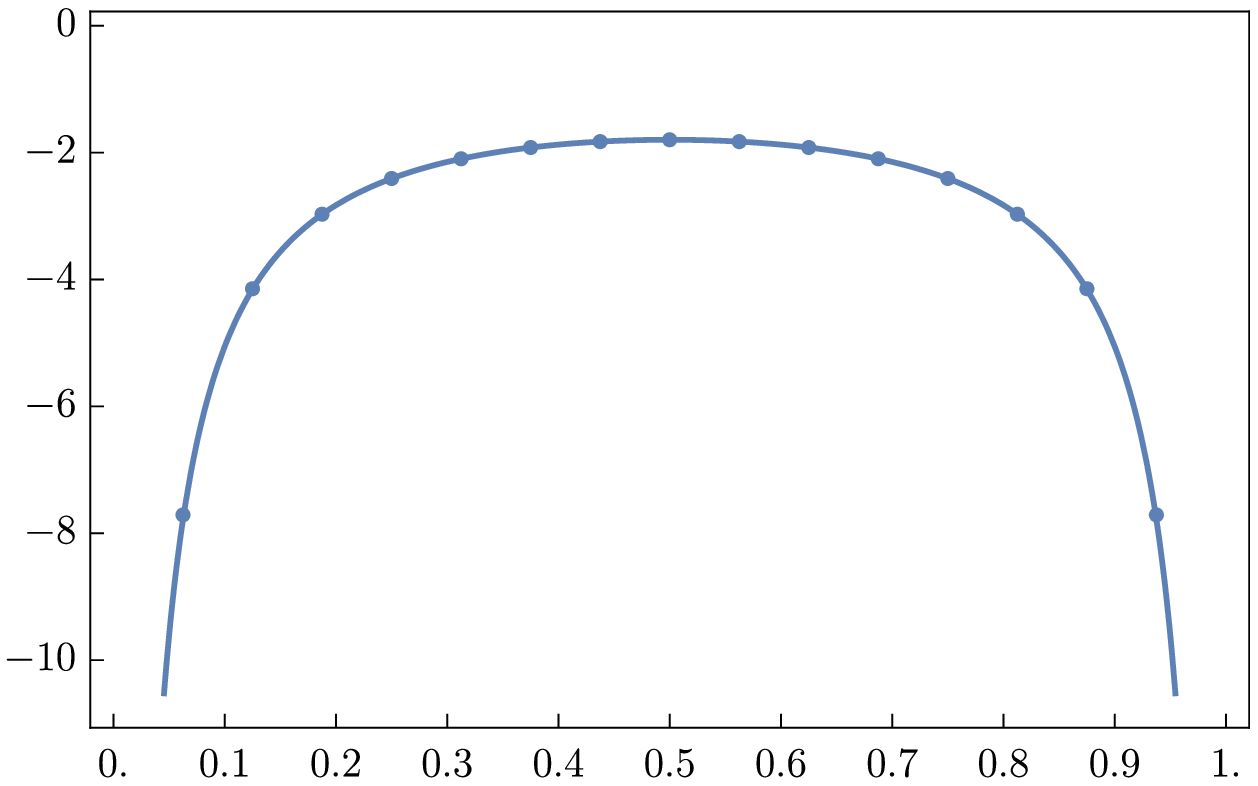}}
\rput(10.5,10.4){$\tilde g_2(x)$}
\rput(15.9,6.8){$x$}
\rput(13.2,8.5){case 3}

\rput(3.95,3){\includegraphics[height=4.5cm]{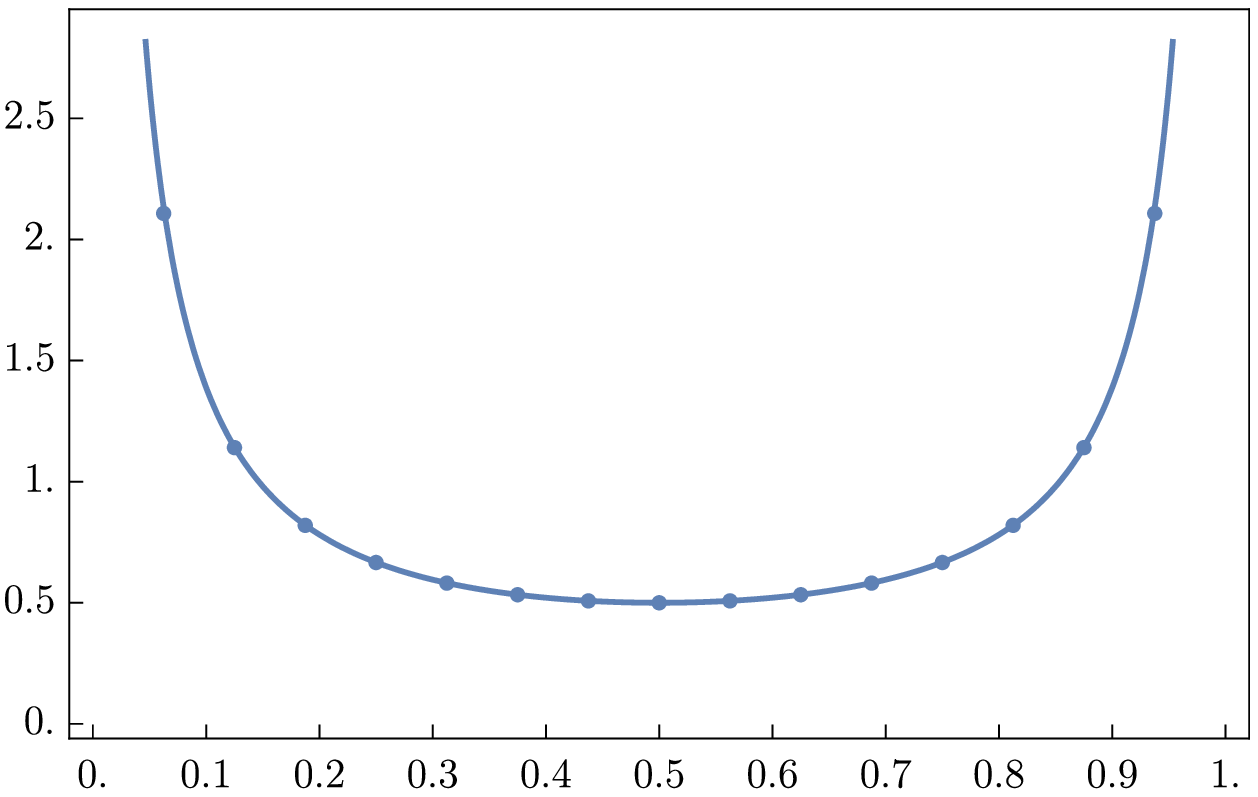}}
\rput(2.1,4.9){$\tilde g_2(x)$}
\rput(7.1,1.3){$x$}
\rput(4.2,4){case 10}

\rput(13.1,3){\includegraphics[height=4.5cm]{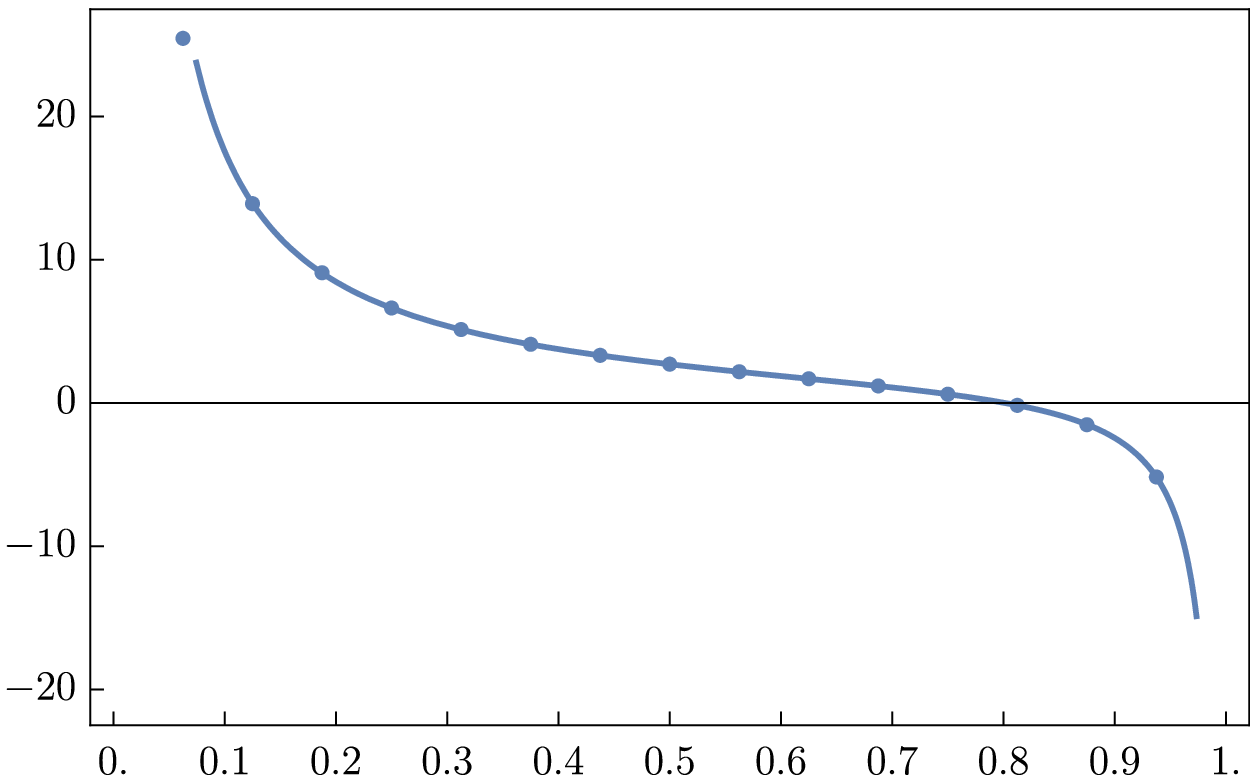}}
\rput(11.7,4.9){$\tilde g_2(x)$}
\rput(16.3,1.3){$x$}
\rput(13.2,4){case 13}
\end{pspicture}
\end{center}
\caption{The function $\tilde g_2(x)$ for the XX spin chain for $\phi=2$ for the cases $2$, $3$, $10$ and $13$. The data points are obtained by fitting \eqref{eq:F.skirt.fit}. The blue lines are the CFT predictions with the conformal dimensions and extrapolation lengths fixed as in \cref{table:ConfDataCasesSkirt}.}
\label{fig:FSkirtNLogN}
\end{figure}

We plot our numerical data for $\tilde g_2(x)$ in \cref{fig:FSkirtNLogN} for the cases $2$, $3$, $10$ and $13$. This data appears alongside the CFT prediction for $\tilde g_2(x)$, with the suitably chosen values of the extrapolation lengths $\Xi_2$ and $\Xi_4$, given in \cref{table:ConfDataCasesSkirt}.

The original derivation of the $N^{-1}\log N$ term in \cite{SD13} was claimed to be valid for domains with an arbitrary number of corners of interior angle $2\pi$. It is based on a perturbative calculation. By varying the position of a boundary near a corner by a distance $\Xi$, one obtains the variation of the free energy as a function of the perturbation, and then integrates it to obtain $\tilde g_2(x)$. On the skirt geometry, there are two corners, each of internal angle $2\pi$, and in this case, one can perform two separate perturbations near those corners. One can then assign an extrapolation length to each of the two slits. In \cref{app:CFT.derivations}, we repeat the derivation of \cite{SD13} while allowing for two such lengths, one for each slit. Because they are assigned to the corners situated at $w=w_2$ and $w=w_4$, we name them $\Xi_2$ and $\Xi_4$. We obtain the more general expression \eqref{eq:g2tilde.skirt.version.2}. 
This generalised result is necessary to correctly reproduce the data for the cases 9 to 13, for which $\Xi_2 \neq \Xi_4$. In the general case, our numerics reveal that the extrapolation lengths are given by the formula 
\be
\Xi_2=1-n_1,\qquad \Xi_4=1-n_2,
\ee
where $n_i$, with $i=1,2$, is the length of $s_i$ defined in \cref{sec:BFSkirt}.

\subsubsection{Critical dense polymers}

We now study the large-$N$ expansion for the bipartite fidelity for the model of dense polymers on the skirt lattice. In \cref{sec:polymers.skirt.BF}, we obtained product expressions for $\mathcal{F}_s^{\alpha}$ for two specialisations: (i)~$x=1/2$ with arbitrary $\phi$, and (ii)~$\phi=0$ with arbitrary $x$. We extract the asymptotic behaviour in these two cases using arguments that are very similar to those presented in \cref{app:Asympt} for the case of the periodic pants lattice. We obtain
\begin{subequations}
\label{eq:skirt.cdp.special.cases}
\begin{alignat}{2}
 \mathcal{F}_s^{\alpha}\big|_{x=\frac 12} = &-\frac 12\log N+\frac{\log 2}{3} +4 \Delta^{\tCDP}_{\phi} \log 2 + \frac{1}{2}(-1+12 \log A - \log \pi +4 \log 2) + \mathcal{O}(N^{-1}),
\\[0.2cm]
 \mathcal{F}_s^{\alpha}\big|_{\phi=0} = &-\frac 12\log N-\frac 1{12}(\tilde G(x)+\tilde G(1-x)) -\frac 12 (x \log x +(1-x)\log(1-x) + 2 \log 2)\nonumber\\[0.1cm]
 & + \frac{1}{2}(-1+12 \log A - \log \pi +4 \log 2) + \mathcal{O}(N^{-1}).
\end{alignat}
 \end{subequations}
We recall that $\tilde G(x)$ and $\Delta^{\tCDP}_{\phi}$ are defined in \eqref{eq:tildeGx} and \eqref{eq:Delta.phi.CDP}, respectively. 

\begin{figure}
\begin{pspicture}(0,0)(17,5)
\rput(4,2.5){\includegraphics[height=4.5cm]{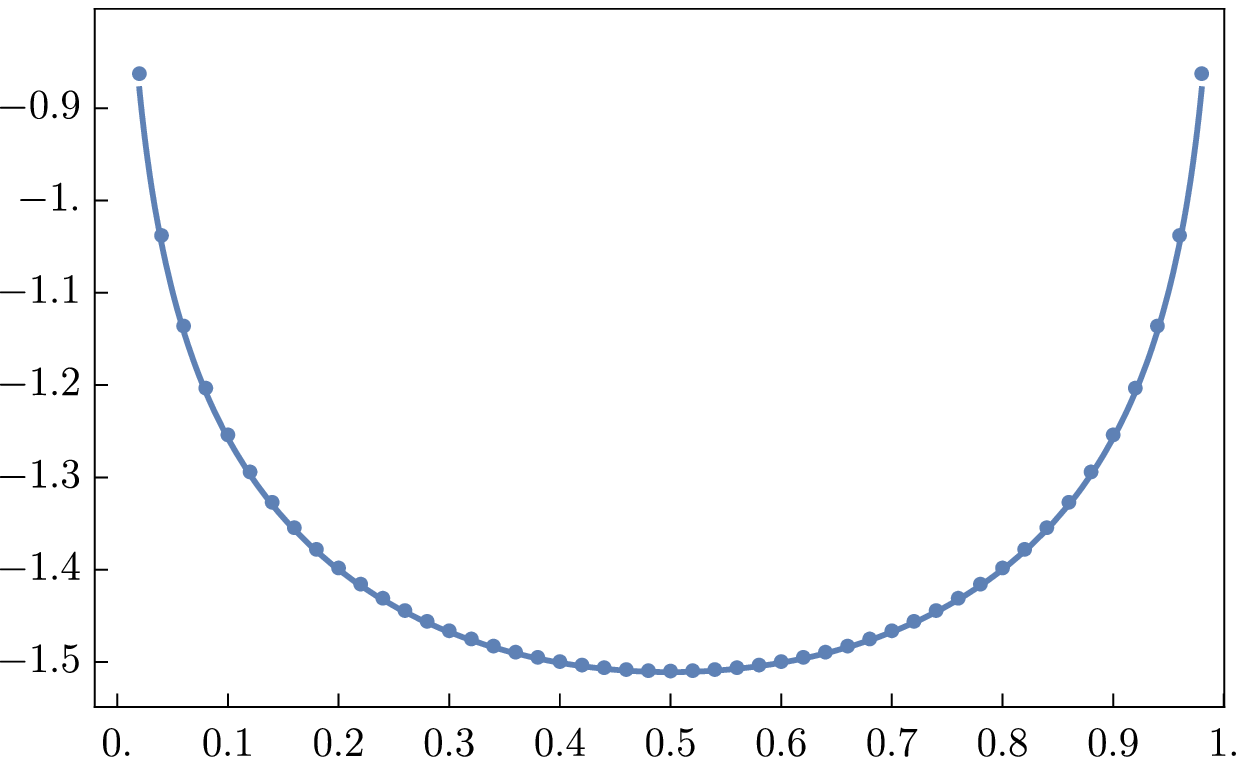}}
\rput(1.7,4.3){$\mathcal{F}_s^{\alpha}$}
\rput(7.2,0.8){$x$}
\rput(13,2.5){\includegraphics[height=4.5cm]{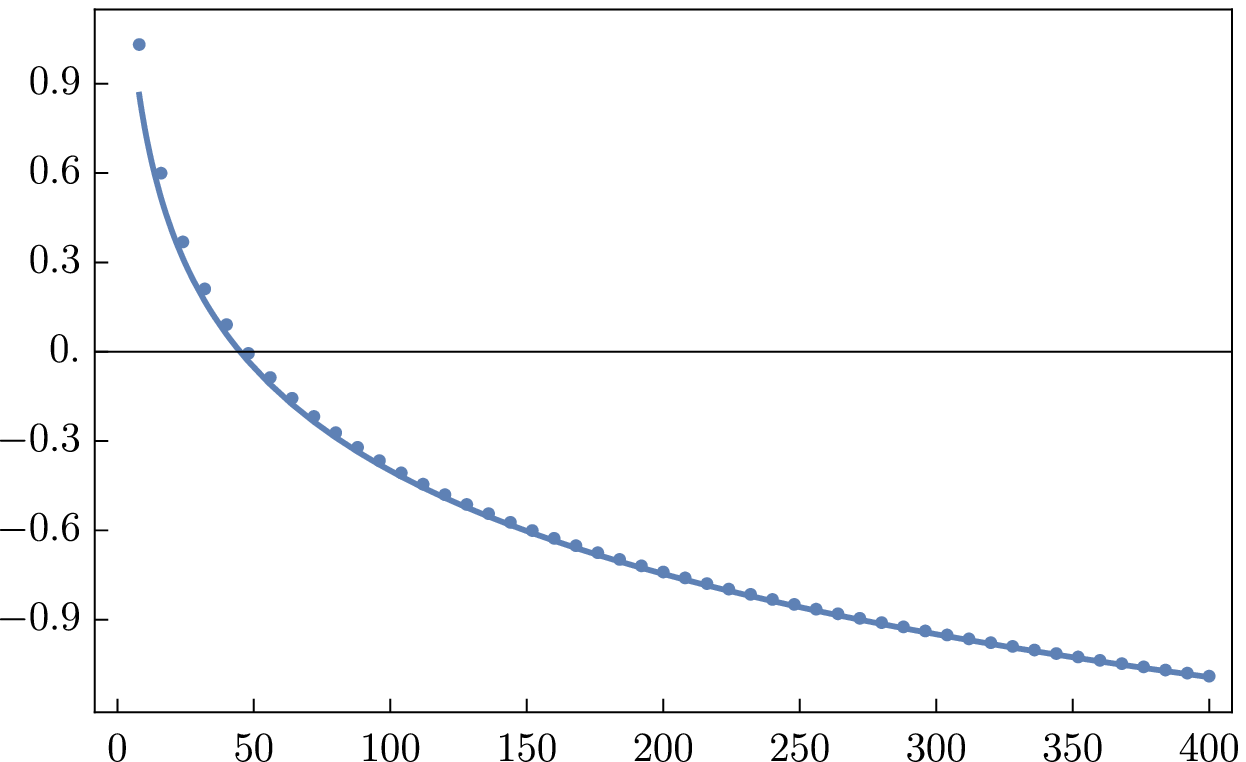}}
\rput(10.7,4.3){$\mathcal{F}_s^{\alpha}$}
\rput(16.2,1.1){$N$}
\end{pspicture}
\caption{The bipartite fidelity $\mathcal{F}_s^{\alpha} $ as a function of $x$ for $N=800$ (left panel) and as a function of $N$ for $x=\frac 14$ (right panel). In both cases, the fugacity is set to $\alpha = 2 \cos (\frac\phi2)$ with $\phi = 2$.
The data points are computed from the determinant \eqref{eq:Fclosed}, whereas the continuous line is \eqref{eq:FSkirtPolygen}.}
\label{fig:FskirtCDP}
\end{figure}

We performed numerical evaluations of the determinant expression \eqref{eq:Fclosed} for values of $x$ and $\phi$ that do not enter the specialisations (i) and (ii) defined in \cref{sec:polymers.skirt.BF}. In \cref{fig:FskirtCDP}, we plot the results for the value $\phi = 2$. From the exact results \eqref{eq:skirt.cdp.special.cases} and our numerical experiments, we conjecture that the general formula is
\begin{equation}
\label{eq:FSkirtPolygen}
 \begin{split}
 \mathcal{F}_s^{\alpha} = &-\frac 12\log N-\frac 1{12}(\tilde G(x)+\tilde G(1-x)) +4 \Delta^{\tCDP}_{\phi} \big(x \log x +(1-x)\log(1-x) + 2 \log 2\big)\\
 & + \frac{1}{2}(-1+12 \log A - \log \pi +4 \log 2) + \mathcal{O}(N^{-1}).
\end{split} 
 \end{equation} 
As illustrated in the example of \cref{fig:FskirtCDP}, this conjecture reproduces the numerical results with great precision. It also coincides with the CFT prediction \eqref{eq:Fs.expansion} for case (i), with $\tilde g_0$ given in \eqref{eq:g0Skirt},
$\tilde g_1(x)$ in \eqref{eq:g1tilde.skirt.4pts}, $\tilde g_2(x)$ in \eqref{eq:g2tilde.skirt.3pts}, the conformal data specified to
\begin{equation}
\label{eq:dimensions.CDP skirt}
c=-2, \qquad 
\Delta_5 = \Delta^\tCDP_{\phi},
\end{equation}
the non-universal constant taking the value
\be
\tilde C' = \frac{1}{2}(-1+12 \log A - \log \pi +4 \log 2),
\ee
and with vanishing extrapolation lengths: $\Xi_2=\Xi_4=0$. Indeed, since the conditions at the two slits' endpoints are identical, we must have $\Xi_2=\Xi_4$. From the exact results \eqref{eq:skirt.cdp.special.cases} and the numerical evaluations, we find that the term proportional to $N^{-1}\log N$ vanishes, namely $\tilde g_2(x)=0$.  A glance at \eqref{eq:g2tilde.skirt.version.2} indicates that this is only possible if both extrapolation lengths vanish.

The values \eqref{eq:dimensions.CDP skirt} are precisely those expected in the conformal description of the model of dense polymers. Indeed, there is no change of boundary condition at the endpoints of the slits. Moreover, the boundary conditions in the legs $A$ and $B$ are known to correspond to the groundstate of the model. Both of these features correspond to identity fields with the dimension $\Delta = 0$. Similarly to what is discussed in \cref{sec:asy.pants.CDP}, the field $\rho(z)$ at $z=z_5$ assigns weights $\alpha \neq 0$ to the non-contractible loops and has the conformal dimension $\Delta^\tCDP_{\phi}$. 

\section{Discussion and conclusion}\label{sec:conclusion}

In this paper, we investigated the bipartite fidelity for critical lattice models on two geometries: the periodic pants domain and the skirt domain. Using arguments of conformal field theory, we obtained predictions for the leading terms in the asymptotic large-$N$ expansions of $\mathcal F$. We compared these with exact and numerical lattice calculations and found a precise match for two lattice models: the XX spin chain and the model of critical dense polymers. These models are known to be described by CFTs with central charges $c=1$ and $c=-2$, respectively.

For the XX spin chain, we considered different instances of the bipartite fidelity with certain spin states of magnetisation $m$ fixed in the domain, or in the presence of twist lines $\phi$ connecting certain points of the domain. In the CFT, these correspond to the insertions of magnetic and electric operators, respectively. The conformal dimensions of these operators are known from the CFT description of the XXZ spin chain in terms of a compact boson \cite{K78,N84,dFHZ87}. The electric operators are also known to be vertex operators whose $n$-point functions are given by the simple formula \eqref{eq:npoint.vertex}. Our analysis of $\mathcal F$ for the XX chain on the periodic pants domain confirms this. The charges of the electric operator are $\alphaC = \bar\alphaC = \frac{\phi}{2\pi}$, and the conformal dimensions are $\Delta = \bar \Delta = \frac{\alphaC^2}2$. If a field marks a transition between two twist lines $\phi$ and~$\phi_A$, then its charge is $\alphaC = \frac{\phi-\phi_A}{2\pi}$. We also investigated the case of an intersection point between three twist lines of twists $\phi$, $\phi_A$ and $\phi_B$, and found that the corresponding charge is $\alphaC = \frac{\phi-\phi_A-\phi_B}{2\pi}$.
%

Our analysis of $\mathcal F$ on the skirt domain goes further, by considering six-point correlators that mix electric and magnetic operators. Our numerical analysis reveals that these mixed correlators also take the simple form \eqref{eq:npoint.vertex} for vertex operators, with the charge of the magnetic operators given by $\alphaC = m$. This is the expected result for a free CFT \cite{YellowCFTbook}, however it is interesting to recover this property from the scaling limit of lattice calculations. In the general case of an interacting CFT, one expects that mixed correlators will involve non-trivial functions of the cross-ratios, sums over conformal blocks, etc, see for instance \cite{EI15}. To push our understanding further, we hope to return to the problem of computing $\mathcal F$ for the XXZ spin chain with anisotropy $\Delta \neq 0$. In that case the factorisation should still occur for $-1 \leq \Delta \leq 1$.

For the model of critical dense polymers, we studied the model with non-contractible loops that are assigned a fugacity $\alpha = 2 \cos (\frac\phi2)$, different from the fugacity $\beta = 0$ of the contractible loops. In the conformal field theory description, this amounts to inserting a bulk operator with conformal dimension $\Delta^{\tCDP}_\phi = \frac{1}{8}\big(\frac{\phi^2}{\pi^2}-1\big)$. Our investigation uses the known map from the loop model to the spin chain, and only covers the case where at most two such operators are inserted in the domain. It is then natural to search for an extension of these results to the case where loops are assigned different weights according to how they encircle the various marked points. However, this appears not to be feasible using the XX representation of the periodic Temperley-Lieb algebra. This hints at the fact that the corresponding conformal fields are not vertex operators.

A most interesting result that we found regards the extrapolation lengths for the terms proportional to $N^{-1}\log N$. In our investigation of the XX spin chain on the skirt domain, we studied special instances of $\mathcal F_s$ where certain spin states $s_1$ and $s_2$ are inserted on the endpoints of the two slits, see \eqref{eq:four.state}. As a result, we found that the CFT prediction of Dubail and St\'ephan \cite{DS11,SD13} had to be generalised to allow for two extrapolation lengths $\Xi_2$ and $\Xi_4$, one for each slit. Moreover, contrary to what these authors claimed, these lengths sometimes take negative values. In fact, for the XX chain on the skirt, we found a linear relation between the extrapolation lengths and the lengths of the inserted states, with the negative values arising when $s_1$ and $s_2$ are non-trivial states. We speculate that this is a consequence of the unusual finite shape of the corner in the corresponding two-dimensional lattice model. Indeed, in the underlying six-vertex model, the endpoints of the slits take the form of two subsequent corners, distant by $n_i$ lattice spacings. In the scaling limit, this shape becomes a corner of internal angle $2 \pi$, and the extrapolation length is modified accordingly. 

On the periodic pants domain, the asymptotic expansion of the bipartite fidelity also has an $N^{-1}\log N$ term. This is an important new result of this paper: for the bipartite fidelity, conical singularities with internal angles of $4\pi$ act similarly to corners with internal angles $2\pi$. Both these features result in $N^{-1}\log N$ contributions to $\mathcal F$. On the pants domain, there is a single conical singularity and therefore a unique extrapolation length. For the XX spin chain, we find that the extrapolation length is $\Xi = -n$, where $n$ is the length of the spin state on the crotch point. Moreover, we remark that the conformal predictions for $g_0(x)$, $g_1(x)$ and $g_2(x)$ are precisely equal to twice the same functions, previously obtained by Dubail and St\'ephan, for the asymptotic expansion of $\mathcal F$ on the flat pants geometry. This can be traced back to the fact that the map $w(z)$ for the flat pants domain is equal to two times $w_p(z)$, with $z$ restricted to the upper half-plane.

While this has not been fully exploited in this paper, we believe that the bipartite fidelity is a useful tool to compute the structure constants $\mathcal C_{123}$ that arise in the three-point functions $\langle\phi_1(z_1,\bar z_1)\phi_2(z_2,\bar z_2)\phi_3(z_3,\bar z_3)\rangle_{\mathbb C}$. Computing a correlation function of a lattice model for a number of arbitrarily located positions in the complex plane is in general a notoriously difficult task, even for two-point functions. In this case, a simple way to measure the conformal dimension $\Delta$ that appears in a two-point correlator is to consider the problem defined on an infinite cylinder and insert the two fields at infinity at the endpoints of this cylinder. The dimension $\Delta$ then appears in the $1/N$ finite-size correction term of the largest eigenvalue of the transfer matrix, and computing it is possible with the usual methods of Yang-Baxter integrability. Similarly for the three-point functions, the pants and skirt domains are useful as they have three points at infinity where the three fields can be inserted. (No field should then be inserted on the crotch point or on the endpoints of the slits.) This idea has the potential to make the lattice computation of $\mathcal C_{123}$ a manageable problem. In the present paper, the only case where we hoped to get a glimpse of the structure constants was in our investigation of $\mathcal F^{\tXX}_p$. Indeed, in \cref{sec:constant.pants}, we set $\alphaC_2 = 0$, calculate the difference $\mathcal F-\mathcal F'$ of two bipartite fidelities for two different sets of fields, and find that the difference of constants $C$ in \eqref{eq:g1.pants.3pts} is simply the logarithm of the ratio of the two structure constants, see \eqref{eq:CC.ratio}. For $\mathcal F^{\tXX}_p$, this analysis applies for the twist $\phi$ specified to $\phi = \phi_A + \phi_B$, so that $\alphaC_2=0$. In \cref{sec:asy.pants.XX}, we however found that varying $\phi$, $\phi_A$ and $\phi_B$ while keeping $\phi - \phi_A - \phi_B$ fixed does not change $C$. This constant instead depends only on $\alphaC_2$, see \cref{fig:Cprime}. We conclude that the structure constant in this case has no non-trivial dependence on the three fields inserted at infinity. This is consistent with the interpretations of these fields as vertex operators, where the structure constant are known to be equal to one.
 
\subsection*{Acknowledgments}

AMD and GP acknowledge the support from the Fonds de la Recherche Scientifique -- FNRS: AMD was supported by the Postdoctoral Research Project CR28075116 and GP was supported by the Aspirant Fellowship FC23367. The authors also acknowledge support from the EOS contract O013018F. The authors thank Y.~Ikhlef, J.L.~Jacobsen and an anonymous referee for useful comments.

%
%

\bigskip
\bigskip
\appendix
%

\section{The bipartite fidelity from conformal field theory calculations}\label{app:CFT.derivations}

In this appendix, we derive the CFT prediction for the leading terms in the $\frac1N$ expansion of the bipartite fidelity, for the periodic pants and the skirt geometry. It closely follows the arguments of St\'ephan and Dubail \cite{DS11,SD13}. 

The outline of this appendix is as follows.  In \cref{sec:perturbation}, we derive a general formula for the constant term of the large-$N$ expansion of $\mathcal{F}$. We find $g_1(x)$ and $\tilde g_1(x)$, namely the constant contributions for the periodic pants and the skirt geometry, in \cref{sec:ppg.contribution,sec:skirt.contribution}, respectively. We evaluate the constants $C$ and $\tilde C$ in \cref{sec:constant.pants,sec:constant.skirt}. Finally, we compute the sub-leading contributions, $g_2(x)$ and $\tilde{g}_2(x)$ in \cref{sec:subleading.pants,sec:subleading.skirt}.

\subsection{Perturbation of the stress-energy tensor}\label{sec:perturbation}

To compute the constant term in the $1/N $ expansion of $\mathcal F$, we follow the strategy of  \cite{DS11,SD13} that consists in varying the aspect ratio $x$ to find $\delta \mathcal F/\delta x$. The general formula derived below applies to both the skirt and the pants geometry and will allow us to compute $g_1(x)$ and $\tilde g_1(x)$. For this reason, we temporarily drop the subscripts of the maps $w$ and the free energies $f$.

We recall the relation between the action and the stress-energy tensor. Under a transformation $w_\mu\mapsto w_\mu + \varepsilon_\mu$ whose support is $A$, the action varies according to
\begin{equation}
\delta S= \frac{1}{2 \pi} \int_A \partial_\mu \varepsilon_\nu T^{\mu \nu} \textrm{d}w_1\textrm{d}w_2.
\end{equation}
The free energy $f$ is defined as minus the logarithm of the partition function: $f=-\log Z$ with $Z= \langle \phi_1 \cdots \phi_n \rangle$. Here, $\phi_i$ is a short-hand notation for $\phi_i(w_{i,1},w_{i,2})$.
Under the perturbation $w_\mu\mapsto w_\mu + \varepsilon_\mu$, the free energy varies as 
\begin{equation}
\label{eqn:deltafdEpsilon}
\delta f= \frac{1}{2 \pi} \int_A \partial_\mu \varepsilon_\nu \frac{\langle T^{\mu \nu} \phi_1 \cdots \phi_n\rangle}{\langle \phi_1 \cdots \phi_n\rangle  } \textrm{d}w_1\textrm{d}w_2.
\end{equation}

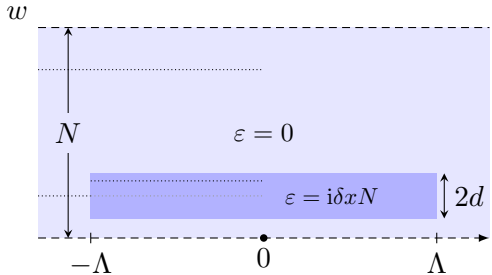
\begin{figure}
\centering
\begin{tikzpicture}
\def\a{3} 
\def\L{2.8}
\def\Lbda{2.3}
\def\offset{0.2}
\def\x{0.4}
\def\deltax{0.2}
\def\d{0.3}
\newcommand{\mybullet}[3]{\fill #1 circle (.05) node#3 {#2}}
\fill[blue!10] (-\a,0) rectangle (\a,\L);
\fill[blue!30] (-\Lbda,\L*\x/2-\d) rectangle (\Lbda,\L*\x/2+\d);
\draw (-\a,\a) node[left]{$w$};
\mybullet{(0,0)}{$0$}{[below]};
\draw[->, >=latex, densely dashed] (-\a,0) to (\a,0);
\draw[densely dashed] (-\a,\L) -- (\a,\L);
\draw[densely dotted] (-\a,{\L*\x/2}) -- (-\Lbda,{\L*\x/2});
\draw[densely dotted, gray!80] (-\Lbda,{\L*\x/2}) -- (0,{\L*\x/2});
\draw[densely dotted] (-\Lbda,{\L*\x/2+\deltax}) -- (0,{\L*\x/2+\deltax});
\draw[densely dotted] (-\a,{\L-\L*\x/2}) -- (0,{\L-\L*\x/2});
\draw[<->,>=stealth] ({-\a+2*\offset},0) -- ({-\a+2*\offset},\L )node[midway, fill=blue!10] {$N$};
\draw (\Lbda, -2pt) node[below] {$\Lambda$} -- (\Lbda, 2 pt);
\draw (-\Lbda, -2pt ) node[below] {$-\Lambda$} -- (-\Lbda, 2 pt);
\draw (5 pt, {\L*\x/2}) node[right, scale=0.8] {$\varepsilon=\ir \delta xN$};
\draw (0, \L/2) node[scale=0.9]{$\varepsilon=0$};
\draw[<->,>=stealth] (\Lbda+0.1,\L*\x/2-\d) --  (\Lbda+0.1 ,\L*\x/2+ \d) node[midway, right] {$2d$};
\end{tikzpicture}
\caption{The perturbation $\varepsilon$. The darkened area is the support $A$ of $\varepsilon$, and the slit is moved upward in $A$.}
\label{fig:PantsPert}
\end{figure}

On the periodic pants and the skirt domain, we perform the perturbation $w=w_1 + \ir w_2 \mapsto w+\varepsilon(w)$ given by 
\begin{equation}\varepsilon(w)=
\begin{cases}
\ir \delta x N &\text{if } |w_2-\tfrac{xN}{2}|<d \text{ and } |w_1|< \Lambda,\\
0 & \text{elsewhere.}
\end{cases}
\end{equation}
The support $A$ of $\varepsilon$ consists of the rectangle where the perturbation takes the value $\ir \delta x N$. It has a width $2d$ and a length $2\Lambda$.  We will take the limit where $d$ tends to zero. In contrast, $\Lambda$ will play the role of a cut-off for certain integrals and will be sent to infinity at the end of the calculation. The effect of the perturbation $\varepsilon$ is to slightly shift the slit in position $\ir xN/2$ by a distance $\ir \delta xN$, as depicted in \cref{fig:PantsPert}. This is equivalent to changing $x$ to $x+\delta x$ while keeping $N$ unchanged.

The derivative of the perturbation yields linear integrals on the boundaries of the support $A$. Because $\partial_1 \varepsilon_{1}=\partial_1\varepsilon_2=0$, the only contributions to the integral \eqref{eqn:deltafdEpsilon} come from $\partial_1\varepsilon_2$ on the width of~$A$, and from $\partial_2\varepsilon_2$ on its length. We take the limit $d\to 0$ in which case only the latter is non-zero:
\begin{equation}
\label{eqn:T22}
\delta  f=\frac{1}{2\pi} \delta x N
\left(  \int_{\mathcal{U}^+}   \frac{\langle T^{22} \phi_1 \cdots \phi_n \rangle}{\langle \phi_1 \cdots \phi_n \rangle}   \textrm{d}w_1 - \int_{\mathcal{U}^-}   \frac{\langle T^{22} \phi_1 \cdots \phi_n \rangle}{\langle \phi_1 \cdots \phi_n \rangle}   \textrm{d}w_1 \right).
\end{equation}
Here $\mathcal U^\pm=[-\Lambda,0] + \ir \tfrac{xN}{2} - \ir 0^\pm$ are segments on the boundary of $A$, on each side of the slit. The integrals along $[0,\Lambda]+ \ir \tfrac{xN}{2} \pm d$ cancel each other in the limit $d\to 0$. 

We apply the change of variable $w=w_1+\ir w_2$, $\bar w=w_1-\ir w_2$ and rewrite the integrand in terms of the non-vanishing components of the stress-energy tensor $T(w)$, $\bar T(\bar w)$ using $T^{22}= -(T(w)+\bar T(\bar w))$. 
We denote the fields after the transformation by $\phi_i (w_i,\bar w_i)$, $i=1,\dots,n$. From here onwards, we only consider the holomorphic part of the expression and write $+c.c.$ for the addition of its complex conjugate:
\begin{multline}
\label{eqn:Tww}
\delta  f=- \frac{1}{2 \pi} \delta x N \Big( \int_{\mathcal{U}^+}   \frac{\langle T(w) \phi_1(w_1,\bar w_1) \cdots \phi_n(w_n,\bar w_n) \rangle}{\langle \phi_1(w_1,\bar w_1) \cdots \phi_n (w_n,\bar w_n) \rangle}   \textrm{d}w \\
- \int_{\mathcal{U}^-} \frac{\langle T(w) \phi_1(w_1,\bar w_1) \cdots \phi_n(w_n,\bar w_n) \rangle}{\langle \phi_1(w_1,\bar w_1) \cdots \phi_n (w_n,\bar w_n) \rangle}   \textrm{d}w
 \Big) + c.c.
\end{multline}

\paragraph{Integral over the real line.} In order to compute these integrals, we pull them back via the inverse of the transformation $w$. As illustrated in \cref{fig:MapPants,fig:MapSkirt}, the pre-image of the integration curves are
\begin{subequations}
\begin{align}
&w^{-1}(\mathcal U^+)=[w^{-1}(-\Lambda+ \ir \tfrac{xN}{2}- \ir 0^+), w^{-1}(\ir \tfrac{xN}{2})] \equiv [1_\Lambda^-,w^{-1}(\ir \tfrac{xN}{2})], \\[0.1cm]
& w^{-1}(\mathcal U^-)=[w^{-1}(-\Lambda+ \ir \tfrac{xN}{2}+ \ir 0^+), w^{-1}(\ir \tfrac{xN}{2})] \equiv [0_\Lambda^+, w^{-1}(\ir \tfrac{xN}{2})],
\end{align}
\end{subequations}
where we introduced the short-hand notations $0^+_\Lambda$ and $1^-_\Lambda$, highlighting the fact that those points tend to $0$ and $1$ for large $\Lambda$, respectively. These values for $0^+_\Lambda$ and $1^-_\Lambda$ are different for $w_p(z)$ and $w_s(z)$, and their leading behaviour for large $\Lambda$ can be computed directly from \eqref{eq:w.pants} and \eqref{eq:w_s}.

We change the variable from $w$ to $z$ in the integrals in \eqref{eqn:Tww} and use the transformation law of the stress-energy tensor $T(w)(w'(z))^2 = T(z)-\tfrac{c}{12}\{w(z),z\}$, where $\{w(z),z\}$ is the Schwarzian derivative. We obtain the following integral:
\begin{equation}
\label{eqn:Tonsegment}
\delta f=\frac{\delta xN}{2\pi}  \int_{0^+_\Lambda }^{1^-_{\Lambda}}  \left(\frac{\dd w}{\dd z}\right)^{-1} \left[ \frac{\langle T(z) \phi_1(z_1,\bar z_1) \cdots \phi_n(z_n,\bar z_n) \rangle}{\langle \phi_1(z_1,\bar z_1) \cdots \phi_n (z_n,\bar z_n) \rangle} -\frac{c}{12} \{w(z),z\} \right]  \textrm{d} z \,+ c.c.
\end{equation}
Finally, we use the conformal Ward identity on the complex plane:
\begin{equation}
\label{eqn:WardIdentity}
\langle T(z) \phi_1(z_1,\bar z_1) \cdots \phi_n(z_n,\bar z_n) \rangle =\sum_{i=1}^n \left[ \frac{\Delta_i}{(z-z_i)^2} 
+ \frac{\partial_i}{z-z_i} \right] \langle \phi_1(z_1,\bar z_1) \cdots \phi_n (z_n,\bar z_n) \rangle
\end{equation}
to write the first part of the integrand in terms of an $n$-point correlator.

\paragraph{Free energy variation of the cylinder and the strip.} The expressions \eqref{eq:Fpdifference} and \eqref{eqn:Fsdifference} for the bipartite fidelity on the periodic pants domain and the skirt domain involve the free energy of the cylinder and rectangle strip geometry. These expressions are standard \cite{YellowCFTbook}. For a cylinder and a rectangle of length $2\Lambda$ with inserted fields of dimension $\Delta$, they read 
\be
f_{c(N)} = -(2\Lambda) \frac{\pi}{N}\Big( \frac{c}{6} - 4\Delta \Big),\qquad
f_{r(N)} = -(2\Lambda) \frac{\pi}{N}\Big( \frac{c}{24}   - \Delta \Big).
\ee
We use these expressions with $N$ replaced by $N_A=x N$, $N_B=N-N_A = (1-x)N$. On the cylinder, the variation of these free energies with respect to $x$ is 
\be
\delta f_{c (N)}=0,\qquad 
\delta f_{c(N_A)}=  (2\Lambda) \frac{\pi\delta x}{x^2 N}\Big( \frac{c}{6} - 4\Delta_1 \Big),\qquad
\delta f_{c(N_B)}= -(2\Lambda) \frac{\pi\delta x}{(1-x)^2 N}\Big( \frac{c}{6} -4 \Delta_3 \Big).
\ee
These free energies tend to infinity with $\Lambda$. However, the differences \eqref{eq:Fpdifference} and \eqref{eqn:Fsdifference} will turn out to be finite.

\subsection{Contribution to the periodic pants geometry}\label{sec:ppg.contribution}

We consider the periodic pants domain and compute the variation $\delta f_p$ from \eqref{eqn:Tonsegment}. The term involving the Schwartzian derivative in the integrand is readily computed. One finds
\be \label{eq:pants.cterm}
 \int_{0^+_\Lambda }^{1^-_{\Lambda}}  \left(\frac{\dd w_p}{\dd z}\right)^{-1} \{w_p(z),z \} \textrm{d}  z 
\,+c .c. = \frac{4 \pi^2}{N^2} \left( \frac{\Lambda}{(1-x)^2} - \frac{\Lambda}{x^2} \right) 
 +\frac{2\pi}{N} \big(b(x) -b(1-x)\big),
\ee
with $b(x)= \frac{(x-2) x}{(1-x)^2}\log x-\frac{1}{2x}$. We now compute the second term involving the fields, for the two special cases (i) and (ii) discussed in \cref{sec:CFT.pants}.

\paragraph{Case (i): Three primary fields.}
There are three primary fields $\phi_1$, $\phi_3$ and $\phi_4$ in the legs but no field at the crotch point. Following the notation of \cref{sec:CFT.pants}, we denote the insertion points by $w_1, w_3$ and $w_4$. These points correspond to $z_1=1$, $z_3=0$ and $z_4=\infty$ in the complex plane via the map $w_p(z)$.

The three-point function is given by 
\begin{equation}
\label{eq:3pt.function}
\Big\langle \prod_{i=1,3,4}\phi_i(z_i,\bar z_i)\Big\rangle_{\mathbb C} = \frac{\mathcal C_{134}} 
{|z_1 - z_3|^{2(\Delta_1 + \Delta_3 - \Delta_4)}
|z_1 - z_4|^{2(\Delta_1 + \Delta_4 - \Delta_3)}
|z_3 - z_4|^{2(\Delta_3 + \Delta_4 - \Delta_1)}},
\end{equation}
where $\mathcal C_{134}$ is the three-point structure constant.
Inserting this in \eqref{eqn:Tonsegment} and using the Ward identity \eqref{eqn:WardIdentity}, we find after some algebra
\begin{multline}
 \int_{0^+_\Lambda }^{1^-_{\Lambda}}  \left(\frac{\dd w_p}{\dd z}\right)^{-1}  \frac{\langle T(z) 
 \phi_1 \phi_3 \phi_4  \rangle}{\langle  \phi_1 \phi_3 \phi_4 \rangle}  \textrm{d}  z \, +c.c.
 =  \frac{8 \pi^2}{N^2}\left(\frac{\Delta_3 \Lambda}{(1-x)^2}- \frac{\Delta_1 \Lambda}{x^2} \right) \\
 +  \frac{4\pi}{N} \left( \Delta_4-\frac{\Delta_3}{(1-x)^2}\right) \log x
  -\frac{4\pi}{N}  \left( \Delta_4-\frac{\Delta_1}{x^2}\right) \log(1-x).
 \end{multline}

We can now take the linear combination \eqref{eq:Fpdifference} of variations of free energies. The terms proportional to the cut-off vanish and we obtain
\be
\frac{\delta \mathcal F_p}{\delta x}=  4 \left( \Delta_4-\frac{\Delta_3}{(1-x)^2}\right) \log x -4  \left( \Delta_4-\frac{\Delta_1}{x^2}\right) \log(1-x) - \frac{c}{6} (b(x) -b(1-x)).
\ee
Integrating with respect to $x$, we find \eqref{eq:g1.pants.3pts}.

\paragraph{Case (ii): Four vertex operator fields.} We now consider the case (ii) where the four fields $\phi_i$, $i = 1, \dots, 4$, are vertex operators with charges $\alphaC_i = \bar \alphaC_i$. Their positions in the complex plane are $z_1=1$, $z_2 = 1-x$, $z_3=0$ and $z_4=\infty$. The charges are constrained to satisfy the neutrality condition $\sum_{i}\alphaC_i=0$. In this case, the $n$-point function has the simple form \eqref{eq:npoint.vertex}. The same correlator vanishes if the sum of the charges is non-zero. Using the Ward identity, we find the simple formula
\begin{equation}\label{eqn:MagicTverex}
\frac{\langle T(z) \prod_{i=1}^n\phi_i(z_i,\bar z_i)\rangle}{\langle  \prod_{i=1}^n\phi_{i}(z_i,\bar z_i) \rangle} = \frac{1}{2} \left( \sum_{i=1}^n\frac{\alphaC_i}{z-z_i} \right)^2.
\end{equation}
This expression allows us to compute the first term of the integral \eqref{eqn:Tonsegment}. As in the case (i), we obtain two terms that depend on the cut-off as well as an expression that is regular as $\Lambda$ tends to infinity. We take the linear combination  \eqref{eq:Fpdifference} so that the divergences vanish, integrate with respect to $x$ and find \eqref{eq:g1.pants.4pts}.

We note that in both cases (i) and (ii), the term $g_1(x)$ of the fidelity is calculated up to an additive constant that cannot be obtained from the perturbative argument presented here and involving $\delta \mathcal F/\delta x$. These overall constants will be discussed further in \cref{sec:constant.pants,sec:constant.skirt} for the periodic pants domain and the skirt domain, respectively.

\subsection{Contribution to the skirt geometry}\label{sec:skirt.contribution}

Let us focus on the skirt geometry. First, we compute in \eqref{eqn:Tonsegment} the term proportional to $c$ in $\delta f_p$. We have
\begin{alignat}{2}
 \int_{0^+_\Lambda }^{1^-_{\Lambda}}  \left(\frac{\dd w_s}{\dd z}\right)^{-1} \{w_s(z),z \} \textrm{d}  z\ + c.c.
 &= 
\frac{\pi^2}{N^2}
  \left( \frac{\Lambda}{(1-x)^2} - \frac{\Lambda}{x^2} \right) 
\\ \nonumber & + 
\frac{\pi(2x-1)}{2N}
\bigg(\frac{2x-3}{(1-x)^2}\log x - \frac{2x+1}{x^2}\log(1-x) + \frac2{x (1-x)} \bigg).
\end{alignat}
The other term in \eqref{eqn:Tonsegment} depends on the fields that are inserted. We compute it for the cases~(i) and~(ii) defined in \cref{sec:CFT.skirt}.

\paragraph{Case (i): One bulk field.} We apply \eqref{eqn:Tonsegment} in the case where one primary field $\phi_5$ is inserted at $w_5 = +\infty$. The corresponding position on the upper half-plane is $z_5=(1+\ir)/2$. To compute the variation of free energy, we apply the method of images to express the one-point function on the upper half-plane as a two-point function on the complex plane. We introduce the image field $\phi_6$ of dimension $\Delta_6 = \Delta_5$ at the position $z_6=(1-\ir)/2$: $\langle \phi_5(z_5,\bar z_5) \rangle_{\mathbb H}= \langle \phi_5(z_5) \phi_6(z_6) \rangle_{\mathbb C}.$
After some algebra, we obtain
\be
 \int_{0^+_\Lambda }^{1^-_{\Lambda}}  \left(\frac{\dd w_s}{\dd z}\right)^{-1}  \frac{\langle T(z) 
 \phi_5 \phi_6  \rangle}{\langle \phi_5 \phi_6  \rangle}  \textrm{d}  z \, + c.c.= \frac{4\pi}{N} (\log x-\log(1-x)) \Delta_5.
\ee
The absence of divergences in this case can be traced back to the fact that there are no fields present in the legs $A$ and $B$. We take the linear combination of terms \eqref{eqn:Fsdifference} and integrate over $x$ to find \eqref{eq:g1tilde.skirt.4pts}.

\paragraph{Case (ii): Five vertex operator fields.} We investigate the case where fields $\phi_2,\phi_4$ are inserted on the endpoints of the slits, $\phi_1,\phi_3$ in the legs $A$ and $B$, respectively, and $\phi_5$ is inserted at $+\infty$. The corresponding points on the upper half-plane are 
\be
z_1=1,\qquad  z_2=\frac{\sqrt{x-x^2}+x-1}{2 x-1},\qquad z_3=0,\qquad z_4=\frac{-\sqrt{x-x^2}+x-1}{2 x-1}, \qquad z_5=\frac{1+\ir}{2}.
\ee
As in the previous case, we introduce the image field $\phi_6$ at the position $z_6=(1-\ir)/2$ in the complex plane. We make the assumption that each field $\phi_i$ is a vertex operator with charge $\alphaC_i$. 

In the case of boundary fields, the ratio of $n$-point functions that appears in \eqref{eqn:Tonsegment} is similar 
to \eqref{eqn:MagicTverex} and reads
\begin{equation}
\frac{\langle T(z) \prod_{i=1}^n\phi_i(z_i)\rangle}{\langle \prod_{i=1}^n\phi_{i}(z_i) \rangle} = \frac{1}{2} \left( \sum_{i=1}^n\frac{\alphaC_i}{z-z_i}
\right)^2.
\end{equation}
We insert the right-hand side of this equality in the integral \eqref{eqn:Tonsegment}.
We evaluate this integral and obtain the real part of the contributions to $\delta f_s$. 
It is invariant under the simultaneous transformation $x\rightarrow 1-x, \alphaC_1\leftrightarrow\alphaC_3$. It also contains terms that diverge for $\Lambda \to \infty$. These divergences however cancel out in the linear combination \eqref{eqn:Fsdifference}. We integrate over $x$ and obtain
 \begin{multline}
  \label{eq:g1tilde.skirt.6ptsGeneric}
\tilde g_1(x) = \frac{c}{24} \tilde G(x)+ \left( x (\alphaC^2_5 + \alphaC_6^2) +\frac12 (\alphaC_2^2+\alphaC_4^2)-\frac{x}{2(1-x)}\alphaC_3^2 - \frac{(\alphaC_3+\alphaC_5+\alphaC_6)^2}{2} \right)  \log x \\
 +\{ x\rightarrow 1-x, \alphaC_1\leftrightarrow \alphaC_3 \}.
  \end{multline}
Applying the method of images to the case where $\phi_5$ is an electric operator, the charge of the image field $\phi_6$ is $\alphaC_6 = -\alphaC_5$. In this particular case, the expression \eqref{eq:g1tilde.skirt.6ptsGeneric} is further simplified and can be expressed solely in terms of the dimensions $\Delta_i=\alphaC_i^2/2$, yielding \eqref{eq:g1tilde.skirt.6pts}.

\subsection[The constant \texorpdfstring{$C$}{C} for the periodic pants geometry]{The constant $\boldsymbol C$ for the periodic pants geometry}\label{sec:constant.pants}

In the two previous sections, we computed the correction to the free energy using a perturbation $\delta x$ of the aspect ratio $x$. These involve additive constants
 $C$ and $C'$ that do not depend on $x$ and are not fixed by the perturbative argument. Let us investigate this further for case (i), defined in \cref{sec:CFT.pants}. In this case, the bipartite fidelity $\mathcal{F}_p$ can be understood in terms of conformal correlation functions as
\be
\label{eq:Fp.CFT}
\mathcal F_p= - \log 
\bigg| 
\frac{
\big\langle \prod_{i=1,3,4} \phi_i(w_i,\bar{w}_i)\big\rangle^2_p
}{\langle\phi_1^-\phi_1^+\rangle_{c(Nx)}\langle\phi_3^-\phi_3^+\rangle_{c(N-Nx)}\langle\phi_4^-\phi_4^+\rangle_{c(N)}}\bigg|
\ee
where
\be
\langle \phi_i^-\phi_i^+\rangle_{c(P)}  = \langle \phi_i(w_-,\bar w_-)\phi_i(w_+,\bar w_+)\rangle_{c(P)}.
\ee
The correlators $\langle\phi_i^-\phi_i^+\rangle_{c(P)}$ are evaluated on the cylinder of perimeter $P$, and the fields $\phi_i$ are inserted at the positions $w_\pm=\pm\Lambda$ where $\Lambda$ is a cut-off. The three-point function in the numerator of \eqref{eq:Fp.CFT} is evaluated on the periodic pants geometry. We set the positions of the three fields to be functions of the cut-off $\Lambda$:
\be
w_1 = (\tfrac{\ir x N}2)^- - \Lambda,
\qquad w_3 = (\tfrac{\ir x N}2)^+ - \Lambda, \qquad w_4 = \Lambda.
\ee
For large $\Lambda$,
the corresponding positions in the complex plane are
\be
\label{eq:z.coordinates}
z_1 \simeq 1-\eE^{-\frac{2\pi(\Lambda + K_x)}{N x}}, \qquad 
z_3 \simeq \eE^{-\frac{2\pi(\Lambda + K_x)}{N (1-x)}}, \qquad 
z_4 \simeq\eE^{\frac{2\pi(\Lambda - K_x)}{N}}.
\ee

Let us also consider the bipartite fidelity $\mathcal F'_p$ where the fields $\phi_1'$, $\phi_3'$ and $\phi_4'$ are different fields. The conformal dimensions of these new fields are $\Delta'_i=\bar\Delta_i'$, $i=1,3,4$. We assume in the following that all these fields are primary. The difference $\mathcal F_p- \mathcal F'_p$ is
 \be 
\mathcal F_p- \mathcal F'_p = - \log 
\left| 
\frac{\big\langle \prod_{i=1,3,4} \phi_i(w_i,\bar{w}_i)\big\rangle^2_p}
	   {\big\langle \prod_{i=1,3,4} \phi_i'(w_i,\bar{w}_i)\big\rangle^2_p}
\times \frac{\langle\phi'_1{}^-\phi'_1{}^+\rangle_{c(Nx)}\langle\phi_3'{}^-\phi_3'{}^+\rangle_{c(N-Nx)}\langle\phi_4'{}^-\phi_4'{}^+\rangle_{c(N)}}{\langle\phi_1^-\phi_1^+\rangle_{c(Nx)}\langle\phi_3^-\phi_3^+\rangle_{c(N-Nx)}\langle\phi_4^-\phi_4^+\rangle_{c(N)}}
\right|.
\ee

We recall that under a map $w(z)$, correlation functions of primary fields $\phi_i$ of dimensions~$\Delta_i=\bar\Delta_i$ transform as 
\begin{equation}\label{eqn:transfoPrimary}
\Big\langle \prod_{i=1}^n \phi_i(w_i,\bar{w}_i)\Big\rangle= \prod_{i=1}^n \left|\frac{\dd w}{\dd z} \right|_{w=w_i}^{-2 \Delta_i} \Big\langle \prod_{i=1}^n \phi_i(z_i,\bar{z}_i)\Big\rangle.
\end{equation}
For the two-point functions, we use the map $w_c(z)=\frac{P}{2\pi}\log z$ from the complex plane to the cylinder of perimeter $P$, and find
\begin{equation}
\langle \phi_i^-\phi_i^+\rangle_{c(P)}=\left|\frac{\dd w_c}{\dd z}\right|^{-2\Delta_i}_{w=w_-} \left|\frac{\dd w_c}{\dd z}\right|_{w=w_+}^{-2\Delta_i}\langle \phi_i(z_-,\bar z_-)\phi_i(z_+,\bar z_+)\rangle_{\mathbb{C}} \simeq\left|\frac{P}{2\pi}\right|^{-4\Delta_i} \eE^{-8 \pi \Lambda \Delta_i/P},
\end{equation}
where $z_\pm=\eE^{\pm 2\pi \Lambda/P}$ is such that $w_c(z_\pm)=w_\pm$. The symbol $\simeq$ means that the equality holds at the leading order in $\Lambda$. Furthermore, in the last equality, we used the known value of the two-point function $\langle \phi_i(z_-,\bar z_-)\phi_i(z_+,\bar z_+)\rangle_{\mathbb{C}} =|z_--z_+|^{-4\Delta_i}$. 

The three-point function is given in \eqref{eq:3pt.function}. For the values \eqref{eq:z.coordinates}, we have
\begin{equation}
\Big\langle \prod_{i=1,3,4}\phi_i(z_i,\bar z_i)\Big\rangle_{\mathbb C} \simeq \mathcal C_{134} \,\eE^{-8 \pi \Delta_4(\Lambda-K_x)/N}.\end{equation}
Using \eqref{eqn:transfoPrimary}, we find
\be
\Big\langle \prod_{i=1,3,4}\phi_i(w_i,\bar w_i)\Big\rangle_{p} \simeq \mathcal C_{134}\,\frac{\big(\tfrac{2\pi}N\big)^{2(\Delta_1+\Delta_3+\Delta_4)}}{x^{2\Delta_1}(1-x)^{2 \Delta_3}}
\eE^{-\frac{4 \pi \Delta_1(\Lambda + K_x)}{Nx}}\eE^{-\frac{4 \pi \Delta_3(\Lambda + K_x)}{N(1-x)}}\eE^{-\frac{4 \pi \Delta_4(\Lambda - K_x)}{N}}.
\ee
After some simplifications, we find that the difference of bipartite fidelities is finite for $\Lambda \to \infty$. The result is
\be
\mathcal F_p- \mathcal F'_p= 4\Big(\delta_4 - \frac{\delta_1}{1-x}-\frac{\delta_3}{1-x}\Big)(x \log x + (1-x)\log x)-2 \log\bigg(\frac{\mathcal C_{134}}{\mathcal C'_{134}}\bigg) 
\ee
where $\delta_i = \Delta_i - \Delta_i'$ and $\mathcal C'_{134}$ is the structure constant for the correlator $\langle \phi'_1\phi'_3\phi'_4 \rangle$. Comparing with \eqref{eq:g1.pants.3pts}, we see that the difference of constants $C$ between $\mathcal F_p$ and $\mathcal F'_p$ is simply 
\be
\label{eq:CC.ratio}
C-C'=-2 \log(\mathcal C_{134}/\mathcal C'_{134}).
\ee

\subsection[The constant \texorpdfstring{$\tilde C$}{C tilde} for the skirt geometry]{The constant $\boldsymbol {\tilde C}$ for the skirt geometry}\label{sec:constant.skirt}

In \cref{sec:skirt.contribution}, we found an explicit expression for the function $\tilde g_1(x)$ for the skirt domain. The result is given in \eqref{eq:g1tilde.skirt.4pts} and \eqref{eq:g1tilde.skirt.6pts} up to a non-universal constant, which a priori can depend on the dimension of the inserted fields. In this section, we extract from this additive constant the dependence on $\Delta_5$.

First, we investigate this constant in greater detail for $\mathcal{F}_{s}$ in the special case (i) where the only non-trivial field is the bulk field $\phi_5(z_5,\bar z_5)$. We consider a second realisation $\mathcal{F}'_s$ where a primary field $\phi'_5(z_5,\bar z_5)$ replaces the field $\phi_5(z_5,\bar z_5)$ and is of conformal dimensions $\Delta_5'=\bar\Delta_5'$.
 In the conformal interpretation, the difference of bipartite fidelities is
\be
\label{eqn:FalphaFalphaPrime}
\mathcal{F}_{s}- \mathcal{F}'_s
=-\log \left| \frac{Z^2_{s}}{Z_c} \frac{Z'_{c}}{(Z'_{s})^2} \right| 
= -\log \left| \frac{\langle \phi_{5}(w_5,\bar w_5)\rangle^2_{s } }{\langle \phi_{5}^-\phi_5^+\rangle_{c(N)} }\cdot
\frac{ \langle \phi_{5}'{}^-\phi_5'{}^+\rangle_{c(N)} }{\langle \phi'_{5}(w_5,\bar w_5)\rangle^2_{s} }  \right|.
\ee
Here, the two-point functions are evaluated on the cylinder of perimeter $N$ and height $\Lambda$. Their expression is given in the previous section in the large-$\Lambda$ limit. Moreover, we calculate the one-point function $\langle \phi'_{5}(w_5,\bar w_5)\rangle_{s}$ for $\text{Re}(w_5)=\Lambda$ and $\Lambda$ large. We compute it using the map to the upper half-plane and the transformation law \eqref{eqn:transfoPrimary}:
\be
\langle \phi_5(w_5,\bar w_5)\rangle_{s} =  \left|\frac{\dd w_s}{\dd z}\right|_{w=w_5}^{-2\Delta_5}
\langle \phi_5(z_5,\bar z_5)\rangle_{\mathbb{H}}.
\ee
In the large-$\Lambda$ limit, $z_5$ tends to $(1+\ir)/2$. The one-point correlator is calculated using the method of images and equals $1$ in this limit. As for the partial derivatives, they diverge for $w_1 \to \infty$. Hence we parameterise $w_5(m) =w_{s}(z(m))$, with $z(m)=\frac{1+\ir}{2}(1-\eE^{-2 m \pi /N})$, to obtain for large $m$
\begin{equation}
\langle \phi_5(w_5(m),\bar w_5(m))\rangle_{s} \simeq \bigg|\frac{\sqrt{2} \pi}{N} \bigg|^{2\Delta_5} (\eE^{2 m\pi/N})^{-2\Delta_5}.
\end{equation}
With this parameterisation, the position of $w_5(m)$ along the skirt is obtained using the explicit form of $w_s(z)$, and is of order $m$:
\begin{equation}
\text{Re}(w_5(m)) = m + K_x-\frac{3N}{4\pi}\log 2 +\mathcal O(1/m).
\end{equation}
Hence we set $m=\Lambda-K_x+\frac{3N}{4\pi}\log 2$ to compare the two partition functions.
The evaluation for large $\Lambda$ yields
\begin{equation}
\mathcal{F}_s =4^{-2\Delta_5}(\eE^{-2 \pi K_x/N})^{-2\Delta_5}.
\end{equation}
Inserting in \eqref{eqn:FalphaFalphaPrime}, we find
\begin{equation}
\mathcal{F}_{s}- \mathcal{F}'_s = 4(\Delta_5-\Delta_5') \left[ x\log x+ (1-x)\log(1-x) + 2\log 2 \right].
\end{equation}
We deduce that the additive constant in the function $\tilde g_1(x)$ has a term proportional to $8 \Delta_5 \log 2$, as given in \eqref{eq:g1tilde.skirt.4pts}. The remaining constant $\tilde C$ in this equation is then independent of $\Delta_5$.

Second, we focus on the case (ii), where five vertex operators are present. We consider a second instance of $\mathcal{F}_s'$ where the field $\phi_5$ is replaced by $\phi_5'$ with conformal dimensions $\Delta_5'=\bar{\Delta}_5'$. The four other fields remain unchanged. The difference of bipartite fidelities reads
\begin{equation}\label{eqn:FalphaFalphaPrime12345}
\mathcal{F}_s-\mathcal{F}_s'=-\log\left| \frac{
\langle \phi_1(w_1)\phi_2(w_2)\phi_3(w_3)\phi_4(w_4)\phi _{5}(w_5,\bar w_5)\rangle^2_{s }}{
\langle \phi_1(w_1)\phi_2(w_2)\phi_3(w_3)\phi_4(w_4)\phi'_{5}(w_5,\bar w_5)\rangle^2_{s }}\cdot \frac{ \langle \phi_{5}'{}^-\phi_5'{}^+\rangle_{c(N)}}{
\langle \phi_{5}    ^-\phi_5    ^+\rangle_{c(N)}}
\right| .
\end{equation} 
We map the ratio of five-point functions into the upper half-plane and obtain
\be
\frac{
\langle \phi_1(w_1)\phi_2(w_2)\phi_3(w_3)\phi_4(w_4)\phi _{5}(w_5,\bar w_5)\rangle_{s }}{
\langle \phi_1(w_1)\phi_2(w_2)\phi_3(w_3)\phi_4(w_4)\phi'_{5}(w_5,\bar w_5)\rangle_{s }}
=  \left|\frac{\dd w_s}{\dd z}\right|_{w=w_5}^{-2(\Delta_5-\Delta_5')}
\frac{
\langle \phi_1(z_1)\phi_2(z_2)\phi_3(z_3)\phi_4(z_4)\phi _{5}(z_5,\bar z_5)\rangle_{\mathbb{H}}}{
\langle \phi_1(z_1)\phi_2(z_2)\phi_3(z_3)\phi_4(z_4)\phi'_{5}(z_5,\bar z_5)\rangle_{\mathbb{H}}}.
\ee
We apply the method of images to express this ratio in terms of a six-point function, where we introduce the new coordinate $z_6 = \bar z_5$. We then use the generic expression for $n$-point functions of products of vertex operators \eqref{eq:npoint.vertex} and find
\be 
\begin{split}
\frac{
\langle \phi_1(z_1)\phi_2(z_2)\phi_3(z_3)\phi_4(z_4)\phi _{5}(z_5,\bar z_5)\rangle_{\mathbb{H}}}{
\langle \phi_1(z_1)\phi_2(z_2)\phi_3(z_3)\phi_4(z_4)\phi'_{5}(z_5,\bar z_5)\rangle_{\mathbb{H}}}
&= \frac{
\langle \phi_1(z_1)\phi_2(z_2)\phi_3(z_3)\phi_4(z_4)\phi _{5}(z_5)\phi _{5}(z_6)\rangle_{\mathbb{C}}}{
\langle \phi_1(z_1)\phi_2(z_2)\phi_3(z_3)\phi_4(z_4)\phi'_{5}(z_5)\phi'_{5}(z_6)\rangle_{\mathbb{C}}} \\
&=\frac{\prod_{i=1}^4|z_i-z_5|^{\alphaC_i\alphaC_5}|z_i-z_6|^{\alphaC_i\alphaC_6}}{
			\prod_{i=1}^4|z_i-z_5|^{\alphaC_i\alphaC_5'}|z_i-z_6|^{\alphaC_i\alphaC_6'}}.
			\end{split}
\ee
We exploit the fact that $z_1,z_2,z_3$ and $z_4$ lie on the real line while $z_6$ is the complex conjugate of $z_5$. Hence, $|z_i-z_6|=|z_i-z_5|$ for each $i=1,\dots,4$. We finally obtain
\be 
\frac{
\langle \phi_1(z_1)\phi_2(z_2)\phi_3(z_3)\phi_4(z_4)\phi _{5}(z_5,\bar z_5)\rangle_{\mathbb{H}}}{
\langle \phi_1(z_1)\phi_2(z_2)\phi_3(z_3)\phi_4(z_4)\phi'_{5}(z_5,\bar z_5)\rangle_{\mathbb{H}}}
= \frac{\prod_{i=1}^4|z_i-z_5|^{\alphaC_i(\alphaC_5+\alphaC_6)}}{
			\prod_{i=1}^4|z_i-z_5|^{\alphaC_i(\alphaC_5'+\alphaC_6')}}
= \frac{\prod_{i=1}^4|z_i-z_5|^{-\alphaC_i(\alphaC_1+\alphaC_2+\alphaC_3+\alphaC_4)}}{
			\prod_{i=1}^4|z_i-z_5|^{-\alphaC_i(\alphaC_1+\alphaC_2+\alphaC_3+\alphaC_4)}}
=1,
\ee
where we used the neutrality condition $\sum_i \alphaC_i=0$. 

The rest of the computation of the difference 
$\mathcal{F}_s-\mathcal{F}_s'$ is similar to the case (i) and yields a term proportional to $8\Delta_5 \log 2$ in the additive constant in $\tilde{g}_1(x)$, as stated in \eqref{eq:g1tilde.skirt.6pts}.

\subsection{Sub-leading term for the periodic pants geometry}\label{sec:subleading.pants}

The computation of the sub-leading correction $g_2(x)$ for the periodic pants domain follows closely the same computation by St\'ephan and Dubail \cite{SD13} for the flat pants domain. Here we only state the main differences with their proof.

For a conical singularity of internal angle $\theta$, the geometry is that of a simply connected domain with a {\it conical corner} of angle $\theta$. The two edges leaving this corner are endowed with periodic boundary conditions. If the corner is at the origin, the map $w(z)$ from the complex plane to this domain behaves as 
\be
\label{eq:map.to.corner}
w(z) = z^{\theta/2\pi}(1+\kappa_1 z + \kappa_2 z^2 + \dots).
\ee 
The preimages in the $z$-plane of the edges leaving the corner are two lines leaving the origin at angles $0$ and $2\pi$. In comparison, for the case considered in \cite{SD13}, the domain of the transformation is the upper half-plane, the origin is mapped to a corner that lies on the boundary of the domain, and the prefactor in \eqref{eq:map.to.corner} is instead $z^{\theta/\pi}$.

The rest of the computation is similar to \cite{SD13} and we have
\be
g_2(x) =\Xi \times \frac{N}{2\pi}
\text{Res}\left[ \eE^{\ir\varphi_c}\left(\frac{\dd w }{\dd z} \right)^{-1} \left( \frac{c}{12} \{ w(z),z\} - \frac{\langle T(z) \phi_1\cdots \phi_n \rangle}{\langle \phi_1 \cdots \phi_n \rangle} \right)\!,\, z=z_c\right]+ c.c.
\ee
where $\varphi_c = \arg \left( w''(z_c)\right)$. Computing the residue with  $w(z) = w_p(z)$, four arbitrary primary fields $\phi_1, \dots, \phi_4$ inserted in the positions $z_1 = 0$, $z_2 = 1-x$, $z_3 = 1$ and $z_4 = \infty$, and with $z_c = z_2$, we obtain \eqref{eq:g2.pants}. This is exactly twice the result obtained in \cite{SD13}. This factor of $2$ can be traced back to $w_p(z)$ being exactly half of the map for the flat pants geometry considered in \cite{PMDR19}.

\subsection{Sub-leading term for the skirt geometry}\label{sec:subleading.skirt}

In this subsection, we compute the sub-leading corrections to the free-energy $\tilde{g}_2(x)$ for the skirt domain, see \eqref{eq:Fs.expansion}. For a geometry with a boundary and a given number of corners of internal angle $2\pi$, this function is
\begin{equation}
\label{eq:g2.skirt.general}
\tilde{g}_2(x) = \frac{N}{2 \pi}\sum_{2\pi \text{\,corners} } \Xi_c \times \text{Res}\left[ \eE^{\ir\varphi_c}\left(\frac{\dd w }{\dd z} \right)^{-1} \left( \frac{c}{12} \{ w(z),z\} - \frac{\langle T(z) \phi_1\cdots \phi_n \rangle}{\langle \phi_1 \cdots \phi_n \rangle} \right)\!,\, z=z_c\right] + c.c.
\end{equation}
Here, the sum runs over all the corners of internal angle $2\pi$. The phase $\varphi_c$ is $\arg \left( w''(z_c)\right)$. The $\Xi_c$ is the extrapolation length corresponding to the corner situated at $z=z_c$. This formula is a generalisation of the one given in \cite{SD13}, where the authors take the same extrapolation length for all corners, $\Xi_c=\Xi$. However, our lattice analysis of $\mathcal F^{\tXX}_s$ in \cref{sec:AsymptoticsSkirtXX} includes cases where spin states of different lengths are assigned to the endpoints of the slits, and this requires a CFT analysis with non-equal extrapolation lengths.

We now apply the formula to the skirt geometry. The phase verifies $\eE^{\ir \varphi_c}=-1$ for both corners at $z=z_2$ and $z=z_4$. We write $\Xi_2$ and $\Xi_4$ for the corresponding extrapolation lengths.

For case (i), we use the method of images to express the one-point correlator as a two-point function and readily find \eqref{eq:g2tilde.skirt.3pts}. In the case (ii), there is a vertex operator $\phi_i$ at the position~$z_i$ on the upper half-plane, with $i = 1, \dots, 5$. We use the method of images as well as \eqref{eqn:MagicTverex} to find the real part of the sub-leading contribution to the free-energy:
\begin{alignat}{1}
\tilde{g}_2(x)
&= \Xi_2 \times \left[\frac{c (1-2 x)^2}{48 (1-x) x} + \frac{\alphaC _2^2-\alphaC _4^2}{16 x(1-x)}-\frac{\alphaC _1 \left(\alphaC _1-\alphaC _2+\alphaC _4\right)}{4 x} -\frac{ \alphaC _3 \left(\alphaC_3-\alphaC _2+\alphaC _4\right)}{4(1-x)}+\frac{\alphaC_5^2+\alphaC _6^2}{2}\right]\\
&+\{ \Xi_2\to \Xi_4, \alphaC_2\leftrightarrow \alphaC_4\}. \nonumber
\end{alignat}
We note that this formula greatly simplifies if the two extrapolation lengths are identical. In that case, the result does not depend on $\alphaC_2$ or $\alphaC_4$. 
We particularise our result to the case $\alphaC_6=-\alphaC_5$ and obtain \eqref{eq:g2tilde.skirt.version.2}.

\section{Asymptotics}\label{app:Asympt}

In this appendix, we derive the asymptotic expansion of $\mathcal F$ on the periodic pants domain, for the XX chain \eqref{eq:LBFXXFinal} and for the model of dense polymers \eqref{eq:LBFaFinal}. The starting point is the closed-form expressions \eqref{eq:B} and \eqref{eq:FaFXX} for finite $N$.

\subsection{Toolbox}
We define the mathematical tools and functions used in the computation. 

\paragraph{Exact expressions.} The building blocks of the calculations are the functions
\begin{subequations}
\label{eq:sXYA}
\begin{alignat}{2} 
&s[x] = \frac{\sin x}{x}, \\
&X(a,b,N) = \sum \limits_{k=0}^{N/2} \log s\Big[\frac{2 \pi a}{N}(k+b)\Big] ,\\
&Y(a,b,N,x) = \sum \limits_{k=0}^{Nx/2}\sum \limits_{k'=0}^{N/2} \log s\Big[\frac{2 \pi a}{N}\big(\frac kx-k'+b\big)\Big], \\
&Z(a,b,N,x) =  \sum \limits_{k=1}^{Nx/2}\sum \limits_{k'=1}^{N/2} \log \left| k'-a-\frac{k-b}{x} \right|.
\end{alignat}
\end{subequations}

\paragraph{Asymptotics for $\boldsymbol{X}$ and $\boldsymbol{Y}$.} 
To compute the asymptotic expansion of the $X$ and $Y$ functions, we use the Euler-Maclaurin formula
\be
\label{eq:EmL}
\sum_{i=a}^b f(i) = \int_a^b f(x) dx + \frac{f(a) + f(b)}{2} + \sum_{k=1}^\infty \frac{B_{2k}}{(2k)!}(f^{2k-1}(b)-f^{2k-1}(a))
\ee
where $B_k$ is the $k$-th Bernoulli number. The results read
\begin{subequations}
\label{eq:XYasymp}
\begin{alignat}{2}
X(a,b,N) &= \frac{N}{2}\mathcal{I}_1(\pi a , 0)+ \Big(b+\frac 12\Big)\log s[\pi a]+\mathcal{O}\big(N^{-1}\big),
\\[0.15cm]
Y(a,b,N,x) &= \frac{N^2x}{4} \mathcal{I}_2(\pi a,-\pi a) + \frac{N x}{2} \Big(-b+\frac 12\Big)\mathcal{I}_1(-\pi a,\pi a)+ \frac{N x}{2} \Big(b+\frac 12\Big)\mathcal{I}_1(-\pi a,0)\nonumber\\
&+ \frac N4 (\mathcal{I}_1(\pi a,-\pi a)+\mathcal{I}_1(\pi a,0)) + \log s[\pi a] \Big\{\frac 12 + xb^2 + \frac x6 + \frac{1}{6x}\Big\}+\mathcal{O}\big(N^{-1}\big),
\end{alignat}
\end{subequations}
with
\begin{equation}
\label{eq:Is}
\mathcal{I}_1(a,b) = \int_0^1 \dd y \log s[y a + b], \qquad  \mathcal{I}_2(a,b) = \int_0^1\hspace{-0.1cm} \int_0^1\dd w \dd z \log s[ wa + bz].
\end{equation}

\paragraph{Asymptotics for $\boldsymbol{Z}$.} 
The function $Z(a,b,N,x)$ is written in terms of Gamma functions as
\begin{equation}
\label{eq:Am}
\begin{split}
Z(a,b,N,x) &= -\frac{N x}{2} \log \pi + \underbrace{\sum \limits_{k=1}^{Nx/2} \log \Big|\sin\Big[\pi \big(\frac{-k+b}{x}-a\big)\Big]\Big|}_{K(a,b,N,x)} \\
&+\underbrace{\sum \limits_{k=1}^{Nx/2} \left\{\log \Gamma \Big(\frac{k}{x}+ \frac{b-1}{x}+1-a\Big)+\log\Gamma \Big(\frac{k}{x}- \frac{b}{x}+a\Big)\right\}}_{\tilde{Z}(a,b,N,x)}.
\end{split}
\end{equation}
Let us focus on the function $\tilde{Z}(a,b,N,x)$. We use the integral representation for the logarithm of the Gamma function: 
\begin{equation}
\log \Gamma(z) = \int_0^\infty \frac{dt}t \Big\{(z-1)\eE^{-t} - \frac{\eE^{-t} - \eE^{-zt}}{1 - \eE^{-t}}\Big\}, \qquad {\rm Re}\, z >0,
\end{equation}
and find 
\begin{alignat}{2}
\tilde{Z}(a,b,N,x) = \sum \limits_{k=1}^{Nx/2}\int_0^\infty \frac{dt}t \left[\Big(\frac{2k}{x}-\frac 1x-1\Big)\eE^{-t}- \frac{2 \eE^{-t}-\eE^{-\frac{kt}{x}}\Big(\eE^{-\big(\frac{b-1}{x}+1-a\big)t}+\eE^{-\big(\frac{-b}{x}+a\big)t}\Big)}{1-\eE^{-t}}\right]& \nonumber\\[0.2cm]
=\int_0^\infty \frac{dt}t \left[\frac{N x}{2}\Big(\frac N2-1\Big)\eE^{-t}-Nx\frac{\eE^{-t}}{1-\eE^{-t}}+\frac{\eE^{-\frac{Nt}{2}}-1}{1-\eE^{\frac tx}}\cdot \frac{\eE^{-\big(\frac{b-1}{x}+1-a\big)t}+\eE^{-\big(\frac{-b}{x}+a\big)t}}{1-\eE^{-t}}\right]&.
\label{eq:Atildedef}
\end{alignat}
To proceed further, we use the relations \cite{PMDR19} 
\begin{subequations}
\label{eq:int123}
\begin{alignat}{2}
i_1(n)&\equiv\int_\eps^\infty \; \frac{dt}t \eE^{-nt} = -\log{\eps} - \log{n} - \gamma + \mathcal{O}(\eps),\\[0.1cm]
i_2(n)&\equiv\int_\eps^\infty \; \frac{dt}{t^2} \eE^{-nt} = \frac 1\eps + n\log\eps + n\log{n} +n(\gamma-1) + \mathcal{O}(\eps),\\[0.1cm]
i_3(n)&\equiv\int_\eps^\infty \; \frac{dt}{t^3} \eE^{-nt} = \frac 1{2\eps^2} - \frac n\eps - \frac{n^2}2 \log{\eps} -\frac{n^2}2 \Big(\log{n} - \frac 32 + \gamma\Big) + \mathcal{O}(\eps),\\[0.1cm]
j(\alpha)& \equiv \int_\eps^\infty \; \frac{dt}t \frac{1}{\eE^{\alpha t}-1} = \frac{1}{\alpha \eps}+\frac{1}{2}\log( \alpha \eps)+ \frac{\gamma}{2}-\frac{1}{2}\log(2\pi )+\mathcal{O}(\eps),
\end{alignat}
\end{subequations}
valid for $n>0$ and $\alpha>0$. Here $\gamma \simeq 0.57722$ is the Euler-Mascheroni constant. In the following, we change the lower bounds of the integrals from $0$ to $\eps$ and take the limit $\epsilon \to 0$ at the end of the computation. The first part of the integral \eqref{eq:Atildedef} becomes
\begin{equation}
\label{eq:Atilde1}
\int_\eps^\infty \frac{dt}t \left[\frac{N x}{2}\Big(\frac N2-1\Big)\eE^{-t}-Nx\frac{\eE^{-t}}{1-\eE^{-t}}\right] = \frac{N x}{2}\Big(\frac N2-1\Big)i_1(1)-Nx \ j(1).
\end{equation}
For the second part, we first consider the term with a factor of $\eE^{-Nt/2}$. We keep the exponentials in the numerator unchanged and expand the denominator around $t=0$. We only need to keep term up to order $t^{-1}$. Indeed, for $k\geq 0$, the integral $\int_0^\infty t^k \eE^{-N t}$ vanishes as $N^{-(k+1)}$ for $N\to \infty$. Hence, we have
\begin{align}
\int_\eps^\infty \frac{dt}t \bigg[\frac{\eE^{-\frac{Nt}{2}}}{1-\eE^{\frac tx}}\cdot \frac{\eE^{-\big(\frac{b-1}{x}+1-a\big)t}+\eE^{-\big(\frac{-b}{x}+a\big)t}}{1-\eE^{-t}}\bigg]& = \frac{3-\frac 1x -x}{12}\bigg[i_1\Big(\frac N2+\frac{b-1}{x}+1-a\Big)+i_1\Big(\frac N2-\frac{b}{x}+a\Big)\bigg] \nonumber \\[0.1cm]
&+ \frac{1-x}{2}\bigg[i_2\Big(\frac N2+\frac{b-1}{x}+1-a\Big)+i_2\Big(\frac N2-\frac{b}{x}+a\Big)\bigg] \nonumber \\[0.1cm]
& -x\bigg[i_3\Big(\frac N2+\frac{b-1}{x}+1-a\Big)+i_3\Big(\frac N2-\frac{b}{x}+a\Big)\bigg]\nonumber \\[0.1cm]
&+\mathcal{O}(N^{-1}). \label{eq:Atilde2}
\end{align}

The last integral in \eqref{eq:Atildedef} is also divergent. Changing the lower bound to $\epsilon$, we simplify the integral by adding and subtracting a function $m(a,b,x,t)\eE^{-t}$ from the integrand. We choose $m(a,b,x,t)$ so that (i) the leading terms in its expansion around $t=0$ are identical to those of the original integrand, and (ii) it can be easily integrated in terms of the integrals $i_1(n)$, $i_2(n)$ and $i_3(n)$ introduced in \eqref{eq:int123}. We find
\begin{alignat}{2}
\label{eq:Atilde3}
\int_\eps^\infty \frac{dt}t &\left[-\frac{1}{1-\eE^{\frac tx}}\cdot \frac{\eE^{-\big(\frac{b-1}{x}+1-a\big)t}+\eE^{-\big(\frac{-b}{x}+a\big)t}}{1-\eE^{-t}}\right] =  \int_\eps^\infty \frac{dt}t\left[J(a,b,x,t)-m(a,b,x,t)\eE^{-t}\right] \\
&\hspace{-0.3cm} +2x(i_3(1)+i_2(1))+\Big(b+\frac 16\Big(\!\!-\!3+\frac 1x\Big)+ \frac{(-1+b)b}{x}+\frac{7x}{6}+a^2x-a(-1+2b+x)\Big)i_1(1)\nonumber
\end{alignat}
with
\begin{subequations}
\label{eq:Atilde4}
\begin{alignat}{2}
&J(a,b,x,t) = -\frac{1}{1-\eE^{\frac tx}}\cdot \frac{\eE^{-\big(\frac{b-1}{x}+1-a\big)t}+\eE^{-\big(\frac{-b}{x}+a\big)t}}{1-\eE^{-t}},\label{eq:Jabxt}\\
&m(a,b,x,t) = 2x\Big(\frac{1}{t^2}+\frac 1t\Big) + \Big(b+\frac 16\Big(\!\!-\!3+\frac 1x\Big)+ \frac{(-1+b)b}{x}+\frac{7x}{6}+a^2x-a(-1+2b+x)\Big).
\end{alignat}
\end{subequations}

We obtain the leading terms in the expansion of $\tilde{Z}$ by combining \eqref{eq:Atilde1}, \eqref{eq:Atilde2},  \eqref{eq:Atilde3} and \eqref{eq:Atilde4}. Using \eqref{eq:int123}, we find that the divergent contributions cancel as expected. The result has a well-defined $\epsilon \to 0$ limit:
\begin{alignat}{2}
\tilde{Z}(a,b,N,x)&=\int_0^\infty \frac{dt}t\left[J(a,b,x,t)-m(a,b,x,t)\eE^{-t}\right]- \frac{N^2x}{8}(3+2\log2-2\log N)+\frac{Nx}{2}\log(2\pi)\nonumber\\
&+\log N\ \frac{1+6b^2-6b(1-x+2ax)+x(-3+x+6a(1-x+ax))}{6x}\nonumber\\
&-\ \frac{3x^2+(1+x^2-3x+6(b-ax)(-1+b+x-ax))\log 2}{6x}+\mathcal{O}(N^{-1}).
\label{eq:AtildeAsympt}
\end{alignat}
We note that the integral $\int_0^\infty \frac{dt}t\left[J(a,b,x,t)-m(a,b,x,t)\eE^{-t}\right]$ is convergent. 

\subsection[Asymptotics for \texorpdfstring{$P_1(N)$}{P1N}]{Asymptotics for $\boldsymbol{P_1(N)}$}\label{app:P1}
To derive the large-$N$ expansion of $P_1(N)$, we start from the definition \eqref{eq:Ps} and rewrite each sine factor using $\sin x=x\,s[x]$. Each product then splits into two products, the first involving the function $s[x]$ and the second involving the arguments $x$. The former products are expressed in terms of the functions $X$ and $Y$ in \eqref{eq:sXYA}. The latter are written in terms of Barnes' $G$-function. The result is
\begin{equation}
\log P_1(N) = -\frac{N(N-2)}{8}(\log N - \log \pi)+ \log G\Big(\frac N2+1\Big)+\frac 12 Y\Big(\frac 12,0,N,1\Big)-X\Big(\frac 12,0,N\Big).
\end{equation}
We recall the asymptotics of the Barnes' G-function: 
\begin{equation}
\label{eq:AsymptoticsGammaG}
 \log G(z) = \left(\frac{(z-1)^2}{2} -\frac{1}{12}\right) \log (z-1)-\frac{3(z-1)^2}{4} +\frac{z-1}{2} \log (2 \pi )+\frac{1}{12}-\log A+ \mathcal{O}(z^{-1}),
 \end{equation}
where $A \simeq 1.282427$ is the Glaisher-Kinkelin constant. Combining this with the asymptotic expansions \eqref{eq:XYasymp} of $X$ and $Y$, we obtain the large-$N$ expansion of $P_1$ up to order $\mathcal O(N^{-1})$.
 
\subsection[Asymptotics for \texorpdfstring{$P_2(N_1,N_2,\phi_1,\phi_2)$}{P2(N1,N2,phi1,phi2)}]{Asymptotics for $\boldsymbol{P_2(N_1,N_2,\phi_1,\phi_2)}$}\label{app:P2}
We follow the same strategy for $P_2$ and find
 \begin{alignat}{2}
 \log P_2(N_1,N_2,\phi_1,\phi_2) &= -\frac{N_1N_2}{4}\log \frac{N_2}{\pi}+ \log s\Big[\frac{\pi}{N_2}\Big(-\frac{1}{2z}+\frac{1}{2}-\frac{\phi_1}{2\pi z}+\frac{\phi_2}{2\pi}\Big) \Big]\nonumber\\
 &+Z\Big(\frac 12 + \frac{\phi_2}{2\pi}, \frac 12 + \frac{\phi_1}{2\pi}, N_2,z\Big)+Y\Big(\frac 12, -\frac{1}{2z} - \frac{\phi_1}{2\pi z}+\frac 12 + \frac{\phi_2}{2\pi},N_2,z\Big)\\
 &-X\Big(\frac 12,z\big( -\frac{1}{2z} - \frac{\phi_1}{2\pi z}+\frac 12 + \frac{\phi_2}{2\pi}\big),N_1\Big)-X\Big(\frac 12, \frac{1}{2z} + \frac{\phi_1}{2\pi z}-\frac 12 - \frac{\phi_2}{2\pi},N_2\Big)\nonumber
 \end{alignat}
 where $z\equiv \frac{N_1}{N_2}$. The term $\log s\Big[\frac{\pi}{N_2}\Big(-\frac{1}{2z}+\frac{1}{2}-\frac{\phi_1}{2\pi z}+\frac{\phi_2}{2\pi}\Big) \Big]$ does not contribute to the leading orders in the large-$N$ expansion. Using \eqref{eq:XYasymp}, we obtain the large-$N$ expansion of $P_2$ up to order $\mathcal O(N^{-1})$.
  
\subsection{Combinations of non-trivial terms}
We consider the combination of $\log P_1$ and $\log P_2$ corresponding to \eqref{eq:B}. In doing so, we find that all the terms involving the integrals $\mathcal{I}_1$ and $\mathcal{I}_2$  that appear in the large-$N$ expansion \eqref{eq:XYasymp} of $X$ and $Y$ cancel out. Let us now investigate the remaining terms. 

\paragraph{Combination of $\boldsymbol{\tilde{Z}}$.} We are interested in the combination
\begin{equation}
\label{eq:CombiA}
\tilde{Z}\big(r_B,r_A,N(1-x),\frac{x}{1-x}\big)-\tilde{Z}(r,r_A,N,x)-\tilde{Z}(r,r_B,N,1-x)
\end{equation}
with $r_{A,B} = \frac{1}{2}+ \frac{\phi_{A,B}}{2 \pi}$ and $r = \frac{1}{2}+ \frac{\phi}{2 \pi} = r_A+r_B-1$. This last constraint follows from \eqref{eq:restrictions} with $\ell = 0$. The only non-trivial combination involves the (convergent) integral $\int_0^\infty \frac{dt}t\left[J(a,b,x,t)-m(a,b,x,t)\eE^{-t}\right]$. From \eqref{eq:Jabxt}, we observe that 
\begin{equation}
J\Big(r_B,r_A, \frac{x}{1-x},t\Big) - J\Big(r_A+r_B-1,r_A, x,t\Big) - J\Big(r_A+r_B-1,r_B, 1-x,t \frac{1-x}{x}\Big)=0.
\end{equation}
Hence, in computing \eqref{eq:CombiA}, we make the change of variable $t \rightarrow t \frac{1-x}{x}$ in the integral of the third term. The integrals involving the functions $J$ cancel and the resulting combination
\begin{multline}
\int_0^\infty \frac{dt}t\left[  -m\Big(r_B,r_A, \frac{x}{1-x},t\Big)\eE^{-t} +m\Big(r_A+r_B-1,r_A, x,t\Big)\eE^{-t}\right. \\  \left.  +m\Big(r_A+r_B-1,r_B, 1-x,t \frac{1-x}{x}\Big) \eE^{-t \frac{1-x}{x}}  \right]
\end{multline}
can be written in terms of the functions $i_1(n)$, $i_2(n)$ and $i_3(n)$ defined in \eqref{eq:int123}. Including the remaining terms in \eqref{eq:AtildeAsympt}, we find that the result simplifies to
\begin{equation}
\begin{split}
&\tilde{Z}\big(r_B,r_A,N(1-x),\frac{x}{1-x}\big)-\tilde{Z}(r_A+r_B-1,r_A,N,x)-\tilde{Z}(r_A+r_B-1,r_B,N,1-x)= \\&\frac{1}{8} N^2 \left\{2 (1-x) x \Big(\log N+\log(1-x)\Big)-2 \log N+\Big((x-1) x+1\Big) (3+\log 4)\right\}+\frac{1}{2} N (x-1) \log (2 \pi )\\
&+\frac{6 r_A^2 (x-1)^2+6 r_A (x-1) (2 r_B x-3 x+1)+x \Big(6r_B (r_B x-3x +1)+13 x-9\Big)+1}{6  x} \log(1-x)\\
   &+\frac{6 r_A^2 (x-1)^2+6 r_A (x-1) (2 r_B x-3 x+2)+x \Big(6 r_B(r_B x-3x+2)+13 x-17\Big)+5}{6 (1-x) } \log x+\mathcal{O}(N^{-1}).
   \end{split}
\end{equation}

\paragraph{Combination of trigonometric functions.}
The logarithm of the product of cosines in \eqref{eq:B} and the functions $K$ appearing in the definition \eqref{eq:Am} of $Z$ combine to give the identity
\begin{equation}
\begin{split}
&K\Big(r_B,r_A, N(1-x),\frac{x}{1-x}\Big) - K\Big(r_A+r_B-1,r_A,N, x\Big) - K\Big(r_A+r_B-1,r_B, N,1-x \Big) \\
&+ \sum \limits_{k'=1}^{N/2} \log \left|\cos \Big[\pi x k'-\frac{\phi_A(x-1)+x \phi_B }{2}\Big] \right| = -\frac{1}{2}N x \log 2 ,
\end{split}
\end{equation}
whose proof is straightforward.

We now have all the ingredients needed to compute the large-$N$ expansion of $\mathcal F_p^\tXX|_{\phi = \phi_A + \phi_B - \pi}$, defined in \eqref{eq:B}. Simplifying the result, we find \eqref{eq:LBFXXFinal}. 

\subsection[Asymptotics for \texorpdfstring{$Q(N,\phi)$}{Q(N,phi)}]{Asymptotics for $\boldsymbol{Q(N,\phi)}$}

We obtain the large-$N$ expansion \eqref{eq:LBFaFinal} of $\mathcal F_p^\alpha$ from \eqref{eq:FaFXX} and \eqref{eq:LBFXXFinal} by computing the asymptotic expansion of the function $Q$, defined in \eqref{eq:Q}. Using the same strategy as for the functions $P_1$ and $P_2$, we find
\begin{equation}
Q(N,\phi) = \log \Gamma \Big(\frac N2 +\frac 12 -\frac{\phi}{2\pi} \Big)-\log \Gamma \Big(\frac 12 -\frac{\phi}{2\pi} \Big)-X\Big(\frac 12, -\frac 12 +\frac{\phi}{2\pi},N \Big)-\Big\{\phi \rightarrow -\phi\Big\}.
\end{equation}
We use
\begin{equation}
\log \Gamma(z) = z \log z -\frac 12 \log z + \frac 12 \log (2\pi) + \mathcal{O}(z^{-1})
\end{equation}
and \eqref{eq:XYasymp}, and find
\begin{equation}
\label{eq:QAsympt}
Q(N,\phi) = -\frac{\phi}{\pi} \log \frac{N}{\pi} + \log \frac{\Gamma \Big(\frac 12 +\frac{\phi}{2\pi}\Big)}{\Gamma \Big(\frac 12 -\frac{\phi}{2\pi}\Big)}+\mathcal{O}\big(N^{-1}\big).
\end{equation}
This indeed yields \eqref{eq:LBFaFinal}.

\begin{thebibliography}{10}

\bibitem{VPRK97}
V.~Vedral, M.B. Plenio, M.A. Rippin, and P.L. Knight.
\newblock Quantifying entanglement.
\newblock {\em Phys.~Rev.~Lett.}, 78:2275, 1997.
\newblock \href{https://arxiv.org/abs/quant-ph/9702027}{\textsf{arXiv:9702027
  [quant-ph]}}.

\bibitem{VP98}
V.~Vedral and M.B. Plenio.
\newblock Entanglement measures and purification procedures.
\newblock {\em Phys.~Rev.~A}, 57:1619, 1998.
\newblock \href{https://arxiv.org/abs/quant-ph/9707035}{\textsf{arXiv:9707035
  [quant-ph]}}.

\bibitem{OAFF02}
A.~Osterloh, L.~Amico, G.~Falci, and R.~Fazio.
\newblock {Scaling of entanglement close to a quantum phase transition}.
\newblock {\em Nature}, 416:608--610, 2002.
\newblock \href{http://arxiv.org/abs/quant-ph/0202029}{\textsf{arXiv:0202029
  [quant-ph]}}.

\bibitem{ON02}
T.J. Osborne and M.A. Nielsen.
\newblock {Entanglement in a simple quantum phase transition}.
\newblock {\em Phys.~Rev.~A}, 66:032110, 2002.
\newblock \href{http://arxiv.org/abs/quant-ph/0202162}{\textsf{arXiv:0202162
  [quant-ph]}}.

\bibitem{VLRK03}
G.~Vidal, J.I. Latorre, E.~Rico, and A.~Kitaev.
\newblock {Entanglement in quantum critical phenomena}.
\newblock {\em Phys.~Rev.~Lett.}, 90:227902, 2003.
\newblock \href{http://arxiv.org/abs/quant-ph/0211074}{\textsf{arXiv:0211074
  [quant-ph]}}.

\bibitem{HLW94}
C.~Holzhey, F.~Larsen, and F.~Wilczek.
\newblock Geometric and renormalized entropy in conformal field theory.
\newblock {\em Nucl.~Phys.~B}, 424:443--467, 1994.
\newblock \href{https://arxiv.org/abs/hep-th/9403108}{\textsf{arXiv:9403108
  [hep-th]}}.

\bibitem{CC04}
P.~Calabrese and J.L. Cardy.
\newblock {Entanglement entropy and quantum field theory}.
\newblock {\em J.~Stat.~Mech.}, 2004:P06002, 2004.
\newblock \href{http://arxiv.org/abs/hep-th/0405152}{\textsf{arXiv:0405152
  [hep-th]}}.

\bibitem{AFOV08}
L.~Amico, R.~Fazio, A.~Osterloh, and V.~Vedral.
\newblock {Entanglement in many-body systems}.
\newblock {\em Rev.~Mod.~Phys.}, 80:517, 2008.
\newblock \href{http://arxiv.org/abs/quant-ph/0703044}{\textsf{arXiv:0703044
  [quant-ph]}}.

\bibitem{ECP08}
J.~Eisert, M.~Cramer, and M.B. Plenio.
\newblock {Area laws for the entanglement entropy -- a review}.
\newblock {\em Rev.~Mod.~Phys.}, 832:277, 2010.
\newblock \href{http://arxiv.org/abs/0808.3773}{\textsf{arXiv:0808.3773
  [quant-ph]}}.

\bibitem{CCD09}
P.~Calabrese, J.L. Cardy, and B.~Doyon.
\newblock {Entanglement entropy in extended quantum systems}.
\newblock {\em J.~Phys.~A: Math.~Theor.}, 42:500301, 2009.

\bibitem{LR10}
J.I. Latorre and A.~Riera.
\newblock {A short review on entanglement in quantum spin systems}.
\newblock {\em J.~Phys.~A: Math.~Theor.}, 42:504002, 2009.
\newblock \href{http://arxiv.org/abs/0906.1499}{\textsf{arXiv:0906.1499
  [cond-mat.stat-mech]}}.

\bibitem{CC05}
P.~Calabrese and J.L. Cardy.
\newblock {Evolution of entanglement entropy in one-dimensional systems}.
\newblock {\em J.~Stat.~Mech.}, 2005:P04010, 2005.
\newblock \href{http://arxiv.org/abs/cond-mat/0503393}{\textsf{arXiv:0503393
  [cond-mat.stat-mech]}}.

\bibitem{FC08}
M.~Fagotti and P.~Calabrese.
\newblock {Evolution of entanglement entropy following a quantum quench:
  Analytic results for the XY chain in a transverse magnetic field}.
\newblock {\em Phys.~Rev.~A}, 78:010306, 2008.
\newblock \href{https://arxiv.org/abs/0804.3559}{\textsf{arXiv:0804.3559
  [cond-mat.stat-mech]}}.

\bibitem{AC17}
V.~Alba and P.~Calabrese.
\newblock Entanglement and thermodynamics after a quantum quench in integrable
  systems.
\newblock {\em Proc.~Natl. Acad.~Sci.}, 114:7947--7951, 2017.
\newblock \href{https://arxiv.org/abs/1608.00614}{\textsf{arXiv:1608.00614
  [cond-mat.str-el]}}.

\bibitem{ABF19}
V.~Alba, B~Bertini, and M.~Fagotti.
\newblock Entanglement evolution and generalised hydrodynamics: interacting
  integrable systems.
\newblock {\em SciPost Phys.}, 7, 2019.
\newblock \href{https://arxiv.org/abs/1903.00467}{\textsf{arXiv:1903.00467
  [cond-mat.stat-mech]}}.

\bibitem{vN55}
J.~von Neumann.
\newblock {\em {Mathematische Grundlagen der Quantenmechanik}}.
\newblock Princeton University Press, 1955.

\bibitem{S93}
M.~Srednicki.
\newblock {Entropy and area}.
\newblock {\em Phys.~Rev.~Lett.}, 71:666, 1993.
\newblock \href{http://arxiv.org/abs/hep-th/9303048}{\textsf{arXiv:9303048
  [hep-th]}}.

\bibitem{J94}
R.~Jozsa.
\newblock {Fidelity for Mixed Quantum States}.
\newblock {\em J.~Mod.~Opt.}, 41(12):2315--2323, 1994.

\bibitem{ZP06}
P.~Zanardi and N.~Paunkovi\`c.
\newblock {Ground state overlap and quantum phase transitions}.
\newblock {\em Phys.~Rev.~E}, 74:031123, 2006.
\newblock \href{http://arxiv.org/abs/quant-ph/0512249}{\textsf{arXiv:0512249
  [quant-ph]}}.

\bibitem{ZB08}
H.-Q. Zhou and J.P. Barjaktarevi{\v c}.
\newblock {Fidelity and quantum phase transitions}.
\newblock {\em J.~Phys.~A: Math.~Theor.}, 41:412001, 2008.
\newblock \href{http://arxiv.org/abs/cond-mat/0701608}{\textsf{arXiv:0701608
  [cond-mat.stat-mech]}}.

\bibitem{S10}
J.~Sirker.
\newblock {Finite temperature fidelity susceptibility for one-dimensional
  quantum systems}.
\newblock {\em Phys.~Rev.~Lett.}, 105:117203, 2010.
\newblock \href{http://arxiv.org/abs/1006.2522}{\textsf{arXiv:1006.2522
  [cond-mat.str-el]}}.

\bibitem{G10}
S.-J. Gu.
\newblock {Fidelity approach to quantum phase transitions}.
\newblock {\em Int.~J.~Mod.~Phys.~B}, 24:4371, 2010.
\newblock \href{http://arxiv.org/abs/0811.3127}{\textsf{arXiv:0811.3127
  [quant-ph]}}.

\bibitem{DS11}
J.~Dubail and J.-M. St\'ephan.
\newblock {Universal behavior of a bipartite fidelity at quantum criticality}.
\newblock {\em J.~Stat.~Mech.}, 2011:L03002, 2011.
\newblock \href{http://arxiv.org/abs/1010.3716}{\textsf{arXiv:1010.3716
  [cond-mat.str-el]}}.

\bibitem{SD13}
J.-M. St\'ephan and J.~Dubail.
\newblock {Logarithmic corrections to the free energy from sharp corners with
  angle $2\pi$}.
\newblock {\em J.~Stat.~Mech.}, 2013:P09002, 2013.
\newblock \href{http://arxiv.org/abs/1303.3633}{\textsf{arXiv:1303.3633
  [cond-mat.stat-mech]}}.

\bibitem{Baxterbook}
R.J. Baxter.
\newblock {\em {Exactly Solved Models in Statistical Mechanics}}.
\newblock Academic Press, 1982.

\bibitem{HL17}
C.~Hagendorf and J.~Li\'enardy.
\newblock {Open spin chains with dynamic lattice supersymmetry}.
\newblock {\em J.~Phys.~A: Math.~Theor.}, 50:185202, 2017.
\newblock \href{http://arxiv.org/abs/1612.02951}{\textsf{arXiv:1612.02951
  [math-ph]}}.

\bibitem{PMDR19}
G.~Parez, A.~Morin-Duchesne, and P.~Ruelle.
\newblock Bipartite fidelity of critical dense polymers.
\newblock {\em J.~Stat.~Mech.}, 2019:103101, 2019.
\newblock \href{https://arxiv.org/abs/1902.02246}{\textsf{arXiv:1902.02246
  [cond-mat.stat-mech]}}.

\bibitem{PR07}
P.A. Pearce and J.~Rasmussen.
\newblock Solvable critical dense polymers.
\newblock {\em J.~Stat.~Mech.}, 2007:P02015, 2007.
\newblock \href{http://arxiv.org/abs/hep-th/0610273}{\textsf{arXiv:0610273
  [hep-th]}}.

\bibitem{PRV10}
P.A. Pearce, J.~Rasmussen, and S.P. Villani.
\newblock {Solvable critical dense polymers on the cylinder}.
\newblock {\em J.~Stat.~Mech.}, page P02010, 2010.
\newblock \href{https://arxiv.org/abs/0910.4444}{\textsf{arXiv:0910.4444
  [hep-th]}}.

\bibitem{CP88}
J.L. Cardy and I.~Peschel.
\newblock {Finite-size dependence of the free energy in two-dimensional
  critical systems}.
\newblock {\em Nucl.~Phys.~B}, 300:377, 1988.

\bibitem{MDJ18}
A.~Morin-Duchesne and J.L. Jacobsen.
\newblock Two-point boundary correlation functions of dense loop models.
\newblock {\em SciPost Phys.}, 4:034, 2018.
\newblock \href{https://arxiv.org/abs/1712.08657}{\textsf{arXiv:1712.08657
  [cond-mat.stat-mech]}}.

\bibitem{L91}
D.~Levy.
\newblock {Algebraic structure of translation-invariant spin-$\frac12$ XXZ and
  $q$-Potts quantum chains}.
\newblock {\em Phys.~Rev.~Lett.}, 67:1971--1974, 1991.

\bibitem{MS93}
P.~Martin and H.~Saleur.
\newblock {On an algebraic approach to higher dimensional statistical
  mechanics}.
\newblock {\em Commun.~Math.~Phys.}, 158:155--190, 1993.

\bibitem{GL98}
J.J. Graham and G.I. Lehrer.
\newblock {The representation theory of affine Temperley-Lieb algebras}.
\newblock {\em Enseign.~Math.}, 44:173--218, 1998.

\bibitem{G98}
R.M. Green.
\newblock {On representations of affine Temperley-Lieb algebras}.
\newblock {\em J.~Algebra}, 60:498--517, 1997.

\bibitem{EG99}
K.~Erdmann and R.M. Green.
\newblock {On representations of affine Temperley-Lieb algebras, II}.
\newblock {\em Pac.~J.~Math.}, 191:243--274, 1999.
\newblock \href{https://arxiv.org/abs/math/9811017}{\textsf{arXiv:math/9811017
  [math.RT]}}.

\bibitem{PS90}
V.~Pasquier and H.~Saleur.
\newblock {Common structures between finite systems and conformal field
  theories through quantum groups}.
\newblock {\em Nucl.~Phys.~B}, 330:523, 1990.

\bibitem{MDSA13}
A.~Morin-Duchesne and Y.~Saint-Aubin.
\newblock {A homomorphism between link and XXZ modules over the periodic
  Temperley-Lieb algebra}.
\newblock {\em J.~Phys.~A: Math.~Theor.}, 46:285207, 2013.
\newblock \href{https://arxiv.org/abs/1203.4996}{\textsf{arXiv:1203.4996
  [math-ph]}}.

\bibitem{MDPR13}
A.~Morin-Duchesne, P.A. Pearce, and J.~Rasmussen.
\newblock {Modular invariant partition function of critical dense polymers}.
\newblock {\em Nucl.~Phys.~B}, 874:312--357, 2013.
\newblock \href{http://arxiv.org/abs/1303.4895}{\textsf{arXiv:1303.4895
  [hep-th]}}.

\bibitem{dFHZ87}
P.~Di Francesco, H.~Saleur, and J.B. Zuber.
\newblock {Relations between the Coulomb gas picture and conformal invariance
  of two-dimensional critical models}.
\newblock {\em J.~Stat.~Phys.}, 49:57--79, 1987.

\bibitem{HQB87}
C.J. Hamer, G.R.W. Quispel, and M.T. Batchelor.
\newblock {Conformal anomaly and surface energy for Potts and Ashkin-Teller
  quantum chains}.
\newblock {\em J.~Phys.~A: Math.~Gen.}, 20:5677--5693, 1987.

\bibitem{KBP91}
A.~Kl\"umper, M.T. Batchelor, and P.A. Pearce.
\newblock {Central charges of the $6$- and $19$-vertex models with twisted
  boundary conditions}.
\newblock {\em J.~Phys.~A: Math.~Gen.}, 24:3111--3133, 1991.

\bibitem{MDJ19}
A.~Morin-Duchesne and J.L. Jacobsen.
\newblock Logarithmic correlation functions for critical dense polymers on the
  cylinder.
\newblock {\em SciPost Phys.}, 7:70, 2019.
\newblock \href{https://arxiv.org/abs/1907.05499}{\textsf{arXiv:1907.05499
  [cond-mat.stat-mech]}}.

\bibitem{N84}
B.~Nienhuis.
\newblock {Critical behavior of two-dimensional spin models and charge
  asymmetry in the Coulomb gas}.
\newblock {\em J.~Stat.~Phys.}, 34:731--761, 1984.

\bibitem{TL71}
H.~Temperley and E.~Lieb.
\newblock {Relations between the ``percolation'' and ``colouring'' problem and
  other graph-theoretical problems associated with regular planar lattices:
  Some exact results for the ``percolation'' problem}.
\newblock {\em Proc.~Roy.~Soc.~Lond.~A}, 322:251, 1971.

\bibitem{J83}
V.F.R. Jones.
\newblock {Index for subfactors}.
\newblock {\em Invent.~Math.}, 72:1, 1983.

\bibitem{M91}
P.~Martin.
\newblock {\em {Potts models and related problems in statistical mechanics}}.
\newblock World Scientific, 1991.

\bibitem{GW93}
F.~Goodman and H.~Wenzl.
\newblock {The Temperley-Lieb algebra at roots of unity}.
\newblock {\em Pacific J.~Math.}, 161:307, 1993.

\bibitem{W95}
B.W. Westbury.
\newblock {The representation theory of the Temperley-Lieb algebras}.
\newblock {\em Math.~Zeit.}, 219:539, 1995.

\bibitem{RSA14}
D.~Ridout and Y.~Saint-Aubin.
\newblock Standard modules, induction and the {Temperley}-{Lieb} algebra.
\newblock {\em Adv.~Theor.~Math.~Phys.}, 18:957, 2014.
\newblock \href{http://arxiv.org/abs/1204.4505}{\textsf{arXiv:1204.4505
  [\mbox{hep-th}]}}.

\bibitem{MDRR16}
A.~Morin-Duchesne, J.~Rasmussen, and P.~Ruelle.
\newblock {Integrability and conformal data of the dimer model}.
\newblock {\em J.~Phys.~A: Math.~Theor.}, 49:174002, 2016.
\newblock \href{http://arxiv.org/abs/1507.04193}{\textsf{arXiv:1507.04193
  [hep-th]}}.

\bibitem{K78}
L.P. Kadanoff.
\newblock {Lattice Coulomb gas representations of two-dimensional problems}.
\newblock {\em J.~Phys.~A: Math.~Gen.}, 11:1399, 1978.

\bibitem{YellowCFTbook}
P.~Di Francesco, P.~Mathieu, and D.~S\'en\'echal.
\newblock {\em {Conformal field theory}}.
\newblock Springer, 1997.

\bibitem{EI15}
B.~Estienne and Y.~Ikhlef.
\newblock {Correlation functions in loop models}.
\newblock 2015.
\newblock \href{https://arxiv.org/abs/1505.00585}{\textsf{arXiv:1505.00585 
  [math-ph]}}.


\end{thebibliography}

\end{document}